\documentclass[lettersize,journal]{IEEEtran}
\usepackage{amsmath,amsfonts}
\usepackage{algorithm}
\usepackage{algpseudocode}

\usepackage{subeqnarray}
\usepackage{array}
\usepackage[caption=false,font=normalsize,labelfont=sf,textfont=sf]{subfig}
\usepackage{tabularx}      % 引入 tabularx 包
\usepackage{textcomp}
\usepackage{stfloats}
\usepackage{url}
\usepackage{verbatim}
\usepackage{graphicx}
\usepackage{cite}
\usepackage{bm}
\usepackage{dsfont}
\usepackage{pifont}
\usepackage{amssymb}
\usepackage{caption} 
\usepackage{comment} 
\usepackage{makecell}  
\usepackage{xcolor} % 加载xcolor宏包，用于定义颜色  
\usepackage{colortbl} % 加载colortbl宏包，用于给表格添加颜色  
\usepackage{booktabs} 
\usepackage{cellspace}  
\usepackage{multirow}
\usepackage{diagbox}
\renewcommand{\algorithmicrequire}{\textbf{Given:}} 
\renewcommand{\algorithmicensure}{\textbf{Return:}}

\newcommand{\RNum}[1]{\uppercase\expandafter{\romannumeral #1\relax}}	

\renewcommand{\algorithmicrequire}{\textbf{Input:}}
\renewcommand{\algorithmicensure}{\textbf{Output:}}
\begin{document}

\title{Precoding Matrix Indicator in the 5G NR Protocol:\\ A Tutorial on 3GPP Beamforming Codebooks}
\author{Boyu Ning, Haifan Yin, Sixu Liu, Hao Deng, Songjie Yang, Yuchen Zhang, Weidong Mei,\\
David Gesbert,  Jaebum Park, Robert W. Heath Jr., and Emil Bj{\"o}rnson

\thanks{The work of Weidong Mei was supported in part by the National Key Research and Development Program of China under Grant 2024YFB2907900, and in part by the Natural Science Foundation of Sichuan Province under Grant 2025ZNSFSC0514. The work of David Gesbert was partially supported by the 3IA Côte d’Azur Investments (ANR) with reference number ANR-23-IACL-0001. The work of Emil Bj{\"o}rnson was supported by the Grant 2022-04222 from the Swedish Research Council.
}

}        

% The paper headers
\markboth{IEEE COMMUNICATIONS SURVEYS \& TUTORIALS, January, 2026}%
{Shell \MakeLowercase{\textit{et al.}}: A Sample Article Using IEEEtran.cls for IEEE Journals}

\maketitle

\begin{abstract}
The evolution of 5G-advanced (5G-A) systems relies heavily on advanced beamforming technologies to achieve high spectral efficiency and network capacity. Although abundant theoretical research has been devoted to optimizing beamforming based on full channel state information (CSI), a major gap remains with the codebook-based beamforming defined in 3rd Generation Partnership Project (3GPP) standards for practical deployment. This disconnect is exacerbated by the often complex and elusive nature of the protocol descriptions themselves. This paper bridges this critical gap by providing a systematic examination of the beamforming codebook technology, i.e., precoding matrix indicator (PMI), in the 5G NR from theoretical, standardization, and implementation perspectives. We begin by introducing the background of beamforming in multiple-input multiple-output (MIMO) systems and the signaling procedures for codebook-based beamforming in practical 5G systems. Then, we establish the fundamentals of regular codebooks and port-selection codebooks in 3GPP standards. Next, we provide rigorous technical analysis of 3GPP codebook evolution spanning Releases 15-18, with particular focus on:  1) We elucidate the core principles underlying codebook design, 2) provide clear physical interpretations for each symbolic variable in the codebook formulas, summarized in tabular form, and 3) offer intuitive visual illustrations to explain how codebook parameters convey information. These essential pedagogical elements are almost entirely absent in the often-obscure standardization documents.  Through mathematical modeling, performance benchmarking, feedback comparisons, and scenario-dependent applicability analysis, we provide researchers and engineers with a unified understanding of beamforming codebooks in real-world systems. Furthermore, we identify future directions and other beamforming scenarios for ongoing research and development efforts. This work serves as both an informative tutorial and a guidance for future research, facilitating more effective collaboration between academia and industry in advancing wireless communication technologies.
\end{abstract}

\section{Introduction}
The deployment of the fifth-generation (5G)-advanced (5G-A) wireless communication systems has brought about transformative changes in the telecommunications landscape, with hallmark technologies such as multiple-input multiple-output (MIMO) playing a pivotal role in achieving the promised high throughput rates, low latency, and high reliability\cite{Heath2018mimo, 5GATutorial}. Massive MIMO, characterized by the use of a large number of antennas at the base station (BS)\cite{New5G}, has emerged as a cornerstone of 5G and 5G-A new-radio (NR) systems, offering significant improvements in spectral efficiency \cite{Emilmimo}. By leveraging spatial diversity and advanced signal processing techniques, massive MIMO enables the simultaneous transmission of multiple data streams to multiple users, thereby significantly enhancing network capacity and user experience\cite{rappaportmmWavebook, rappaportmmWave5G, emilten,shaomimo}. 
There is hope that enhancements in the operation in the millimeter-wave frequency band will make practical deployments more attractive\cite{mmWaveSuv}, but it is coupled with major challenges for the RF architecture and algorithmic processing. Firstly, in order to reduce hardware overhead in devices, a compromise solution has been proposed in the form of a hybrid digital and analog architecture\cite{heathcova}.  In addition to minimizing the overhead of channel estimation and ensuring stable initial access, beam management techniques have been developed\cite{BeamManagementSuv}. These techniques aim to manage initialized beams to balance communication performance and overhead\cite{heathmmwave}. To reduce the computational overhead of beamforming, a plethora of low-complexity algorithms, such as compressed sensing algorithms \cite{Heath2015ICASSP, HeathOMP}, have been introduced. These algorithms aim to efficiently handle high-dimensional data by exploiting the signal's sparse nature in some domain \cite{HeathSIP}.  In the future, wireless communication systems are expected to adopt larger-scale multi-antenna technology\cite{WangZheMag}, such as Gigantic MIMO in the upper mid-band\cite{GiganticMIMO} and
ultra massive MIMO in terahertz bands\cite{ningmimo}, in conjunction with advanced channel estimation and beamforming techniques to effectively utilize the spatial domain to enhance communication quality and capacity. These techniques allow the system to focus transmitted energy in the direction of intended users while minimizing interference, thus optimizing the overall network performance\cite{WangzheSuv}.

The choice between time-division duplexing (TDD) and frequency-division duplexing (FDD) schemes in wireless communication systems has a profound impact on channel estimation and beamforming. In TDD bands, one can leverage channel reciprocity to directly estimate the downlink channel from uplink channel measurements, thereby reducing the overhead associated with channel state information (CSI) acquisition \cite{Marzetta2010TDD,TDD}. This reciprocity is particularly advantageous in scenarios with good uplink signal-to-noise ratio (SNR) because one can then acquire high-quality CSI using the theoretically minimal amount of uplink reference signaling. In contrast, FDD systems rely on the user equipment (UE) to estimate the downlink channel based on downlink reference signaling and feedback implicit channel information, such as channel quality indicator (CQI), precoding matrix indicator (PMI), and rank indicator (RI)\cite{choi2014FDD, FDD1}. Among these, PMI, also known as the \emph{beamforming codebook}, plays a critical role in downlink beamforming by providing the BS with information about the preferred precoding matrix for transmission\cite{FDD2}. In academic research, the term ``beamforming codebook'' typically refers to a set of predefined beamforming vectors or beams, where UEs feed back the index of the best predefined beam to the BS. This approach simplifies the feedback process by limiting the amount of information that needs to be transmitted, but it also imposes constraints on the flexibility and adaptability of the beamforming process\cite{DLovefeedback}. However, the beamforming codebook specified in the 3rd Generation Partnership Project (3GPP) Release 18, which marks the initiation of 5G-A\cite{5gA2}, builds upon the foundational frameworks established in Releases 15-17 \cite{Release15,Release16,Release17}, delivering a more advanced and flexible feedback framework for downlink beamforming. It encompasses a sophisticated mechanism for reconstructing beamforming vectors, evolving from Type I to Type II codebooks, by the combination of spatial, spectral, and temporal bases\cite{38.211,38.214}. This evolution reflects the increasing complexity and capability of 5G systems to handle diverse deployment scenarios and user requirements, enabling more efficient and effective communication in a wide range of environments.

Despite the widespread adoption of PMI technologies in real-world MIMO systems, there exists a significant gap between the theoretical study of codebooks in academia and their practical implementation in industry standards. This is because the description of PMI in the 5G NR protocol is elusive, lacking detailed insights and explanations. While some scholars have provided detailed interpretations of the standards and analyzed the function of codebooks, their work has not yet garnered widespread attention in the academic community. For instance, the authors in \cite{heathMag} offered a comprehensive overview of MIMO across mobile standards, spotlighting the novel beam-based feedback system in 5G NR. They also expounded on how beamforming, codebooks, and feedback enable the deployment of larger arrays. In addition, the authors in \cite{YinChina} traced the evolution of the codebook in the 5G NR standard, briefly explaining their motivation, the corresponding feedback mechanism, and the format of the indicators. Unfortunately, despite these efforts, these studies have yet to stir significant interest or discussion within the scholarly society, and academic papers often focus on idealized scenarios and simplified models. This gap can hinder the effective translation of research findings into practical implementations, as the practical challenges and constraints faced by industry practitioners are not always adequately addressed in academic studies. 

Currently, there is no comprehensive tutorial or holistic overview of beamforming codebooks that effectively bridges the gap between theory and practice. This lack of clarity motivates a thorough exploration of PMI and its evolution in 5G and 5G-A. Only by closing this gap can academia and industry collaboratively develop beamforming codebooks for sixth-generation (6G) systems that can deliver the MIMO advantages demonstrated in visionary work on the MIMO evolution. This paper aims to address these challenges by providing a comprehensive tutorial on beamforming codebooks, with a focus on their evolution, design principles, and practical applications in 5G and 5G-A systems. The specific contributions of this work are as follows:
\begin{enumerate}
    \item \emph{Accurate Synopsis on Beamforming}: We provide a concise summary of beamforming techniques, spanning from full CSI-based beamforming to codebook-based beamforming. We introduce the basics of orthogonal frequency division multiplexing (OFDM) MIMO systems and systematically summarize classical linear precoding schemes. By reviewing this tutorial, readers will gain insights into state-of-the-art beamforming techniques and efficiently replicate simulations in future research works by referring to Appendix \ref{appdA}.
   
    \item \emph{Bridging Academia and Industry}: We compare codebook-based beamforming schemes discussed in academia and industry, highlighting the significance of PMI codebooks and their practical implementation in the 5G NR systems. This includes an explanation of standardized procedures in Appendix \ref{resource} and \ref{reporting} such as CSI-RS resource allocation for channel state acquisition and reporting configurations defining periodicity and triggering mechanisms. We further dissect the relationship between codebook matrix structures and practical antenna configurations. 
    
    \item \emph{Insightful Protocol Analysis}: We offer a clear and insightful analysis of the PMI descriptions in 5G NR, from R15 to R18, as well as Type I to Type II, demystifying the often-elusive terminology and providing a deeper understanding of their functionality. To facilitate the comprehension of the 3GPP specifications, some tables from the original documents are explicitly retained. However, different from directly reading the protocol text, this tutorial has been methodically reorganized with a clearer logical narrative. This is supplemented by extensive original figures that elucidate the physical principles, along with precise parameter interpretations. Readers can quickly grasp the core concepts of codebooks in Sec. \ref{funds} and find concise mathematical modeling and comparative analysis of the codebooks in Sec. \ref{comparisonandapp}. Detailed discussions can be navigated via the subsection headings provided in Sec. \ref{SecPMI}.

    \item \emph{Holistic Tutorial and Future Directions}: This work serves as a comprehensive tutorial for researchers, engineers, and practitioners, offering a unified perspective on PMI and its critical role in massive MIMO systems. Furthermore, we outline future research directions for beamforming codebooks, encompassing the design of their bases, the overarching framework, hierarchical structures, and fully analog codebooks. Beyond these refinements, we also envision the application of codebook technology to new beamforming scenarios, emerging MIMO-related functionalities, novel frequency bands, and advanced array geometries in future 6G systems.
\end{enumerate}

The rest of the paper is organized as follows. Sec. \ref{Background} introduces the background on beamforming in MIMO systems. Sec. \ref{3gppcodebook} introduces the background on codebook-based beamforming within the 3GPP standard. Sec. \ref{funds} covers the codebook fundamentals, including both regular and port-selection codebooks. Sec. \ref{SecPMI} provides a detailed account of the beamforming codebook evolution, encompassing various codebooks from R15 to R18. Sec. \ref{comparisonandapp} discusses the various 5G codebooks in terms of compact model, performance, feedback overhead, and applicable scenarios. Future research directions and other scenarios are discussed in Sec. \ref{futuredirection}. Finally, we conclude the paper in Sec. \ref{conclusion}. 

\begingroup
\allowdisplaybreaks

\newcounter{MYtempeqncn}
\begin{figure*}[!b]
\normalsize
\setcounter{MYtempeqncn}{\value{equation}}
\setcounter{equation}{1}
\hrulefill
\begin{equation}\label{ar-per-ue}
R_k= \sum_{m=1}^{M} \log_2\det\left[\mathbf{I}_{N_r}+\left(\sum_{i\neq k}\mathbf{H}_i\left[m\right]\mathbf{W}_i\left[m\right]\mathbf{W}_i^H\left[m\right]\mathbf{H}_i^H\left[m\right]+\sigma_k^2\mathbf{I}_{N_r}\right)^{-1}\mathbf{H}_k\left[m\right]\mathbf{W}_k\left[m\right]\mathbf{W}_k^H\left[m\right]\mathbf{H}_k^H\left[m\right]\right]
\end{equation}
\setcounter{equation}{\value{MYtempeqncn}}
\end{figure*} 

\section{Background on Beamforming}\label{Background}

MIMO communication systems make use of beamforming to enhance the received signal power, multiplex data streams with different spatial directivity, and protect those data streams from mutual interference. The beamforming design is instrumental in making efficient use of the spatial degrees of freedom offered by MIMO, making it worthwhile to use antenna arrays at both the transmitters and receivers in practical systems.
 In this section, focusing on downlink transmission, we present a concise overview of prevalent beamforming schemes. Based on the schemes of CSI acquisition, these beamforming schemes are broadly categorized into a) Full CSI-based beamforming and b) codebook-based beamforming. We first introduce the system model of OFDM MIMO, underlining the role of multi-user beamforming on the achievable rates. Following this, we will delve into both full CSI-based beamforming and codebook-based beamforming, with the latter incorporating the use of codebook technology.

\subsection{Downlink OFDM  MIMO Systems}
Consider a downlink OFDM MIMO system with $M$ subcarriers, where a BS is equipped with $N_t$ antennas to serve $K$ users, each with $N_r$ antennas. The received signal at the $k$-th user on the $m$-th subcarrier can be written as
\begin{align}\label{mu-re-sig}
\mathbf{y}_k\left[m\right] = & \mathbf{H}_k\left[m\right] \mathbf{W}_k\left[m\right] \mathbf{s}_k\left[m\right] \notag \\
& + \underbrace{\sum_{i=1, i \neq k}^K \mathbf{H}_k\left[m\right] \mathbf{W}_i\left[m\right] \mathbf{s}_i\left[m\right]}_{\text{Inter-user interference}} + \mathbf{n}_k\left[m\right],
\end{align}
where $\mathbf{H}_k[m] \in \mathbb{C}^{N_r \times N_t}$ is the channel matrix between the BS and the $k$-th user, ${\mathbf{s}}_k[m] \sim \mathcal{CN}(\mathbf{0}, \mathbf{I}_{v_k})$ denotes the data stream intended for the $k$-th user, with $v_k$ being the number of data streams\footnote{The spatially multiplexed data streams are called ``spatial layers'' in 3GPP.}, $\mathbf{W}_k[m] \in \mathbb{C}^{N_t \times v_k}$ is the beamforming matrix for transmitting ${\mathbf{s}}_k[m]$, and $\mathbf{n}_k[m] \sim \mathcal{CN}(\mathbf{0}, \sigma_k^2 \mathbf{I}_{N_r})$ represents the additive white complex Gaussian noise (AWGN) with $\sigma_k^2$ being the noise power. Furthermore, it is assumed that the transmitted data vectors for different users are independent of each other, as are the noise vectors.

The achievable rate of the $k$-th user, accumulated over $M$ subcarriers, is expressed in \eqref{ar-per-ue} at the bottom of this page. This rate is achieved by the assumed linear beamforming, treating inter-user interference as additional noise, and using channel codes developed for AWGN channels.
On each subcarrier, the BS can employ beamforming techniques to either: 1) better match a single data stream to the channel characteristics, or 2) simultaneously transmit multiple data streams while maintaining inter-stream orthogonality. In both cases, the beamforming must also be designed to limit the inter-user interference.
In the following, we will analyze the beamforming design for each subcarrier independently, and thus omit the subcarrier index for notational simplicity.

\subsection{Full CSI-Based Beamforming}

Downlink CSI is essential for the BS in designing beamforming strategies that enable the transmission of multiple data streams while controlling mutual interference.
In TDD systems, uplink-downlink channel reciprocity allows the BS to derive downlink CSI from uplink pilot estimates, e.g., using sounding reference signal (SRS) in the 3GPP protocol, enabling direct computation of the most desirable beamforming vectors. This approach relies on accurate channel estimation and perfect calibration between uplink and downlink hardware chains. With the acquired CSI, beamforming can be optimized at the BS to maximize the achievable sum data rate, e.g., by running the weighted minimum mean-square error (WMMSE) algorithm \cite{wmmse}. 

In this case, the UE transmits uplink pilot signals, and no explicit feedback is required. However, as the BS's beamforming is not constrained by protocols and is independent of the codebooks, it is necessary to send downlink demodulation pilot signals using the selected beamforming matrices so each UE can estimate the product $\mathbf{H}_k\left[m\right] \mathbf{W}_k\left[m\right]$ between its channel and beamforming matrix. This pilot is known as demodulation reference signal (DMRS) in 3GPP. While CSI-based beamforming is not the primary focus of this tutorial, it is a significant component of MIMO systems and generally recommended by textbooks on massive MIMO \cite{Marzetta2016a,massivemimobook}. We have summarized several representative beamforming methods for both single-user MIMO (SU-MIMO) and multi-user MIMO (MU-MIMO) systems in Appendix A, which interested readers are encouraged to read.

\subsection{Codebook-Based Beamforming}

In FDD systems, the absence of channel reciprocity due to frequency band asymmetry necessitates explicit CSI feedback from each UE, incurring a large overhead for massive MIMO or mmWave systems with large antenna arrays. It can also be desirable to use alternative methods in TDD systems due to hardware calibration errors, low SNR on the uplink pilots (due to the substantially lower uplink power), or dynamic environments that require more frequent pilot signaling than the 3GPP protocol permits. To address these limitations, codebook-based beamforming emerges as a pragmatic alternative. Instead of explicit CSI feedback, the UEs determine an effective precoding vector from a predefined codebook and feed back its corresponding indices, which drastically reduces the feedback overhead if the codebook is compact. This framework circumvents the need for full channel knowledge at the BS while maintaining reasonable beamforming performance, making it indispensable for 5G-A and beyond, particularly in scenarios with limited CSI acquisition capabilities.
However, the key to success is that the codebook design provides efficient CSI compression.

There are three primary mechanisms that underpin codebook-based beamforming, i.e., exhaustive beam training, multi-stage beam training, and codebook indicator feedback.
\subsubsection{Exhaustive Beam Training}  
This method finds the optimal beam alignment by testing all predefined beams in a codebook. The BS sequentially transmits pilots using each beam, and the UE measures the received signal energy for each one. The UE then selects the beam index with the highest energy and feeds it back to the BS. While straightforward and robust, this process requires testing all beams, leading to significant training overhead.

\subsubsection{Multi-stage Beam Training}  
This approach reduces training overhead by using a hierarchical codebook with multiple levels of beam resolution. The process starts with wide beams that cover large areas in the initial stage. Based on the UE's feedback from each stage, the BS and UE progressively narrow the search space by testing finer, more directional beams in subsequent stages. This iterative refinement avoids an exhaustive search with narrow beams from the beginning, making the beam training process more efficient.

\subsubsection{Codebook Indicator Feedback}  
In this mechanism, both the BS and UE share a predefined, structured codebook of precoding matrices. The BS transmits pilot signals, and the UE estimates the channel to select the optimal precoding matrix from this codebook. Instead of sending full channel information, the UE only feeds back the index of this matrix, known as the Precoding Matrix Indicator (PMI). This leverages channel sparsity and structured codebooks to minimize feedback overhead, making it highly efficient for systems with limited feedback resources\cite{huang2006performance, mo2015limited, inoue2007kerdock}. 

Each mechanism introduces distinct research challenges. For \emph{exhaustive beam training}, the design of beam patterns and solutions is critical. Far-field systems adopt discrete fourier transform (DFT)-based beams for full spatial coverage \cite{xiaocodebook, dreifuerst2022massive}, while near-field scenarios require distance-aware ``beam focusing'' to address spherical wavefronts \cite{nearfild,kangjian,XL2}. Wideband systems must mitigate beam squint via frequency-dependent beam codebooks \cite{ningcodebookwide,codebooksquint,chensquint}, and hybrid analog-digital architectures impose constant-modulus constraints on beamformer design \cite{ningexh,hybridcode,chenhybrid}. \emph{Multi-stage beam training} focuses on protocol optimization, including parallel training\cite{paralel},  hierarchical training \cite{qichenhao}, adaptive training\cite{adaptive}, one-sided training\cite{oneside}, etc. For \emph{codebook indicator feedback}, the design remains underexplored in academia, despite their widespread adoption in 3GPP standards.

\section{Background on 3GPP Standard}\label{3gppcodebook}

This section provides the background and context for codebook-based beamforming within the 3GPP standard.  As shown in Fig. \ref{flowchart}, the BS---called the next generation nodeB (gNB)---initiates the process by sending resource configuration information and reporting configuration information to the UE through radio resource control (RRC) signals. Resource configuration information pertains to the details of the measurement resources and is set up in the protocol using a three-level structure: \emph{CSI-ResourceConfig, ResourceSet}, and \emph{Resource}\footnote{The information elements defined in the protocol are italicized.}. Reporting configuration information refers to the details concerning the reporting of measurement results, which is configured  via \emph{CSI-ReportConfig}. Then, the gNB sends CSI-RS, and the UE measures the quality of the wireless channel based on CSI-RS and performs channel estimation. These measurements are essential for the UE to determine the best beamforming vectors for the current channel conditions. The UE then provides feedback to the gNB, with the beamforming information in the form of a PMI. This feedback enables the gNB to construct beamforming vectors based on the codebook structure for subsequent transmissions. 

\begin{figure}[t]
	\centering
	\includegraphics[width=0.45\textwidth]{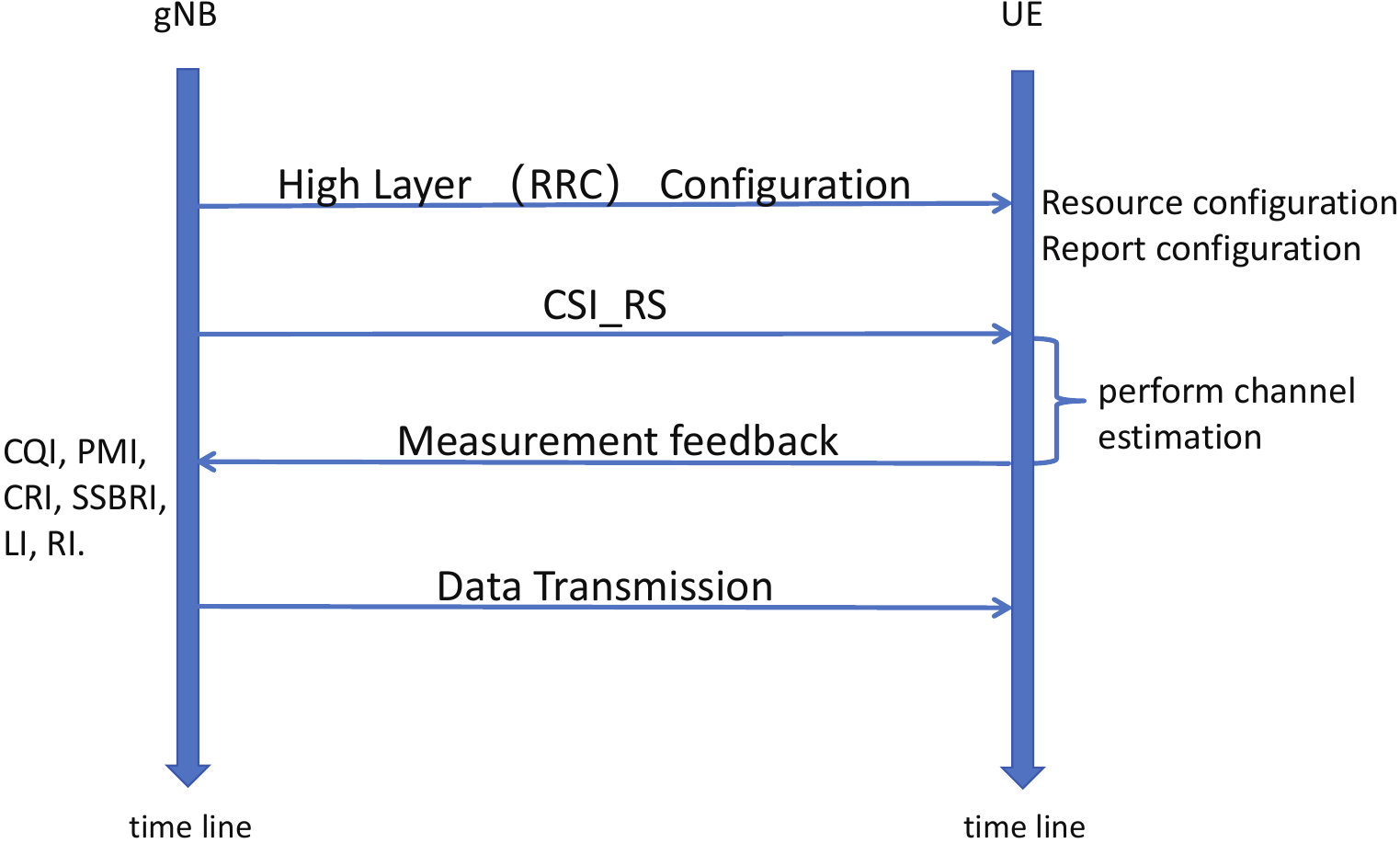}
 
	\caption{CSI-RS measurement and feedback flowchart.}
	\label{flowchart}
\end{figure}
\subsection{CSI-RS Measurement and Feedback} \label{flow}
In 5G NR, CSI-RS measurement and PMI feedback are critical components for enhancing the performance of massive MIMO systems. The general flowchart that outlines the CSI-RS measurement and PMI feedback process is depicted in Fig. \ref{flowchart}, which can be summarized as follows.

(i)  {\emph{High Layer Configuration (RRC Configuration)}}: The gNB sends an RRC configuration to the UE. This configuration includes the resource configuration for CSI-RS and the reporting configuration, which specifies the CSI-RS signals and pilot periodicity used by the gNB, and how/when the UE should report measurements back to the gNB.

(ii)  {\emph{Transmission of CSI-RS}}: The gNB transmits CSI-RS to the UE. These reference signals are used by the UE to measure the conditions of the wireless channel. The configuration of the CSI-RS, such as periodicity and time-frequency resources mapping, is based on the RRC configuration previously sent.

(iii) {\emph{Measurement Feedback}}: The UE measures the conditions of wireless channel using the CSI-RS and compiles a report that may include 1) channel quality indicator (CQI) that indicates the quality of the channel; 2) PMI that suggests the best precoding matrix for spatial multiplexing; 3) CSI-RS resource indicator (CRI) that identifies the CSI-RS resources used for the measurements; 4) SS/PBCH block resource indicator (SSBRI) that indicates the associated synchronization signal block (SSB) for cell access; 5) Layer indicator (LI) that indicates which column of the precoder matrix of the reported PMI corresponds to the strongest stream; 6) Rank indicator (RI) that indicates the effective rank of the channel matrix, which corresponds to the number of spatial streams the channel can support. The UE sends this measurement feedback to the gNB via the physical uplink control channel (PUCCH) or physical uplink shared channel (PUSCH). Typically, control information like CQI/PMI is sent via PUCCH, but if the UE also has data to send, it may be included in PUSCH.

(iv) {\emph{Downlink Data Transmission}}: Based on the feedback received from the UE, the gNB adjusts its transmission parameters and may use the selected precoding matrix, i.e., codebook-based beamforming vectors, to send data to the UE via the physical downlink shared channel (PDSCH). This ensures that the data is transmitted using the efficient beamforming for the current channel conditions.

The CSI-RS configuration in 5G NR systems involves a multi‑layered procedure. The CSI‑RS reporting framework bundles the configured resources into resource sets, indicating which resources the UE should measure and how to report the results. These protocol configurations do not affect the codebook technique itself. Readers interested in these details are advised to refer to Appendices \ref{resource} and \ref{reporting} when needed.

\subsection{Codebook Matrix and Antenna Configuration}
Now, we shed light on the beamforming codebooks utilized in 5G NR systems, by which the precoding matrices are constructed based on the information fed back through the PMI. The term ``matrix'' implies multiple streams transmitted in the systems, whereas the matrix simplifies to a vector when only a single stream is carried. The composition and underlying technologies of these PMI beamforming codebooks will be detailed in Sec. \ref{SecPMI}. In this subsection, we primarily introduce the common characteristics of these codebooks to help readers better understand the codebook framework.

The BS is equipped with a physical antenna array in the shape of a uniform planar array (UPA) with dual-polarized antennas. A dual-polarized antenna consists of two co-located antenna elements that transmit/receive signals with opposite polarization, typically configured with slanted $\pm 45$ polarization to achieve similar propagation conditions in both polarizations.\footnote{If one instead uses horizontal and vertical polarization, these polarization might experience substantially different fading characteristics due to interactions with the ground plane and vertical buildings.}
The number of antennas in the physical antenna array is determined by the effective isotropic radiated power (EIRP) that is desirable in the considered deployment scenario. However, it can be mapped to a smaller logical antenna array, based on hardware constraints and the desire to reduce the reference signaling.

The configuration of CSI-RS ports is based on a logical antenna array, whose relationship to the physical antenna array can be understood through two typical cases, as illustrated in Fig. \ref{s2}. In case 1, the logical antenna array shares identical dimensions with the physical antenna array—both being $N_1 \times N_2$. Here, the CSI-RS measures the full channel individually across all physical antennas, and the beamforming vector derived from the PMI codebook is applied directly to these antennas. In case 2, the logical antenna array is of smaller dimension than the physical array, denoted as $N_h \times N_v$, where $N_h$ and $N_v$ are integer multiples of $N_1$ and $N_2$, respectively. A dimension-increasing linear mapping—referred to as port external beamforming (PEB)—is then applied from the logical to the physical antenna array. PEB can be implemented through digital, analog, or hybrid beamforming architectures. In digital beamforming, logical ports correspond to virtual ports while antenna ports are digital; in analog beamforming, logical ports are directly implemented as digital ports; and in hybrid beamforming, logical ports may represent either digital ports or virtual ports between digital ports and data streams. This mapping from a smaller logical array of size $2N_1N_2$ to a larger physical array of size $2N_hN_v$ is designed autonomously by the base station and is not standardized in the protocol. Importantly, the UE remains unaware of this mapping operation. Under this configuration, the beamformed CSI-RS measures an effective channel associated with the reduced-dimensional logical array, and the resulting PMI-based beamforming vector is applied to this effective channel.

\begin{figure}[t]
	\centering
	\includegraphics[width=0.4\textwidth]{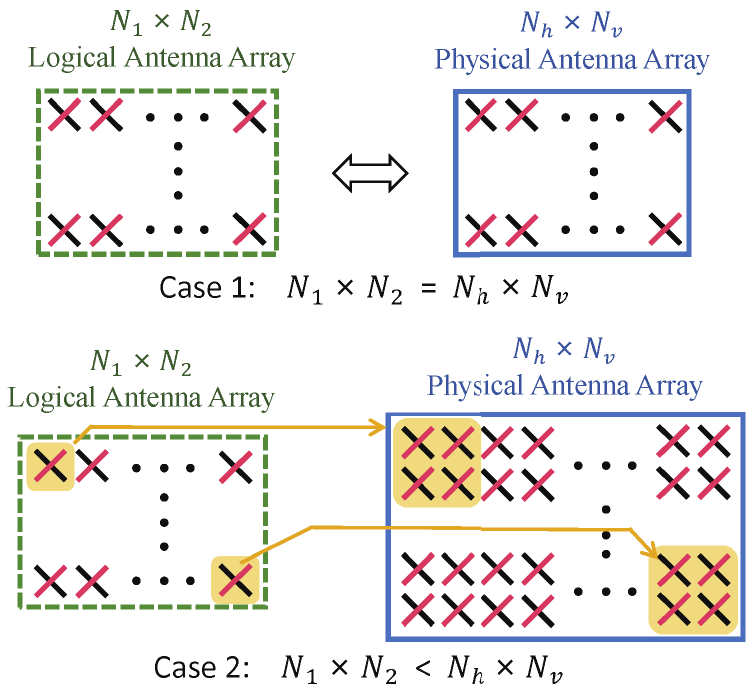}
	\caption{Logical antenna array vs physical antenna array.}\label{s2}
\end{figure}

Note that having more CSI-RS ports means that more measurement occasions are required, and more measurement occasions mean that a greater number of resource elements (REs) need to be occupied, as shown in Fig. \ref{schematic} in the Appendix. In the channel estimation model, it is often assumed that the channels on these REs are the same, with multiple measurements achieved through time, frequency, or code division. \emph{If the number of ports is too large, this assumption may not hold and thus degrade the performance of channel estimation. In light of this, the 5G NR protocol defines the logical antenna array to limit the number of ports supported by gNB for transmitting CSI-RS, so as to support a larger number of antennas.} Currently, the 3GPP Release 18 supports a number of ports including 4, 8, 12, 16, 24, and 32. For each given number of ports, the supported array shapes are also limited, as illustrated in Fig. \ref{logicay}. It can be seen that,  the logical antenna array supports not only UPA but also uniform linear arrays (ULA).

\begin{figure}[t]
	\centering
	\includegraphics[width=0.48\textwidth]{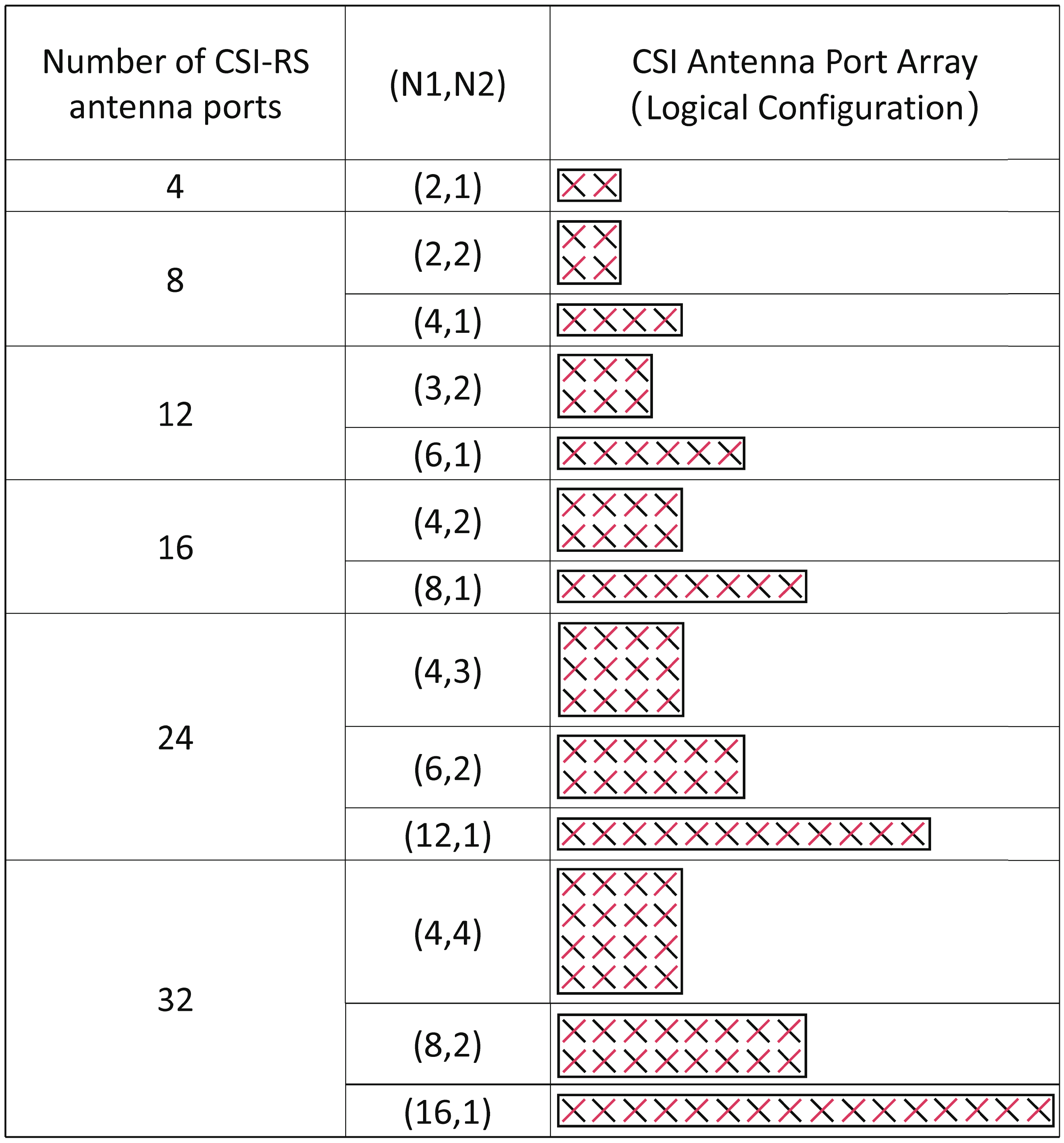}
	\caption{Supported configurations of logical antenna array.}\label{logicay}
\end{figure}

\begin{figure}[t]
	\centering
	\includegraphics[width=0.45\textwidth]{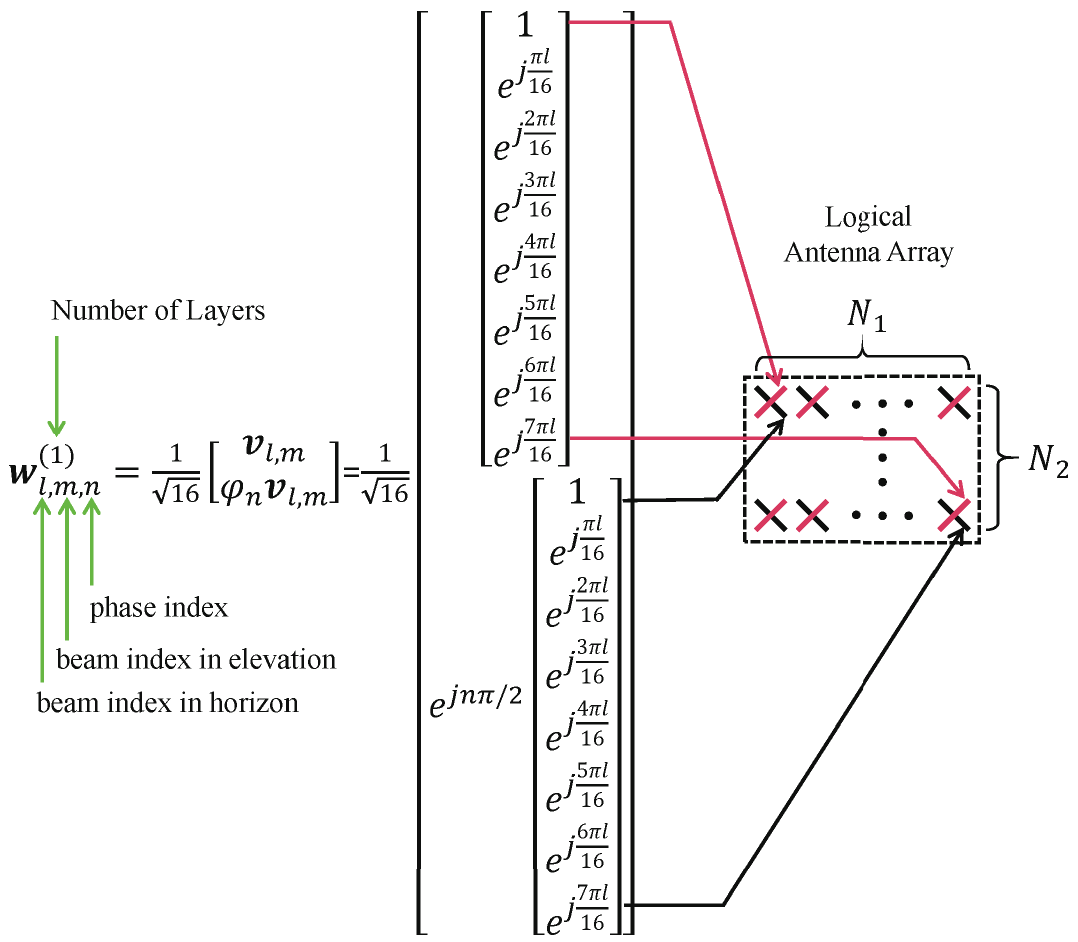}
	\caption{A codebook vector mapped to a logical antenna array.}\label{s1}
\end{figure}

Taking a look at the most basic and simplest PMI codebook, i.e., the Type I codebook for a single data stream in Table 5.2.2.2.1-5 of TS 38.214, the codebook vector is given by 
\begin{equation}\label{ty1}
\mathbf{w}_{\rm{Type \;I}}=\frac{1}{\sqrt{P_{\rm{CSI\text{-}RS}}}}\left[ \begin{array}{c}
	\mathbf{v}_{l,m}\\
	\varphi _n\mathbf{v}_{l,m}\\
\end{array} \right],
\end{equation}
where $P_{\rm{CSI\text{-}RS}}$ is the number of CSI-RS ports, $\mathbf{v}_{l,m}$ is a beam for of one type of polarized antennas, $\varphi _n\mathbf{v}_{l,m}$ is a beam for the opposite type of polarized antennas, and $\varphi _n$ is the phase difference between two types of polarized antennas. ${l,m}$ are the indices that select the beam $\mathbf{v}_{l,m}$ from a beam set,  which will be detailed in Sec. \ref{SecPMI}. Let ${\bf{H}}_1$ and ${\bf{H}}_2$ represent the channels on two types of polarized antennas, and assume the transmitted signal $s={\sqrt{P_{\rm{CSI\text{-}RS}}}}$, then the received signals at the UE can be written as 
\begin{equation}
    \mathbf{y}=\left[ \mathbf{H}_1,\mathbf{H}_2 \right] \mathbf{w}_{\rm{Type \;I}}s=\mathbf{H}_1\mathbf{v}_{l,m}+\varphi _n\mathbf{H}_2\mathbf{v}_{l,m}.
\end{equation}
As can be seen, the channels ${\bf{H}}_1$ and ${\bf{H}}_2$ enjoy the same $\mathbf{v}_{l,m}$, which implies that they basically have a similar spatial structure. In other words, we believe that $\mathbf{v}_{l,m}$ is the optimal beam selected under the channel $\mathbf{H}_1+\varphi _n\mathbf{H}_2$, where $\varphi _n$ is the phase difference between the two channels for maximum combining. The fundamental reason for designing the codebook in this manner is that dual-polarized antennas are used. An example is given in Fig. \ref{s1} with a $16 \times 1$ dimensional Type I codebook, where the number of horizontal logical antenna elements is $N_1 = 8$ and the number of vertical logical antenna elements is $N_2 = 1$.\footnote{The horizontal oversampling factor is $O_1 = 4$ in this example, and the vertical oversampling factor is $O_2 = 1$. The concepts are explained in detail in Sec. \ref{SecPMI}.} The total number of CSI-RS ports is $P_{\rm{CSI\text{-}RS}} = 16$, and the number of layers is 1.

Consequently, the antenna groups of both polarizations have the same dimensions ($N_1 \times N_2$) and physical locations. Thus, their channels are similar and only a phase compensation $\varphi _n$ is needed to merge the power.

\section{Codebook Fundamentals}\label{funds}

\begin{figure}[t]
	\centering
	\includegraphics[width=0.48\textwidth]{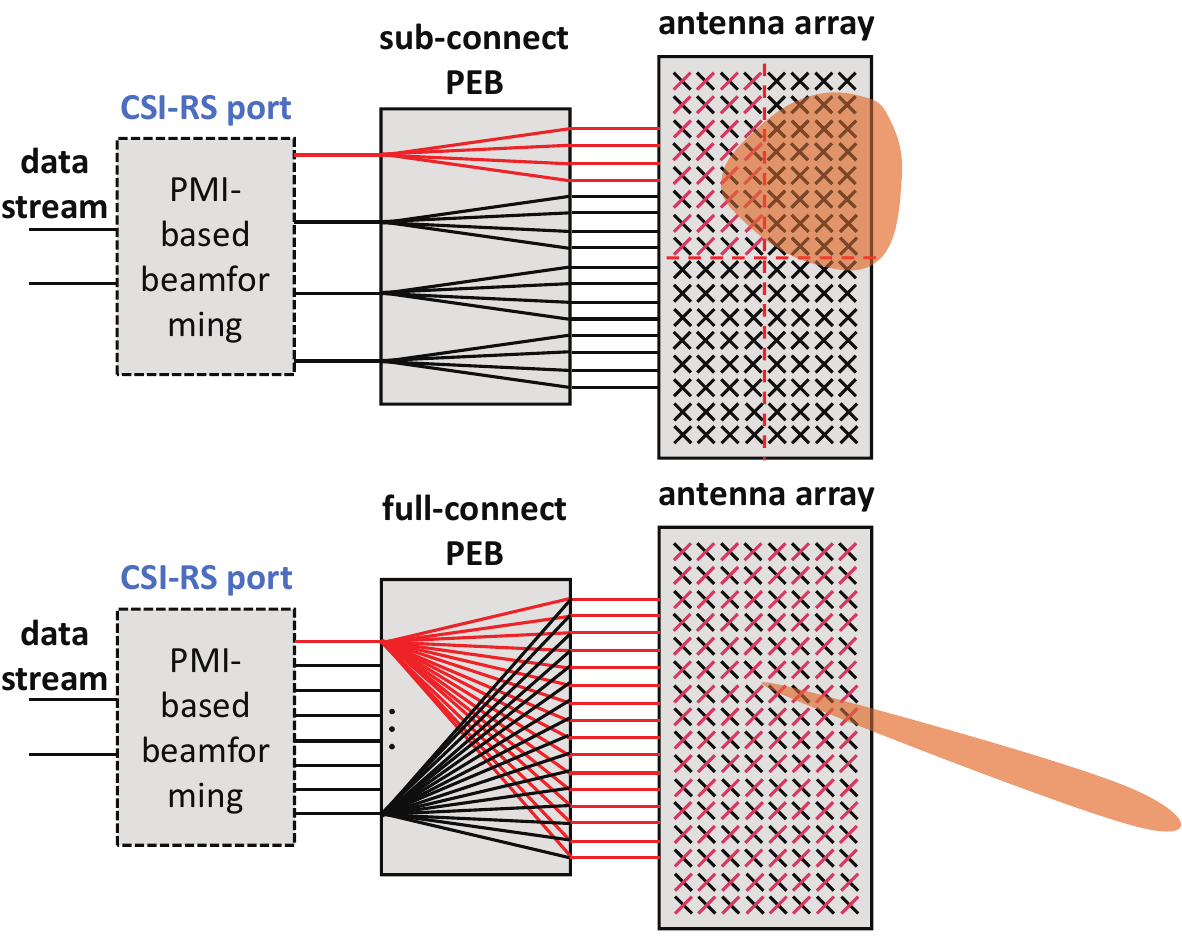}
	\caption{Sub-connect and full-connect PEB structures.}\label{twoPEB}
\end{figure}
The core component of the 3GPP codebook is the bases designed for beamforming matrices, which include spatial, frequency, and temporal domain bases. These bases represent different beams, delay taps, and Doppler shifts, respectively. While the frequency and temporal bases are simply DFT vectors, the spatial domain basis is more complex and is derived from the Kronecker product of two basis vectors based on the PEB and the array's topology. 

\subsection{Port External Beamforming}
The PEB determines whether the UE feeds back using a regular codebook or a port-selection codebook, which have a different spatial domain basis, while the array's topology dictates the dimensions of the two basis vectors. As illustrated in Fig. \ref{twoPEB}, the PEB can be divided into sub-connected and full-connected structures. In the sub-connected structure, each CSI-RS port is independently mapped onto a subarray through the PEB, typically using a DFT vector mapping. The beam pattern of the PEB can be equivalently regarded as the antenna radiation pattern at the CSI-RS ports, leading the UE to measure the logical antenna-domain channel and to use a regular codebook for feedback. In the full-connected structure, each CSI-RS port is mapped onto the entire array via the PEB, where the mapping can be either DFT vectors or the channel's singular vectors. This PEB allows each CSI-RS to carry an independent beam; hence, the UE measures the beam-domain channel and uses a port-selection codebook for feedback.    

\subsection{Regular Codebook Fundamentals}
First, we briefly introduce the spatial basis for the regular codebook and assume no PEB is applied\footnote{The regular codebook is oriented towards BSs that adopt either a sub-connected PEB or no PEB. For full-connect PEB configurations, the UE utilizes a port-selection codebook for PMI feedback.}, i.e., the CSI-RS ports are equivalent to the physical antenna ports.  Considering a UPA with $2N_HN_V$ antennas using dual polarization, where $N_H$ antennas are arranged horizontally and $N_V$ antennas are arranged vertically, the horizontal beam steering vector can be represented as 
\begin{equation}
{\bf{a}}_l=[1, e^{j2\pi\frac{l}{N_H}}, \ldots, e^{j2\pi(N_H-1)\frac{l}{N_H}}]^T,
\end{equation}
where $l=0,1,\ldots,N_H -1$ represents the beam index for different horizontal directions. Note that this is a column in a DFT matrix. With a half-wavelength antenna spacing, the beam points in the horizontal angular direction $\arcsin(x)(2l/N_H)$ for $l \leq N_H/2$ and in the direction $\arcsin(x)(2l/N_H-1)$ for $l>N_H/2$.
Similarly, the vertical beam steering vector is given by  
\begin{equation}
{\bf{u}}_m=[1, e^{j2\pi\frac{m}{N_V}}, \ldots, e^{j2\pi(N_V-1)\frac{m}{N_V}}]^T,
\end{equation}
where $m=0,1,\ldots,N_V-1$ represents the beam index for different vertical directions. $\{{\bf{a}}_l\}_{l=0}^{N_H-1}$ (resp. $\{{\bf{u}}_m\}_{m=0}^{N_V-1}$) is referred to as the orthogonal beam group, which contains orthogonal beams in horizontal (resp. vertical) directions. 

\begin{figure}[t]
	\centering
	\includegraphics[width=0.45\textwidth]{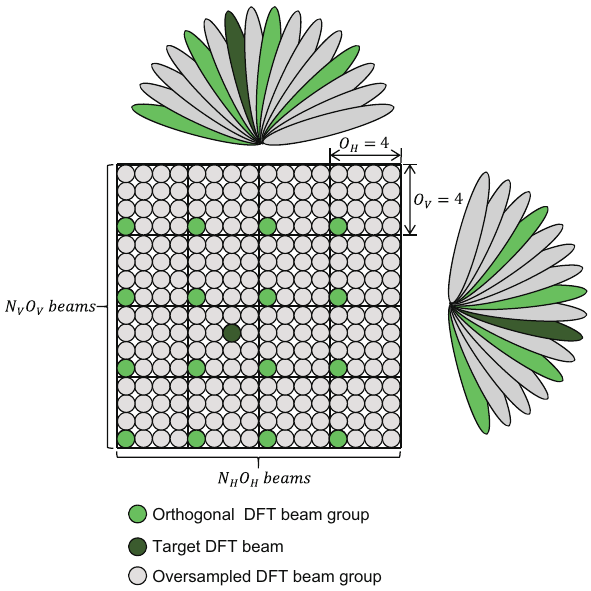}
	\caption{Supported configurations of logical antenna array.}\label{pic_6}
\end{figure}

In 5G codebooks, the number of beams considered in the spatial domain can exceed the number of orthogonal beams, to enable a closer matching to practical channels. Assume that $O_HN_H$ beams are considered in the horizontal direction, where $O_H$ is called the oversampling factor. These beams can be divided into $O_H$ orthogonal DFT beam groups. The supplementary $O_H-1$ beam groups are known as the oversampled DFT beam groups, as shown in Fig. \ref{pic_6}. In this case, the oversampled beams in horizontal and vertical can be respectively expressed as
\begin{equation}
{ \bf{\hat a}}_l=[1, e^{j2\pi\frac{l}{N_HO_H}}, \ldots, e^{j2\pi(N_H-1)\frac{l}{N_HO_H}}]^T
\end{equation}
and 
\begin{equation}
{\bf{\hat u}}_m=[1, e^{j2\pi\frac{m}{N_VO_V}}, \ldots, e^{j2\pi(N_V-1)\frac{m}{N_VO_V}}]^T,
\end{equation}
where $l=0,1,\ldots,N_HO_H-1$, $m=0,1,\ldots,N_VO_V-1$, and $O_V$ is the  vertical oversampling factor. Any horizontal beam combined with a vertical beam can form a two-dimensional beam that serves as a spatial basis for the codebook, i.e.,
\begin{equation}\label{vv}
{{\bf{v}}_{l,m}} = {{\bf{\hat a}}_l} \otimes {{\bf{\hat u}}_m},
\end{equation}
where $\otimes$ is the Kronecker product operator. As shown in Fig.  \ref{pic_6}, there are $O_HO_V=16$ groups of orthogonal basis vectors, each containing $N_HN_V=16$ basis vectors. The basis vectors within each group are orthogonal to each other, e.g., the basis vectors indicated in green. 

For the Type II regular codebook, the UE will choose $L$ beam bases from within the selected orthogonal DFT beam group and linearly combine them to be applied to polarized antennas.  Readers may refer to Fig. \ref{pic_10} in Sec. \ref{regularII} for an intuitive understanding of the physical meaning of this codebook.  The indices of these bases are denoted as  $l(i),m(i)$ for $i=0,\ldots,L-1$. Thus, the codebook vector for a single data stream can be expressed as 
\begin{equation}
 {{\bf{w}}_{{\rm{Type \;II}}}} = \frac{1}{{\sqrt \beta  }}\left[ {\begin{aligned}
{\sum\limits_{i = 0}^{L - 1} {{a_{1,i}}{{\bf{v}}_{l(i),m(i)}}} }\\
{\sum\limits_{i = 0}^{L - 1} {{a_{2,i}}{{\bf{v}}_{l(i),m(i)}}}}
\end{aligned}} \right],
\end{equation}
where $\beta$ denotes the power scaling coefficient. $a_{1,i}$ and $a_{2,i}$ represent the weighting coefficients of the $i$th beam on the first and second polarization directions, respectively. If the transmitted signal is $s={\sqrt{\beta}}$, the received signals at the UE by applying ${{\bf{w}}_{{\rm{Type \;II}}}}$ can be written as 
\begin{equation}
\begin{aligned}\label{regy}
{\bf{y}} &= \left[ {{{\bf{H}}_1},{{\bf{H}}_2}} \right]{{\bf{w}}_{{\rm{Type \; II}}}}s\\
 &= \sum\limits_{i = 0}^{L - 1} {{a_{1,i}}{{\bf{H}}_1}{{\bf{v}}_{l(i),m(i)}}}  + \sum\limits_{i = 0}^{L - 1} {{a_{2,i}}{{\bf{H}}_2}{{\bf{v}}_{l(i),m(i)}}} .
\end{aligned}
\end{equation} 

\subsection{Port-Selection Codebook Fundamentals}
Next, we introduce the spatial basis for the port-selection codebook, assuming that full-connect PEB is applied. This type of codebook is particularly well-suited for mmWave communications, characterized by a large number of antennas and the use of hybrid beamforming instead of fully digital beamforming to reduce power consumption and hardware complexity. The full-connect PEB architecture enables flexible mapping between logical CSI-RS ports and physical antennas, making it ideal for hybrid beamforming, which integrates both analog and digital components. Moreover, mmWave propagation is typically dominated by line-of-sight (LoS) paths and a few strong scatterers, resulting in sparse channel characteristics in the angular domain. With full-connect PEB, the UE observes the channel in this angular domain, and thanks to the inherent sparsity, it only needs to select a subset of the CSI-RS ports with high energy, so a small value of $L$ will be sufficient. These selected ports are then linearly combined to form the beamforming vector between data streams and CSI-RS ports. This approach reduces feedback overhead while remaining compatible with full-connect PEB. In this case, the beam bases become the standard vector bases. Assuming the number of CSI-RS ports is ${P_{{\text{CSI-RS}}}}$, the beam basis ${{\bf{p}}_d} \in {{\mathbb{R}}^{{P_{{\text{CSI-RS}}}}/2}}$ of the port-selection codebook can be represented by\footnote{In the 3GPP protocol, the spatial basis of the port-selection codebook is denoted by ${{\bf{v}}_d}$. To avoid confusion with the basis of the regular codebook, we temporarily represent the spatial basis with ${{\bf{p}}_d}$ in this subsection.}
\begin{equation}
{{\bf{p}}_d} = \left\{ {\begin{split}
&{[1,0\ldots,0]^T,\qquad {\rm{if}}\;d = 0},\\
&{[0,0,\ldots,1]^T,\;\; \quad  {\rm{if}}\;d  = \frac{{P_{{\text{CSI-RS}}}} }{2}- 1},\\
&{[0,\ldots,1,\ldots,0]^T,\; {\rm{otherwise,}}}
\end{split}} \right.
\end{equation}
where $d=0,1,..,{P_{{\text{CSI-RS}}}}/2-1$ and the beam basis ${{\bf{p}}_d}$ contains a value of $1$ in the $(d+1)$-th element and zeros elsewhere. Then, the codebook vector for a single data stream is given by
\begin{equation}
{\bf{w}}_{{\rm{Type \;II}}}^{{\rm{PS}}} = \frac{1}{{\sqrt \beta  }}\left[ {\begin{aligned}
{\sum\limits_{i = 0}^{L - 1} {{a_{1,i}}{{\bf{p}}_{d(i)}}} }\\
{\sum\limits_{i = 0}^{L - 1} {{a_{2,i}}{{\bf{p}}_{d(i)}}}}
\end{aligned}} \right].
\end{equation}
Let $\bf{F}$ denote the full-connect PEB of CSI-RS on each polarization, which maps CSI-RS logical ports to physical antenna ports. The effective channel estimated by the UE can be written as 
\begin{equation}
{{\bf{H}}_{{\rm{eff}}}} = [{{\bf{H}}_1}{\bf{F}},{{\bf{H}}_2}{\bf{F}}],
\end{equation}
where the matrix $\mathbf{F}$ is typically composed of DFT vectors, i.e., ${\bf{F}} = [{{\bf{v}}_0},{{\bf{v}}_1},\ldots,{{\bf{v}}_{{P_{\rm{CSI\text{-}RS}}}-1}}]$ with ${{\bf{v}}_n}$ consistent with (\ref{vv}). If the transmitted signal is $s={\sqrt{\beta}}$, the received signals at the UE by applying ${\bf{w}}_{{\rm{Type \;II}}}^{{\rm{PS}}}$ can be written as
\begin{equation}\label{psy}
\begin{aligned}
{\bf{y}} &= {{\bf{H}}_{{\rm{eff}}}}{\bf{w}}_{{\rm{Type\; II}}}^{{\rm{PS}}}s = \left[ {{{\bf{H}}_1}{\bf{F}},{{\bf{H}}_2}{\bf{F}}} \right]{\bf{w}}_{{\rm{Type\; II}}}^{{\rm{PS}}}s\\
 &= {{\bf{H}}_1}[{{\bf{v}}_0},{{\bf{v}}_1},\ldots,{{\bf{v}}_{{P_{\rm{CSI\text{-}RS}}} - 1}}]\sum\limits_{i = 0}^{L - 1} {{a_{1,i}}{{\bf{p}}_{d(i)}}} \\
 & \qquad \qquad + {{\bf{H}}_2}[{{\bf{v}}_0},{{\bf{v}}_1},\ldots,{{\bf{v}}_{{P_{\rm{CSI\text{-}RS}}} - 1}}]\sum\limits_{i = 0}^{L - 1} {{a_{2,i}}{{\bf{p}}_{d(i)}}} \\
 &= \sum\limits_{i = 0}^{L - 1} {{a_{1,i}}{{\bf{H}}_1}{{\bf{v}}_{d(i)}}}  + \sum\limits_{i = 0}^{L - 1} {{a_{2,i}}{{\bf{H}}_2}{{\bf{v}}_{d(i)}}},
\end{aligned}
\end{equation}
where the noise was omitted for brevity.
By comparing (\ref{regy}) and (\ref{psy}), it can be observed that the regular codebook and the port-selection codebook serve the same role in data transmission. The former directly generates a combination of DFT beams from the data stream to the antenna ports, while the latter selects a combination of DFT beams in PEB, i.e., from the CSI-RS ports to the physical antenna ports.

\begin{table*}[t]  
\centering   
\captionsetup{
  labelsep=newline, % 让标题另起一行
}
\renewcommand{\arraystretch}{2}  %每行的高度为1.5倍
\caption{R15 Type I Codebook Mode 1 (TS 38.214 Table 5.2.2.2.1-6)}  \label{tabmode1}
\begin{tabular}{|Sc|Sc|Sc|Sc|}   
\hline  
\rowcolor{gray!20}
\multicolumn{4}{|c|}{\textbf{\textit{Codebook Mode} = 1}}   \\
\hline 
\rowcolor{gray!20}
$i_{1,1}$ & $i_{1,2}$ & $i_2$ & \  \\
\hline
$0,1, \cdots,N_1O_1-1$ &$ 0,1, \cdots,N_2O_2-1$ &$0,1$&${\bf{W}}^{(2)}_{i_{1,1},i_{1,1}+k_1,i_{1,2},i_{1,2}+k_2,i_2}$ \\
\hline
\multicolumn{4}{|Sc|}{\makecell{Precoding Matrix: ${\bf{W}}^{(2)}_{l,l',m,m',n}=\frac{1}{\sqrt{2P_{\rm{CSI\text{-}RS}}}}\left[ \begin{aligned}
	&\;\;\;\;{\bf{v}}_{l,m} \qquad \ \ \quad {\bf{v}}_{l',m'}  
\\ 
	&\varphi_n{\bf{v}}_{l,m} \quad -\varphi_n{\bf{v}}_{l',m'} \;
\end{aligned} \right] $.  } }\\
\hline
\end{tabular}  
\end{table*}

\section{5G Codebook Evolution}\label{SecPMI}
We have introduced the basic framework and spatial basis of both Type I and Type II through the simplest single-stream feedback assumption, with the aim of enabling readers to quickly grasp the core concepts of the codebooks. In the 3GPP codebook protocol, the codebook also considers other aspects, such as multi-stream feedback, amplitude quantization, phase quantization, subband quantization, designation of the strongest coefficient, bitmap restrictions, etc. Furthermore, following the first version of Type I and Type II codebooks in 3GPP Release 15, subsequent versions have further upgraded Type II codebooks. For instance, Release 16 introduced the Enhanced Type II Codebook for both regular and port-selection conditions, which was designed for spectral bases and allowed for compressed feedback of beamforming vectors across different subbands. In Release 17, the protocol was updated with the Further Enhanced Port-Selection Type II Codebook, which removed the restriction of selecting consecutive ports, allowing for the free selection of ports, and further enhanced the compression in the frequency domain. Release 18 proposed the Enhanced Type II Codebook for Predicted PMI, designing a temporal basis that allows for feedback of future beamforming vectors based on predictions in mobile scenarios. These codebooks are introduced in 3GPP TS 38.214 5.2.2.2, but the protocol documents do not explain the deeper implications and design motivations of the codebooks. In the following subsections, we will define and explain each symbol used in the protocol codebooks, describe the design thoughts and physical significance to provide a more in-depth insight.

\subsection{R15 Type I Regular Codebook Mode 1}

\begin{table}[t]   
\begin{center} 
\centering  
\captionsetup{
  labelsep=newline, % 让标题另起一行
}
\caption{Parameters in the Type I Codebook.}  
\renewcommand{\arraystretch}{1.5}  %每行的高度为1.5倍
\label{codebook1} 
\begin{tabular}{|c|m{6cm}|}   
\hline   \textbf{Parameters} &\bf{Interpretation} \\  
\hline   $N_{1}$, $N_{2}$ & The number of antennas in the horizontal and vertical dimensions, respectively.\\ 
\hline   $O_{1}$, $O_{2}$ & The horizontal and vertical oversampling factors, respectively. \\
\hline   $P_{\rm{CSI\text{-}RS}}$ & The number of CSI-RS ports, which is equal to $2N_{1}N_{2}$.\\ 
\hline   ${\bf{v}}_{l,m}$ & The beam basis of the first data stream.\\
\hline   ${\bf{v}}_{l^{'},m^{'}}$ & The beam basis of the second data stream.\\
\hline   $l$ & The horizontal index of the selected beam in the first data stream.\\ 
\hline   $l^{'}$ & The horizontal index of the selected beam in the second data stream.\\
\hline   $m$ & The vertical index of the selected beam in the first data stream.\\ 
\hline   $m^{'}$ & The vertical index of the selected beam in the second data stream.\\
\hline   $k_1$ & The offset of the horizontal index of the beam in the second data stream relative to the beam in the first data stream.\\
\hline   $k_2$ & The offset of the vertical index of the beam in the second data stream relative to the beam in the first data stream.\\
\hline   \bf{$i_1$} & A PMI vector with 3 elements $i_{1,1}$, $i_{1,2}$ and $i_{1,3}$, which is reported by the UE.\\
\hline   $i_{1,1}$ & %is a part of the indice \textbf{$i_1$} reported by the UE, 
Determines the horizontal index of the beam used by the first data stream and the horizontal index baseline for the second data stream.\\
\hline   $i_{1,2}$ & 
Determines the vertical index of the beam used by the first  data stream and the vertical index baseline for the second data stream.\\
\hline   $i_{1,3}$ & Determines the offset of the beam indices used by the second data stream relative to the index baseline.\\
\hline   $i_2$ & Determines the parameter $n$ in both Codebook Mode 1 and Mode 2, and additionally determines the beam basis in Codebook Mode 2.\\
\hline   $n$ & The polarization phase index.\\
\hline   $\varphi_{n}$  & The polarization phase coefficient between two groups of polarized antennas.\\ 
\hline   ${\bf{u}}_{m}$ & The array response vector of the vertical $N_{2}$ CSI-RS ports, and the Kronecker product of the array response vector of the horizontal $N_{1}$ CSI-RS ports construct the spatial beam basis ${\bf{v}}_{l,m}$ .\\ 
\hline   
\end{tabular}   
\end{center}   
\vspace{-10pt}
\end{table}

The codebook in 3GPP Release 15 is the first standardized codebook of the 5G era, comprehensively defining the structure of Type I and Type II codebooks. Type I consists only of regular codebooks and does not include port-selection codebooks. However, based on the feedback behavior for different subbands, Type I is further classified into codebook mode 1 and mode 2. Compared to Type II, the defining feature of Type I is that it selects only a single beam. If this beam is consistent across the entire bandwidth part (BWP), then the UE uses mode 1 for feedback; otherwise, mode 2 feedback is used. The Type I codebook supports between $1$ and $8$ data streams. The protocol specifies dedicated codebook tables for each number of streams. If the number of streams exceeds 3, the protocol no longer distinguishes between mode 1 and mode 2, but instead refers to them collectively\footnote{Based on the design philosophy of the codebook, the authors consider mode 1-2 to be the extension of mode 1.} as mode 1-2. The expression of the Type I Regular Codebook Mode 1 for two data streams in the protocol is presented in Table \ref{tabmode1}, in which the parameters are interpreted in Table \ref{codebook1}. 

The PMI parameters that are actually fed back to the gNB are $i_1=[i_{1,1},i_{1,2},i_{1,3}]$ and $i_2$, which determine the values of $l,l',m,m'$,  and $n$ in ${\bf{W}}^{(2)}_{l,l',m,m',n}$. 
The size of the logical antenna array $N_1 \times N_2$ and the size of oversampling $O_1 \times O_2$ are configured with a higher layer RRC. Then, as shown in Table \ref{tabmode1}, the specific association can be represented as
\begin{equation} \label{w35}
    {\bf{W}}^{(2)}_{i_{1,1},i_{1,1}+k_1,i_{1,2},i_{1,2}+k_2,i_2} = {\bf{W}}^{(2)}_{l,l',m,m',n},
\end{equation}
where
\begin{equation}
 \begin{aligned}
&{i_{1,1}} \in \{ 0,1, \ldots ,{N_1}{O_1} - 1\}, \\
&{i_{1,2}} \in \{ 0,1, \ldots ,{N_2}{O_2} - 1\}, \\
&{i_{1,3}} \in \{ 0,1,2,3\},\; {i_2} \in \{ 0,1\} .\\
\end{aligned}
\end{equation}
In (\ref{w35}), one can observe that $i_{1,3}$ does not appear explicitly; however, $k_1$ and $k_2$ are indicated through $i_{1,3}$. It is defined in a table\footnote{Part of the mapping of $i_3$ to $k_1$ and $k_2$ are illustrated in Table \ref{tabmap}.}, for instance, in TS 38.214 Table 5.2.2.2.1-3 or 5.2.2.2.1-4.  From the codebook matrix, each column represents a beamforming vector used for each data stream. It is evident that the beams adopted for the pair of polarizations are identical, differing only by a phase factor, i.e.,
\begin{equation}\label{phase}
 \varphi_n =e^{j\pi n /2}, \;\; n=i_2 \in \{0,1\}.
\end{equation}
The two data streams utilize distinct beams, i.e., ${\bf{v}}_{l,m}$ and ${\bf{v}}_{l',m'}$. The beam index for the first data stream is indicated by $l$ and $m$, while for the second data stream, it is indicated by $l'$ and $m'$. Notably, $l$ and $m$ are selected directly from $N_1O_1 \times N_2O_2$ predefined beams, whereas $l'$ and $m'$ are indicated by the relative offsets $k_1$ and $k_2$ based on the first stream's beam index. The $N_1O_1 \times N_2O_2$ predefined beams are given by 
\begin{equation}\label{beamv}
 {\bf{v}}_{l,m}=\left[ {\bf{u}}_m,\; e^{j\frac{2\pi l}{O_1N_1}}{\bf{u}}_m,\;\cdots, \;e^{j\frac{2\pi l(N_1-1)}{O_1N_1}}{\bf{u}}_m \right]^T,
\end{equation}
where $l=0,1,\ldots,N_1O_1-1$ and $m=0,1,\ldots,N_2O_2-1$ with
\begin{equation}
{\bf{u}}_m=\left[ 1, \; e^{j\frac{2\pi m}{O_2N_2}  },\; \cdots,\; e^{j\frac{2\pi m(N_2-1)}{O_2N_2}  }\right].
\end{equation}

The aforementioned content basically supports the report of the beamforming matrix based on the codebook, but the 3GPP standard also considers issues of interference and scheduling for multi-user transmission and has further constrained the PMI feedback. A key concept here is that of a bitmap and bit sequence. During the process of spatial beam selection, the $N_1O_1 \times N_2O_2$ beam bases can be indicated by a bitmap, i.e.,
\begin{equation*}
{\bf{V}} = \left[ {\begin{array}{*{20}{c}}
{{v_{0,0}}}&{{v_{0,1}}}& \cdots &{{v_{0,{N_2}{O_2} - 1}}}\\
{{v_{1,0}}}&{{v_{1,1}}}& \cdots &{{v_{1,{N_2}{O_2} - 1}}}\\
 \vdots & \vdots & \ddots & \vdots \\
{{v_{{N_1}{O_1} - 1,0}}}&{{v_{{N_1}{O_1} - 1,1}}}& \cdots &{{v_{{N_1}{O_1} - 1,{N_2}{O_2} - 1}}}
\end{array}} \right],
\end{equation*}
 where $v_{l,m} \in \{0,1\}$ is associated with the beam ${\bf{v}}_{l,m}$. A bit value of zero in the bitmap indicates that the associated beam is not permitted for use in the codebook. The bit sequence ${\bf{a}}=[a_0,a_1,\ldots,a_{N_1O_1N_2O_2-1}]^T$ is imported from the bitmap by
 \begin{equation}\label{type1map}
{\bf{a}} = {\rm{vec}}({\bf{V}}^T),
 \end{equation}
where $a_{N_2O_2l+m}$ is associated with the beam ${\bf{v}}_{l,m}$. This bit sequence restricts beams in certain directions, thereby avoiding interference with specific users (such as those with weak signals or preferred users) or other BSs.
There is another bit sequence for stream restriction represented as 
\begin{equation}
    {\bf{r}}=[r_0,r_1,\ldots,r_7]^T,
\end{equation}
where $r_0$ is the least significant bit (LSB) and $r_7$ is the most significant bit (MSB). When $r_i$ is set to zero, PMI reporting is prohibited for any precoder corresponding to $i+1$ data streams. For instance, if the bit sequence is $\mathbf{r} = [1,1,0,1,0,0,0,0]^T$, this indicates that the PMI only supports precoders for 1-stream, 2-stream, and 4-stream transmissions.

\begin{table}[t]  
\begin{center} 
\centering   
\captionsetup{
  labelsep=newline, % 让标题另起一行
}
\renewcommand{\arraystretch}{1.5}  %每行的高度为1.5倍
\caption{Mapping of $i_{1,3}$ to $k_1$ and $k_2$}  \label{tabmap}
\begin{tabular}{|c|c|c|c|c|c|c|}   
\hline  
\rowcolor{gray!20}
\multirow{2}{*}{} & \multicolumn{2}{c|}{$N_1>N_2>1$}& \multicolumn{2}{c|}{$N_1=N_2$}&  \multicolumn{2}{c|}{$N_1>2,N_2=1$}   \\ 
\cline{2-7}   
\rowcolor{gray!20}
\multirow{-2}{*}{$i_{1,3}$}&$k_1$&$k_2$&$k_1$&$k_2$&$k_1$&$k_2$\\
\hline 0&0&0&0&0&0&0\\
\hline 1&\ \ $O_1$\ \ &0&\ $O_1$\ &0&\ \ $O_1$\ \ &0\\
\hline 2&0&$O_2$&0&$O_2$& $2O_1$&0\\
\hline 3&$2O_1$&0&$O_1$&$O_2$& $3O_1$&0\\
\hline
\end{tabular}  
\end{center}
\vspace{-10 pt}
\end{table}

\begin{table*}[t]  
\centering   
\captionsetup{
  labelsep=newline, % 让标题另起一行
}
\renewcommand{\arraystretch}{1.5}  %每行的高度为1.5倍
\caption{R15 Type I Regular Codebook Mode 2 (TS 38.214 Table 5.2.2.2.1-6)}  \label{tabmode2}
\begin{tabular}{|Sc|Sc|Sc|Sc|}     
\hline 
\rowcolor{gray!20}
\multicolumn{4}{|Sc|}{\textbf{\textit{Codebook Mode} = 2}, $N_2>1$}   \\
\hline 
\rowcolor{gray!20}
\multirow{2}{*}{} & \multirow{2}{*}{}& \multicolumn{2}{c|}{$i_2$}   \\ 
\cline{3-4}   
\rowcolor{gray!20}
\multirow{-2}{*}{$i_{1,1}$}& \multirow{-2}{*}{$i_{1,2}$} 
 & \textbf{0} &\textbf{1} \\
 \hline 

 $0,1, \cdots,\frac{N_1O_1}{2}-1$ &$ 0,1, \cdots,\frac{N_2O_2}{2}-1$ &${\bf{W}}^{(2)}_{2i_{1,1},2i_{1,1}+k_1,2i_{1,2},2i_{1,2}+k_2,0}$&${\bf{W}}^{(2)}_{2i_{1,1},2i_{1,1}+k_1,2i_{1,2},2i_{1,2}+k_2,1}$\\
\hline 
\rowcolor{gray!20}
\multirow{2}{*}{} & \multirow{2}{*}{}& \multicolumn{2}{c|}{$i_2$}   \\ 
\cline{3-4}   
\rowcolor{gray!20}
\multirow{-2}{*}{$i_{1,1}$}& \multirow{-2}{*}{$i_{1,2}$} 
 & \textbf{2} &\textbf{3} \\
 \hline 
 \multicolumn{1}{|Sc|}{
  $0,1, \cdots,\frac{N_1O_1}{2}-1$} &$ 0,1, \cdots,\frac{N_2O_2}{2}-1$ &${\bf{W}}^{(2)}_{2i_{1,1}+1,2i_{1,1}+1+k_1,2i_{1,2},2i_{1,2}+k_2,0}$&${\bf{W}}^{(2)}_{2i_{1,1}+1,2i_{1,1}+1+k_1,2i_{1,2},2i_{1,2}+k_2,1}$\\
\hline 
\rowcolor{gray!20}
\multirow{2}{*}{} & \multirow{2}{*}{}& \multicolumn{2}{c|}{$i_2$}   \\ 
\cline{3-4}   
\rowcolor{gray!20}
\multirow{-2}{*}{$i_{1,1}$}& \multirow{-2}{*}{$i_{1,2}$} 
 & \textbf{4} &\textbf{5} \\
 \hline 
  $0,1, \cdots,\frac{N_1O_1}{2}-1$&$ 0,1, \cdots,\frac{N_2O_2}{2}-1$ &${\bf{W}}^{(2)}_{2i_{1,1},2i_{1,1}+k_1,2i_{1,2}+1,2i_{1,2}+1+k_2,0}$&${\bf{W}}^{(2)}_{2i_{1,1},2i_{1,1}+k_1,2i_{1,2}+1,2i_{1,2}+1+k_2,1}$\\
 \hline 
\rowcolor{gray!20}
\multirow{2}{*}{} & \multirow{2}{*}{}& \multicolumn{2}{c|}{$i_2$}   \\ 
\cline{3-4}   
\rowcolor{gray!20}
\multirow{-2}{*}{$i_{1,1}$}& \multirow{-2}{*}{$i_{1,2}$} 
 & \textbf{6} &\textbf{7} \\
 \hline 

 $0,1, \cdots,\frac{N_1O_1}{2}-1$ &$ 0,1, \cdots,\frac{N_2O_2}{2}-1$ &${\bf{W}}^{(2)}_{2i_{1,1}+1,2i_{1,1}+1+k_1,2i_{1,2}+1,2i_{1,2}+1+k_2,0}$&${\bf{W}}^{(2)}_{2i_{1,1}+1,2i_{1,1}+1+k_1,2i_{1,2}+1,2i_{1,2}+1+k_2,1}$\\
  \hline 
  \multicolumn{4}{|Sc|}{\makecell{Precoding Matrix: ${\bf{W}}^{(2)}_{l,l',m,m',n}=\frac{1}{\sqrt{2P_{\rm{CSI\text{-}RS}}}}
\left[ \begin{array}{c}\mathbf{v}_{l,m} \qquad \ \mathbf{v}_{l',m'}\\
    \ \varphi_n\mathbf{v}_{l,m}\ \ \ -\varphi_n\mathbf{v}_{l',m'}
\end{array} \right]$. }} \\  
\hline
\end{tabular} 
\end{table*}

\subsection{R15 Type I Regular Codebook Mode 2}
In R15 Type I Regular Codebook Mode 1, the same beam basis is adopted for the precoding vectors across the entire bandwidth, but it allows for different subbands to configure distinct polarization phases through the PMI parameter $i_2$, as exemplified in (\ref{phase}). To provide more flexibility for subband-level precoding vectors, Codebook Mode 2 introduces a novel mechanism that enables different subbands to alter both the beam basis and polarization phase simultaneously through the PMI parameter $i_2$. Specifically, beam selection in Codebook Mode 2 is divided into two steps. The first step involves using the two index parameters $i_{1,1}$ and $i_{1,2}$ from index $i_1$ to choose a beam group that contains 4 beams. The second step employs the parameter $i_2$ to select a specific beam from this group. Notably, parameter $i_2$ also provides phase adjustment options while finalizing the beam selection process.

The expression of the Type I Regular Codebook Mode 2 for two data streams in the protocol is presented in Table \ref{tabmode2}. Compared to Table \ref{tabmode1},
the main distinction is that the ranges of values for the index parameters $i_{1,1}$, $i_{1,2}$, and $i_{2}$ have become 
\begin{equation}
 \begin{aligned}
&{i_{1,1}} \in \{ 0,1, \ldots ,{N_1}{O_1}/2 - 1\}, \\
&{i_{1,2}} \in \{ 0,1, \ldots ,{N_2}{O_2}/2 - 1\}, \\
&{i_2} \in \{ 0,1,\ldots,7\} .\\
\end{aligned}
\end{equation}
That is to say, the index parameters $i_{1,1}$ and $i_{1,2}$ no longer indicate the beam basis, but rather a smaller number of beam group bases. The index $i_{2}$ indicates the beam basis within the selected group and the corresponding polarization phase. As shown in Fig. \ref{pic_9}, a beam group consisting of two horizontal beams and two vertical beams forms a total of four beams. The index of the beam group can be indicated by the wideband PMI parameter $i_{1} = [i_{1,1}, i_{1,2}]$. After the index ${i}_1$ selects the beam group, $i_2$ completes the beam selection for different subbands. As can be seen from the table, the value range of $i_2$ is from $0$ to $7$, encompassing a total of 8 values. Within this range, there are 4 even numbers and 4 odd numbers, which correspond to the two values of the polarization phase coefficient, 0 and 1, respectively. In other words, when $i_2$ is even, the polarization phase coefficient $n$ is equal to 0, and when $i_2$ is odd, the polarization phase coefficient $n$ is equal to 1. The 4 even or odd values represent the 4 beams within a beam group. Through this process, the index $i_2$, by taking a value from 0 to 7, completes the tasks of reporting polarization phase coefficients and selecting beams within the beam group. Mode 1 and Mode 2 for Type I Codebook are illustrated in Fig. \ref{mode1-2}.

\begin{figure}[t]
	\centering
	\includegraphics[width=0.45\textwidth]{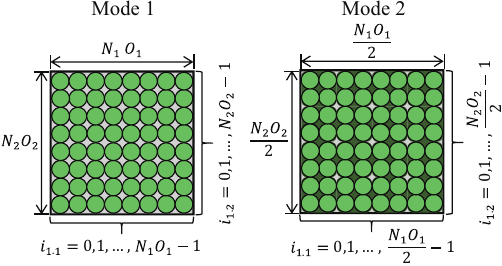}
	\caption{Type I Codebook — Mode 1 and Mode 2 Illustration.}
	\label{mode1-2}
    \vspace{-10pt}
\end{figure}

\begin{table*}[t]  
\centering   
\captionsetup{
  labelsep=newline, % 让标题另起一行
}
\renewcommand{\arraystretch}{2}  %每行的高度为1.5倍
\caption{R15 Type II Regular Codebook (TS 38.214 Table 5.2.2.2.4-1)}  \label{tabtypeii}
\begin{tabular}{|Sc|Sc|}   
\hline  
\rowcolor{gray!20}
\textbf{Layers}  &  \\
\hline   
$v=1$ & ${\bf{w}}^{(1)}_{q_1,q_2,n_1,n_2,p^{(1)}_1,p^{(2)}_1,i_{2,1,1}}
={\bf{w}}^1_{q_1,q_2,n_1,n_2,p^{(1)}_1,p^{(2)}_1,i_{2,1,1}}$  \\
\hline
$v=2$ & $ {\bf{W}}^{(2)}_{q_1,q_2,n_1,n_2,p^{(1)}_1,p^{(2)}_1,i_{2,1,1},p^{(1)}_2,p^{(2)}_2,i_{2,1,2}}
= \frac{1}{\sqrt{2}}
\begin{bmatrix}{\bf{w}}^1_{q_1,q_2,n_1,n_2,p^{(1)}_1,p^{(2)}_1,i_{2,1,1}} \ \ {\bf{w}}^2_{q_1,q_2,n_1,n_2,p^{(1)}_2,p^{(2)}_2,i_{2,1,2}}
\end{bmatrix}$ \\
\hline
\multicolumn{2}{|Sc|}{{\linespread{2}\selectfont  
\makecell{  Precoding Vector: ${\bf{w}}^l_{q_1,q_2,n_1,n_2,p^{(1)}_l,p^{(2)}_l,c_l}
=\frac{1}{\sqrt{N_1N_2\sum^{2L-1}_{i=0}(p_{l,i}^{(1)}p_{l,i}^{(2)})^2}}
\begin{bmatrix}\sum^{L-1}_{i=0} {\bf{v}}_{m_1^{(i)},m_2^{(i)}}p^{(1)}_{l,i}p^{(2)}_{l,i}\varphi_{l,i}\\
\sum^{L-1}_{i=0} {\bf{v}}_{m_1^{(i)},m_2^{(i)}}p^{(1)}_{l,i+L}p^{(2)}_{l,i+L}\varphi_{l,i+L}
\end{bmatrix},\;\; l=1,2.$ \\}
}}\\ 
\hline
\end{tabular}  
\end{table*} 

\subsection{R15 Type II Regular Codebook}\label{regularII}
The Type II codebook is known as the \emph{high-precision codebook}. This is reflected in the fact that the single-polarized beamforming is no longer a single DFT beam, but is obtained through a weighted combination of $L \in \{2,3,4\}$ DFT beam bases. Beamforming with different polarizations uses the same DFT beam bases, but the weights for the combinations can be different. Therefore, the beamforming vectors can be linearly independent within the space of the beam bases, instead of being parallel vectors that only have a phase difference, as in the Type I Codebook.

\begin{table}[t]   
\centering
\captionsetup{
  labelsep=newline, % 让标题另起一行
  justification=centering
}
\renewcommand{\arraystretch}{2}  %每行的高度为1.5倍
\caption{Supported Configurations of $(N_1,N_2)$ and $(O_1,O_2)$}  \label{tabNO}
\begin{tabular}{|>{\centering\arraybackslash}p{1.5cm}|>{\centering\arraybackslash}p{2cm}|>{\centering\arraybackslash}p{2cm}|}
\hline  
\rowcolor{gray!20}
$P_{\rm{CSI-RS}}$ & $(N_1,N_2)$ & $(O_1,O_2)$\\
\hline 
4 & (2,1) & (4,1)\\
\hline 
\multirow{2}{*}{8} & (2,2) & (4,4)\\
\cline{2-3}
 & (4,1) & (4,1)\\
\hline 
\multirow{2}{*}{12} & (3,2) & (4,4)\\
\cline{2-3}
 & (6,1) & (4,1)\\
\hline
\multirow{2}{*}{16} & (4,2) & (4,4)\\
\cline{2-3}
 & (8,1) & (4,1)\\
\hline
\multirow{3}{*}{24} & (4,3) & (4,4)\\
\cline{2-3}
 & (6,2) & (4,4)\\
\cline{2-3}
 & (12,1) & (4,1)\\
\hline
 \multirow{3}{*}{32} & (4,4) & (4,4)\\
\cline{2-3}
 & (8,2) & (4,4)\\
\cline{2-3}
 & (16,1) & (4,1)\\
\hline
\end{tabular}     
\vspace{-2pt}
\end{table}

The expression of the Type II Regular Codebook in the protocol is presented in Table \ref{tabtypeii}. Similar to Type I, the beam bases are selected from $O_1 \times O_2$ oversampling of $N_1 \times N_2$ orthogonal DFT beams across two dimensions, horizontal and vertical, resulting in $N_1O_1 \times N_2O_2$ predefined beams, i.e.,
\begin{equation}\label{vmm}
 {\bf{v}}_{{m_1},m_2}=\left[ {\bf{u}}_{m_2},\; e^{j\frac{2\pi {m_1}}{O_1N_1}}{\bf{u}}_{m_2},\;\cdots, \;e^{j\frac{2\pi {m_1}(N_1-1)}{O_1N_1}}{\bf{u}}_{m_2} \right]^T,
\end{equation}
where ${m_1}=0,1,\ldots,N_1O_1-1$ and ${m_2}=0,1,\ldots,N_2O_2-1$ with
\begin{equation}
{\bf{u}}_{m_2}=\left[ 1, \; e^{j\frac{2\pi {m_2}}{O_2N_2}  },\; \cdots,\; e^{j\frac{2\pi {m_2}(N_2-1)}{O_2N_2}  }\right].
\end{equation}
The values of $N_1$ and $N_2$ are configured with the higher layer parameter \emph{n1-n2-codebookSubsetRestriction}. The values of $O_1$ and $O_2$ are associated with the values of $N_1$ and $N_2$ as shown in Table \ref{tabNO}. The number of CSI-RS ports ${P_{{\text{CSI-RS}}}}$ is equal to $2N_1N_2$ due to the use of two types of polarized antennas, where  ${P_{{\text{CSI-RS}}}} \in \{4,8,12,24,32\}.$ The supported layer number \(v\) is within the set \(\{1,2\}\), where the layer number signifies the number of data streams. Therefore, the UE should not report an RI greater than 2. Let ${\bf{v}}_{m_1^{(i)},m_2^{(i)}}$ represent the $i$-th selected beam from (\ref{vmm}). The precoding matrix based on the above beam bases can be written as 
\begin{equation}\label{a46}
 {\bf{w}}_{{q_1},{q_2},{n_1},{n_2},p_l^{(1)},p_l^{(2)},{c_l}}^l = \frac{1}{{\sqrt \beta _l  }}\left[ {\begin{array}{*{20}{c}}
{\sum\limits_{i = 0}^{L - 1} {a_{l,i}^1 \cdot { {\bf{v}}_{m_1^{(i)},m_2^{(i)}}}} }\\
{\sum\limits_{i = 0}^{L - 1} {a_{l,i}^2 \cdot { {\bf{v}}_{m_1^{(i)},m_2^{(i)}}}} }
\end{array}} \right],
\end{equation}
where ${\beta _l} = {N_1}{N_2}\sum\nolimits_{i = 0}^{2L - 1} {{{(p_{l,i}^{(1)}p_{l,i}^{(2)})}^2}}$ is the power scaling coefficient, $a_{l,i}^1 = p_{l,i}^{(1)}p_{l,i}^{(2)}{\varphi _{l,i}}$ is the combination weight of the beam basis for the first polarization direction, and $a_{l,i}^2 = p_{l,i + L}^{(1)}p_{l,i + L}^{(2)}{\varphi _{l,i + L}}$ is the combination weight of the beam basis for the second polarization direction. The parameters \(p^{(1)}_{l}\) and \(p^{(2)}_{l}\) represent the wideband amplitude coefficients and the subband one, respectively, for the number of data streams \(l \in \{1,2\}\). Both are vectors of length \(2L\), i.e.,
\begin{equation}
\begin{aligned}
p_l^{(1)} = [p_{l,0}^{(1)},p_{l,1}^{(1)}, \ldots ,p_{l,2L - 1}^{(1)}],\\
p_l^{(2)} = [p_{l,0}^{(2)},p_{l,1}^{(2)}, \ldots ,p_{l,2L - 1}^{(2)}],
\end{aligned}
\end{equation}
where the value of $L \in \left\{ {2,3,4} \right\}$  is configured with the higher layer parameter \emph{numberOfBeams}. The UE is configured with the higher layer parameter \emph{subbandAmplitude} set to true or false. The latter implies \(p^{(2)}_{l,i}=1\) for all $l$ and $i$. $p_l^{(1)}$ is reported by the indicator $i_{1,4,l}$ and $p_l^{(2)}$ is reported by the indicator $i_{2,2,l}$. The first \(L\) elements of each vector determine the amplitude coefficient for the selected \(L\) beams of the \(l\)-th data stream in the first polarization direction, while the last \(L\) elements determine that in the second polarization direction. The amplitude coefficient indicators $i_{1,4,l}$ and $i_{2,2,l}$ are composed of 
\begin{equation}\label{a48}
\begin{aligned}
&{i_{1,4,l}} = \left[ {k_{l,0}^{{\rm{(1)}}},k_{l,1}^{{\rm{(1)}}}, \ldots ,k_{l,2L - 1}^{{\rm{(1)}}}} \right],\;k_{l,i}^{{\rm{(1)}}} \in \left\{ {0,1, \ldots ,7} \right\},\\
&{i_{2,2,l}} = \left[ {k_{l,0}^{{\rm{(2)}}},k_{l,1}^{{\rm{(2)}}}, \ldots ,k_{l,2L - 1}^{{\rm{(2)}}}} \right],\;k_{l,i}^{{\rm{(2)}}} \in \left\{ {0,1} \right\}.
\end{aligned}
\end{equation}
The mapping from $k_{l,i}^{{\rm{(1)}}}$ to the amplitude coefficient $p_{l,i}^{(1)}$ is given in Table \ref{tabk1} and the mapping from $k_{l,i}^{{\rm{(2)}}}$  to the amplitude coefficient $p_{l,i}^{(2)}$ is presented in Table \ref{tabk2}. One can observe that the PMI does not directly indicate numerical values; instead, it points to the index corresponding to the values in a table, thereby reducing the overhead of feedback. Furthermore, the wideband amplitude coefficient requires 3 bits for indication, while the subband amplitude coefficient needs only 1 bit. This hybrid approach of wideband and subband indication is designed to make a trade-off between feedback accuracy and overhead. The parameter ${\varphi _{l,i}}$ represents the subband phase coefficient on the $i$-th beam basis for the $l$-th data stream. The phase coefficient indicators are given by
\begin{equation}\label{a49}
{i_{2,1,l}} = \left[ {{c_{l,0}},{c_{l,1}}, \ldots ,{c_{l,2L - 1}}} \right],
\end{equation}
and the mapping from ${c_{l,i}}$ to ${\varphi _{l,i}}$ is 
\begin{equation}
{\varphi _{l,i}} = {e^{j{{2\pi {c_{l,i}}} \mathord{\left/
 {\vphantom {{2\pi {c_{l,i}}} {{N_{{\rm{PSK}}}}}}} \right.
 \kern-\nulldelimiterspace} {{N_{{\rm{PSK}}}}}}}},
 \end{equation}
where ${N_{{\rm{PSK}}}} \in \left\{ {4,8} \right\}$  is configured with the higher layer parameter \emph{phaseAlphabetSize}.

\begin{table}[t]   
\begin{center} 
\centering  
\captionsetup{
  labelsep=newline, % 让标题另起一行
}
\renewcommand{\arraystretch}{2}  %每行的高度为1.5倍
\caption{Mapping of Elements of $i_{1,4,l}$: $k_{l,i}^{(1)}$ to $p_{l,i}^{(1)}$}  \label{tabk1}
\begin{tabular}{|c|c|}   
\hline  
\rowcolor{gray!20}
$k_{l,i}^{(1)}$ &$p_{l,i}^{(1)}$  \\
\hline  
0 &0\\
\hline  1&$\sqrt{1/64}$\\
\hline  2&$\sqrt{1/32}$\\
\hline  3&$\sqrt{1/16}$\\
\hline  4&$\sqrt{1/8}$\\
\hline  5&$\sqrt{1/4}$\\
\hline  6&$\sqrt{1/2}$\\
\hline  7&$1$\\
\hline
\end{tabular}   
\end{center}   
\vspace{-12pt}
\end{table}

\begin{table}[t]   
\begin{center} 
\centering  
\captionsetup{
  labelsep=newline, % 让标题另起一行
}
\renewcommand{\arraystretch}{2}  %每行的高度为1.5倍
\caption{Mapping of Elements of $i_{2,2,l}$: $k_{l,i}^{(2)}$ to $p_{l,i}^{(2)}$}  \label{tabk2}
\begin{tabular}{|c|c|}   
\hline  
\rowcolor{gray!20}
$k_{l,i}^{(2)}$ &$p_{l,i}^{(2)}$  \\
\hline  0&$\sqrt{1/2}$\\
\hline  1&$1$\\
\hline
\end{tabular}   
\end{center}   
\vspace{-10pt}
\end{table}

Equations (\ref{a48}) and (\ref{a49}) illustrate how the UE reports the combination weights of the $L$ beams in (\ref{a46}), i.e., $a_{l,i}^1 = p_{l,i}^{(1)}p_{l,i}^{(2)}{\varphi _{l,i}}$ and $a_{l,i}^2 = p_{l,i + L}^{(1)}p_{l,i + L}^{(2)}{\varphi _{l,i + L}}$ for $i=0,1,\ldots,L-1$, through PMI feedback. Prior to this, the UE must also inform the gNB of which $L$ beams have been selected. To reduce the overhead of beam indication, the process is divided into two steps, as shown in Fig. \ref{pic_9}. This approach results in the beam indices of ${{\bf{v}}_{m_1^{(i)},m_2^{(i)}}}$ being represented as 
\begin{equation}\label{a32}
    \begin{aligned}
        m_{1}^{(i)}&=O_1n_{1}^{(i)}+q_1,\\
        m_{2}^{(i)}&=O_2n_{2}^{(i)}+q_2,
    \end{aligned}
\end{equation}
where 
\begin{equation} 
\begin{aligned}
{q_1} &\in \left\{ {0,1, \ldots ,{O_1} - 1} \right\},\\
{q_2} &\in \left\{ {0,1, \ldots ,{O_2} - 1} \right\},\\
n_1^{(i)} &\in \left\{ {0,1, \ldots ,{N_1} - 1} \right\},\\
n_2^{(i)} &\in \left\{ {0,1, \ldots ,{N_2} - 1} \right\}.
\end{aligned}
\end{equation}
In step 1, the UE selects an orthogonal beam group using $q_1$ and $q_2$, i.e., horizontal and vertical indices, where the beams of different colors represent different orthogonal beam groups in Fig. \ref{pic_9}. In step 2, the UE selects $L$ beams within the selected beam group using horizontal and vertical indices $n_1^{(i)}$ and $n_2^{(i)}$. The selected beam group and associated beams are identified by $i_{1,1}$ and $i_{1,2}$, i.e., 
\begin{equation}
  {i_{1,1}} = [{{q_1}},{{q_2}}], \;\;\;{i_{1,2}} \in \left\{ 0,1, \ldots , C(N_1N_2,L)-1 \right\},
\end{equation}
where values of $C(x,y)$ are given in 38.214 Table 5.2.2.2.3-1. As a matter of fact, $C(x,y)$ represents the number of ways to choose $x$ elements from a set of $y$ elements, which can also be computed by % where values of $C(x,y)$ are given in Table \ref{CombinTab} (also available in 38.214 Table 5.2.2.2.3-1).
\begin{equation}\label{cxy}
 C(x,y) = \frac{{x!}}{{y!(x - y)!}},\;\;x > y.
\end{equation}
It can be seen that \(i_{1,2}\) does not individually indicate the indices of the $L$ beams, but rather indicates the index of a beam combination. This method can reduce the feedback overhead. When the BS receives $i_{1,2}$, it needs to recover $\{n_1^{(i)}, n_2^{(i)}\}_{i=0}^{L-1}$ from it according to an algorithm predefined in the protocol, as shown in Algorithm 1.
\renewcommand{\algorithmicrequire}{\textbf{Input:}}
\renewcommand{\algorithmicensure}{\textbf{Output:}}
\begin{algorithm}
\caption{Mapping of $i_{1,2}$ to  $\{n_1^{(i)}, n_2^{(i)}\}_{i=0}^{L-1}$}
\label{ciawf}       %
\begin{algorithmic}[1]
\Require $L$, $N_1$, $N_2$, $i_{1,2}$
\State \textbf{Initialize:  $s_{-1}=0$ }
\For {$i = 0,1,\ldots, L-1$}
\State Find the largest ${x^*} \in \left\{ {L - 1 - i, \ldots ,{N_1}{N_2} - 1 - i} \right\}$ such that ${i_{1,2}} - {s_{i - 1}} \ge C\left( {{x^*},L - i} \right)$.
\State ${e_i} = C\left( {{x^*},L - i} \right)$
\State ${s_i} = {s_{i - 1}} + {e_i}$
\State ${n^{\left( i \right)}} = {N_1}{N_2} - 1 - {x^*}$
\State $n_1^{\left( i \right)} = {n^{\left( i \right)}}\bmod {N_1}$
\State $n_2^{\left( i \right)} = \left( {{n^{\left( i \right)}} - n_1^{\left( i \right)}} \right)/{N_1}$
\EndFor
\Ensure $\{n_1^{(i)}, n_2^{(i)}\}_{i=0}^{L-1}$
\end{algorithmic}
\end{algorithm}

\begin{figure}[t]
	\centering
	\includegraphics[width=0.4\textwidth]{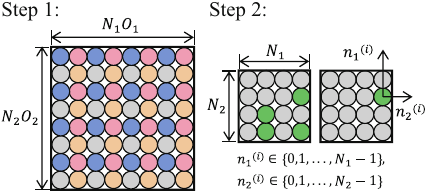}
	\caption{Beam indication in Type II Regular Codebook.}\label{pic_9}
    \vspace{-1pt}
\end{figure}

Algorithm 1 illustrates the mapping relationship between the combination types of \(L\) beams and their indices under a certain set of rules. This mapping is not unique; therefore, Algorithm 1 is merely the one adopted by the 3GPP protocol, but it is not the only algorithm for mapping of $i_{1,2}$ to  $\{n_1^{(i)}, n_2^{(i)}\}_{i=0}^{L-1}$. 

We have now described how a precoding matrix is constructed by feeding back PMI. In summary, the indicator \({i_1}\) is responsible for reporting the indices of the selected beams, the position of the strongest coefficient on each layer, and the wideband amplitude coefficients. The indicator \({i_2}\) is responsible for reporting the indices of subband amplitude coefficients and phase coefficients. The current protocol supports a feedback RI value \(v \in \{1,2\}\), where each PMI value corresponding to $v$ is given by
\[\begin{aligned}
&{i_1} = \left\{ {\begin{aligned}
&{\left[ 
{{i_{1,1}}}\;\;{{i_{1,2}}}\;\;{{i_{1,3,1}}}\;\;{{i_{1,4,1}}}
 \right]},\quad {v  = 1},\\
&{\left[ 
{{i_{1,1}}}\;\;{{i_{1,2}}}\;\;{{i_{1,3,1}}}\;\;{{i_{1,4,1}}}\;\;{{i_{1,3,2}}}\;\;{{i_{1,4,2}}}
\right]},\quad {v  = 2},
\end{aligned}} \right.\\
&{i_2} = \left\{ {\begin{aligned}
&{\left[ {{i_{2,1,1}}} \right]},\quad {{\rm{SA}} = {\rm{false}},\;v  = 1},\\
&{\left[ 
{{i_{2,1,1}}}\;\;{{i_{2,1,2}}}
 \right]},\quad {{\rm{SA}} = {\rm{false}},\;v  = 2},\\
&{\left[ 
{{i_{2,1,1}}}\;\;{{i_{2,2,1}}}
\right]},\quad {{\rm{SA}} = {\rm{true}},\;v  = 1},\\
&{\left[ 
{{i_{2,1,1}}}\;\;{{i_{2,2,1}}}\;\;{{i_{2,1,2}}}\;\;{{i_{2,2,2}}}
 \right]},\quad  {{\rm{SA}} = {\rm{true}},\;v  = 2},
\end{aligned}} \right.
\end{aligned}\]
where SA represents the higher layer parameter \emph{subbandAmplitude}, which can be set to true or false.
It can be observed that \(i_{1,1}\) and \(i_{1,2}\) are independent of the layer, i.e., there are no parameters such as \(i_{1,1,l}\) and \(i_{1,2,l}\), where \(l=1,2\). This indicates that a common set of beam bases is selected across different layers, utilizing different combination coefficients to generate orthogonal precoding vectors, as shown in Fig. \ref{pic_10}.
\begin{figure}[t]
	\centering
	\includegraphics[width=0.48\textwidth]{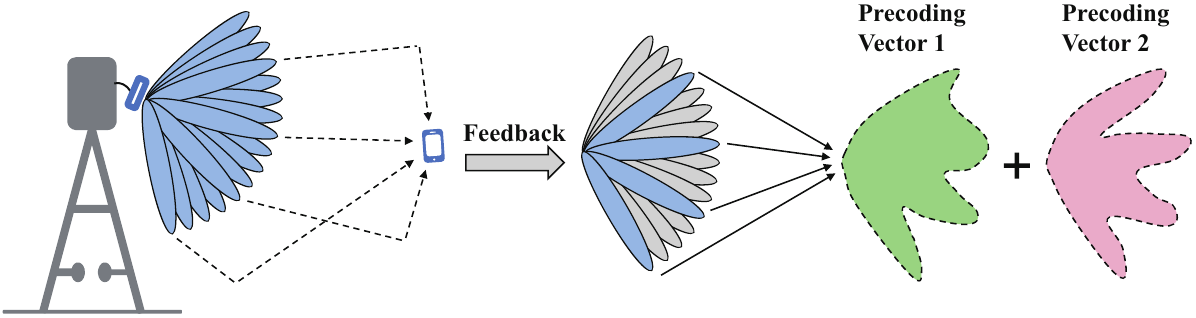}
	\caption{Precoding vectors by Type II Regular Codebook.}\label{pic_10}
    \vspace{-2pt}
\end{figure}

To recapitulate, we have already introduced all the PMI parameters, with the exception of \(i_{1,3,l},\; l=1,2\). This parameter is used to indicate the strongest coefficient on layer \(l\), with the aim of reducing feedback overhead and performing scaling normalization on the amplitude. As a consequence, the default value of indicators $k_{l,{i_{1,3,l}}}^{{(1)}} = 7$, $k_{l,{i_{1,3,l}}}^{{(2)}} = 1$, and ${c_{l,i{}_{1,3,l}}} = 0$, with these parameters not reported for $l = 1, \ldots ,v $.
To further reduce feedback overhead, various strategies have been proposed in the 3GPP protocol. For instance,  when SA is set to false, \(k_{l,i}^{(2)} = 1\) for all \(l\) and \(i\), \({i_{2,2,l}}\) is not reported, and only phase information is reported for subbands. When SA is set to true, the following principles are also used to reduce overhead:
\begin{itemize}
  \item The subband amplitude for the beam corresponding to the strongest coefficient is not reported.
  \item The subband amplitudes and phases are not reported when the associated wideband amplitude is zero.
  \item For \(L=2\) or \(3\), the number of subband amplitudes reported for each layer does not exceed 4.
  \item When \(L=4\), the number of subband amplitudes reported for each layer does not exceed 6.
\end{itemize}
Let ${M_l}$ be the number of elements of ${i_{1,4,l}}$ that satisfy $k_{l,i}^{{(1)}} > 0$. Then, for $l = 1, \ldots ,v$, the elements of ${i_{2,2,l}}$ and ${i_{2,1,l}}$ corresponding to the $\min \left( {{M_l},{K^{(2)}}} \right) - 1$ strongest coefficients (excluding the strongest coefficient indicated by ${i_{1,3,l}}$), are reported, where
\begin{equation}
\begin{aligned}
&k_{l,i}^{{\rm{(2)}}} \in \left\{ {0,1} \right\},\;\;\;\;{c_{l,i}} \in \left\{ {0,1, \ldots ,{N_{{\rm{PSK}}}} - 1} \right\},\\
&{K^{{\rm{(2)}}}} = 4\;{\rm{if }}\;L = 2,3;\;\;\;\;\;{K^{{\rm{(2)}}}} = 6\;{\rm{if }}\;L = 4,
\end{aligned}
\end{equation}
where ${K^{{\rm{(2)}}}}$ is the maximum allowed number of subband amplitudes reported for each layer. The remaining $2L - \min \left( {{M_l},{K^{{\rm{(2)}}}}} \right)$ elements of ${i_{2,2,l}}$ are not reported and are set to $k_{l,i}^{{\rm{(2)}}} = 1$. The elements of ${i_{2,1,l}}$ corresponding to the ${M_l} - \min \left( {{M_l},{K^{{\rm{(2)}}}}} \right)$ weakest non-zero coefficients are reported, where ${c_{l,i}} \in \left\{ {0,1,2,3} \right\}$. The remaining $2L - {M_l}$ elements of ${i_{2,1,l}}$ are not reported and are set to ${c_{l,i}} = 0$. 

In the Type I Codebook, there is a bit sequence ${\bf{a}}$ (see (\ref{type1map})) that restricts beams used in the precoding vector. In the Type II codebook, there is a joint beam and power restriction sequence $B=B_1B_2$ (concatenation) that first selects 4 beam groups with a bitmap of 11 bits, i.e.,
\begin{equation}
 B_1=b_1^{(10)}b_1^{(9)}\ldots b_1^{(0)},
\end{equation}
and then restricting the maximum allowed power value for each beam in the selected beam groups, with an additional bitmap of  \( 8N_1N_2 \) bits, i.e.,
\begin{equation}
\begin{aligned}
 &B_2=B_2^{(0)}B_2^{(1)}B_2^{(2)}B_2^{(3)},\\
 &B_2^{(k)}=b_2^{(k,2N_1N_2-1)},\ldots,b_2^{(k,0)},\\
\end{aligned}
\end{equation}
for $k=0,1,2,3$, where each bitmap $B_2^{(k)}$ has $2N_1N_2$ bits. The bits $b_2^{\left( {k,2\left( {{N_1}{x_2} + {x_1}} \right) + 1} \right)}b_2^{\left( {k,2\left( {{N_1}{x_2} + {x_1}} \right)} \right)}$ indicate the maximum allowed amplitude coefficient $p_{l,i}^{\left( 1 \right)}$ for the vector in group ${g^{\left( k \right)}}$ indexed by ${x_1},{x_2}$, where the maximum amplitude coefficients are presented in Table \ref{tabmaxap}. The bit sequences $B_1$ and $B_2$ are concatenated to construct the full restriction sequence $B$.

\begin{table}[t]   
\centering
\captionsetup{
  labelsep=newline, % 让标题另起一行
}
\renewcommand{\arraystretch}{2}  %每行的高度为1.5倍
\caption{Maximum Allowed Amplitude Coefficients}  \label{tabmaxap}
\begin{tabular}{|>{\centering\arraybackslash}p{4.5cm}|>{\centering\arraybackslash}p{2.5cm}|}
\hline  
\rowcolor{gray!20}
 $b_{2}^{(k,2(N_{1}x_{2}+x_{1})+1)}b_{2}^{(k,2(N_{1}x_{2}+x_{1}))}$ &  $p_{l,i}^{(1)}$ \\
\hline 
00 & 0 \\
\hline 
01 & $\sqrt{1/4}$ \\
\hline 
10 & $\sqrt{1/2}$ \\
\hline 
11 & 1 \\
\hline
\end{tabular}     
\vspace{-5pt}
\end{table}

\begin{figure*}[t]
	\centering
	\includegraphics[width=0.9\textwidth]{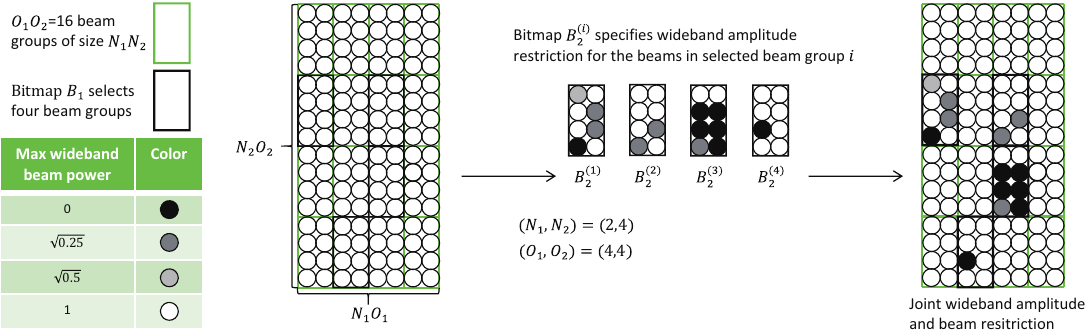}
	\caption{Joint beam and power restriction on beams in Type II Codebook.}\label{pic_11}
\end{figure*}

The 3GPP protocol refers to this as codebook subset restriction, and it was described using highly complex mathematical methods to detail the process. Specifically, the protocol first defines the ${O_1}{O_2}$ vector groups as
\[
G\left( {{r_1},{r_2}} \right) = \left\{ {{v_{{N_1}{r_1} + {x_1},{N_2}{r_2} + {x_2}}}\left| {\begin{aligned}
&{{x_1} = 0,1, \ldots ,{N_1} - 1}\\
&{{x_2} = 0,1, \ldots ,{N_2} - 1}
\end{aligned}} \right.} \right\}\]
for ${r_1} \in \left\{ {0,1, \ldots ,{O_1} - 1} \right\}$ and ${r_2} \in \left\{ {0,1, \ldots ,{O_2} - 1} \right\}$. The UE shall be configured with restrictions for 4 vector groups indicated by $\left( {r_1^{\left( k \right)},r_2^{\left( k \right)}} \right)$ for $k=0,1,2,3$ and identified by the group indices
\begin{equation}\label{a58}
    {g^{\left( k \right)}} = {O_1}r_2^{\left( k \right)} + r_1^{\left( k \right)},
\end{equation} 
where the indices are assigned such that ${g^{\left( k \right)}}$ increases as $k$ increases. ${B_1} = b_1^{\left( {10} \right)} \cdots b_1^{\left( 0 \right)}$ is the binary representation of the integer ${\beta _1}$ with
\begin{equation}\label{a59}
    {\beta _1} = \sum\limits_{k = 0}^3 {C\left( {{O_1}{O_2} - 1 - {g^{\left( k \right)}},4 - k} \right)},
\end{equation}
where $C( {x,y} )$ is defined in (\ref{cxy}). The group indices ${g^{\left( k \right)}}$ and indicators $\big( {r_1^{\left( k \right)},r_2^{\left( k \right)}} \big)$ for $k = 0,1,2,3$ may be found from ${\beta _1}$ by the following Algorithm 2.
\begin{figure}[t] 
    \centering 
    \includegraphics[width=0.45\textwidth]{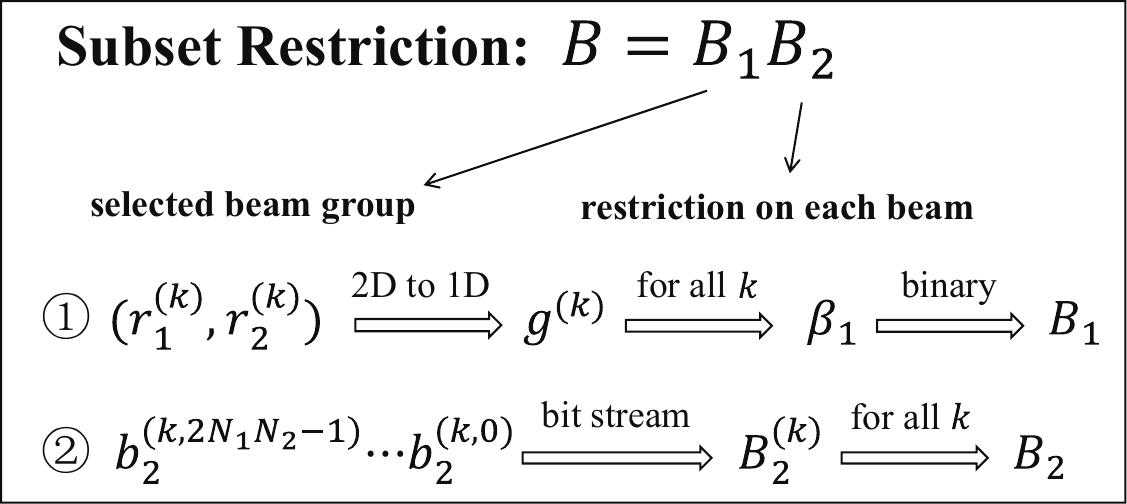} 
    \caption{The symbol relationship in subset restriction.} 
    \label{pic_12} 
    \vspace{-12pt}
\end{figure}

\renewcommand{\algorithmicrequire}{\textbf{Input:}}
\renewcommand{\algorithmicensure}{\textbf{Output:}}
\begin{algorithm}
\caption{Mapping of ${\beta _1}$ to  ${g^{\left( k \right)}}$ and $\big( {r_1^{\left( k \right)},r_2^{\left( k \right)}} \big)$}
\label{ciawf}       %
\begin{algorithmic}[1]
\Require  $O_1$, $O_2$, ${\beta _1}$
\State \textbf{Initialize:  $s_{-1}=0$ }
\For {$k = 0,1,\ldots, 3$}
\State Find the largest ${x^*} \in \left\{ {3 - k, \ldots ,{O_1}{O_2} - 1 - k} \right\}$ such that ${\beta _1} - {s_{k - 1}} \ge C\left( {{x^*},4 - k} \right)$.
\State ${e_k} = C\left( {{x^*},4 - k} \right)$
\State ${s_k} = {s_{k - 1}} + {e_k}$
\State ${g^{\left( k \right)}} = {O_1}{O_2} - 1 - {x^*}$
\State $r_1^{\left( k \right)} = {g^{\left( k \right)}}\bmod {O_1}$
\State $r_2^{\left( k \right)} = \left( {{g^{\left( k \right)}} - r_1^{\left( k \right)}} \right)/{O_1}$
\EndFor
\Ensure ${g^{\left( k \right)}}$ and $\big( {r_1^{\left( k \right)},r_2^{\left( k \right)}} \big)$
\end{algorithmic}
\end{algorithm}

For ease of understanding the protocol details provided above, Fig. \ref{pic_11} illustrates the process of indicating restriction on beam and power through the bit sequence \( B=B_1B_2\), in conjunction with the symbol relationship shown in Fig. \ref{pic_12} to comprehend Algorithm 2. First, we divide all available beams within the \( N_1O_1 \times N_2O_2 \) grid into \( O_1O_2 \) groups, each containing \( N_1N_2 \) beams. $B_1$ selects four of these beam groups, and $B_2$ applies restrictions on beam and power to these selected groups. Since $B_1$ is a bit sequence, we need to first determine the group indices $\big( {r_1^{\left( k \right)},r_2^{\left( k \right)}} \big)$ for \( k=0, 1, 2, 3 \), where \( r_1^{\left( k \right)} \) and \( r_2^{\left( k \right)}  \) represent the horizontal and vertical group indices, respectively. Using (\ref{a58}), we can convert the two-dimensional indices $\big( {r_1^{\left( k \right)},r_2^{\left( k \right)}} \big)$ into a one-dimensional index ${g^{\left( k \right)}}$. Then, (\ref{a59}) allows us to represent the indices of all four beam groups with a single number ${\beta _1}$, and $B_1$ is the binary representation of this number. Therefore, Algorithm 2 essentially involves deducing ${g^{\left( k \right)}}$ from ${\beta _1}$, and then inferring $\big( {r_1^{\left( k \right)},r_2^{\left( k \right)}} \big)$. In $B_2$, the four \( B_2^{(k)} \) individually indicate the beam and power restrictions for the four selected beam groups. Since each beam group has \( N_1N_2 \) beams and each beam requires 2 bits to indicate its restriction, \( B_2^{(k)} \) requires \( 2N_1N_2 \) bits, and thus $B_2$ requires \( 8N_1N_2 \) bits. This completes the tutorial on R15 Type II Codebook and the emerging parameters are interpreted in Table \ref{paratypeii}. 

With the standardization of the first Type II Regular Codebook, Release 15 also froze the first Type II Port-Selection Codebook. The design concept of this codebook is essentially the same as that of the regular codebook, with the only difference being that beam selection is replaced by port selection. Interested readers may refer to Appendix \ref{r15ps}.

\begin{table}[t]   
\begin{center} 
\centering  
\captionsetup{
  labelsep=newline, % 让标题另起一行
}
\renewcommand{\arraystretch}{1.5}  %每行的高度为1.5倍
\caption{New Parameters in Type II Regular Codebook.}  
\label{paratypeii} 
\begin{tabular}{|c|m{6cm}|}   
\hline   \textbf{Parameters} &\textbf{Interpretation} \\  
\hline   $i_{1,1}$ & Indicates the index of the selected orthogonal beam group among the oversampled beam groups. \\
\hline   $i_{1,2}$ & Indicates the index of the combination of $L$ beams within the selected beam group.\\
\hline   $i_{1,3,l}$ &  Indicates the strongest coefficient on layer \(l\).  \\
\hline   $i_{1,4,l}$ & Indicates the wideband amplitude coefficient of the selected beams on layer \(l\). \\
\hline   $i_{2,1,l}$ & Indicates the subband phase coefficient of the selected beams on layer \(l\). \\
\hline   $i_{2,2,l}$ & Indicates the subband amplitude coefficient of the selected beams on layer \(l\). \\

\hline   ${L}$ & The number of selected beams.\\
\hline   $\beta _l$ & The power scaling coefficient on layer \(l\).\\
\hline   ${\bf{v}}_{{m_1},m_2}$ & The beam basis on direction of $({m_1},m_2).$ \\
\hline   $p_{l,i}^{(1)}$ and $k_{l,i}^{{\rm{(1)}}}$ & The wideband amplitude coefficient and associated index of the $i$-th beam on layer \(l\), respectively.\\
\hline   $p_{l,i}^{(2)}$ and $k_{l,i}^{{\rm{(2)}}}$& The subband amplitude coefficient and associated index of the $i$-th beam on layer \(l\), respectively.\\
\hline   ${\varphi _{l,i}}$ and ${c_{l,i}}$ & The subband phase coefficient and associated quantity of the $i$-th beam on layer \(l\), respectively.\\
\hline   ${q_1}$ and ${q_2}$ &  Indicate the selected orthogonal beam group among the oversampled beam groups.\\
\hline   $n_1^{(i)}$ and $n_2^{(i)}$ & Indicate the horizontal and vertical indices of the $i$-th selected beam within a beam group.\\
\hline    ${M_l}$& Indicates the number of non-zero wideband amplitude coefficients. \\
\hline    ${K^{{\rm{(2)}}}}$ & Indicates the maximum allowed number of subband amplitudes reported for each layer. \\
\hline    $B_1$ & The bit sequence for selecting four restricted beam groups among $O_1O_2$ beam groups.\\
\hline    $B_2$ & The bit sequence to detail the restrictions on beam and power of the selected four beam groups.\\
\hline   
\end{tabular}   
\end{center}   
\vspace{-12pt}
\end{table}

\begin{table*}[t]  
\centering   
\captionsetup{
  labelsep=newline, % 让标题另起一行
}
\renewcommand{\arraystretch}{2}  
\caption{R16 Enhanced Type II Codebook (TS 38.214 Table 5.2.2.2.5-5)}  \label{tabesII}
\begin{tabular}{|Sc|Sc|}   
\hline  
\rowcolor{gray!20}
\textbf{Layers}  &  \\
\hline   
$v=1$ & ${\bf{w}}^{(1)}_{q_1,q_2,n_1,n_2,n_{3,1},p^{(1)}_1,p^{(2)}_1,i_{2,5,1},t}
={\bf{w}}^1_{q_1,q_2,n_1,n_2,n_{3,1},p^{(1)}_1,p^{(2)}_1,i_{2,5,1},t}$  \\
\hline
% $v=2,3,4$& ... \\
$v=2$ & $ {\bf{W}}^{(2)}_{q_1,q_2,n_1,n_2,n_{3,1},p^{(1)}_1,p^{(2)}_1,i_{2,5,1},n_{3,2},p^{(1)}_2,p^{(2)}_2,i_{2,5,2},t}
= \frac{1}{\sqrt{2}}
\begin{bmatrix} {\bf{w}}^1_{q_1,q_2,n_1,n_2,n_{3,1},p^{(1)}_1,p^{(2)}_1,i_{2,5,1},t} \ \ {\bf{w}}^2_{q_1,q_2,n_1,n_2,n_{3,2},p^{(1)}_2,p^{(2)}_2,i_{2,5,2},t}
\end{bmatrix}$ \\
\hline
$v=3$ & {\linespread{2}\selectfont\makecell{$ {\bf{W}}^{(3)}_{q_1,q_2,n_1,n_2,n_{3,1},p^{(1)}_1,p^{(2)}_1,i_{2,5,1},n_{3,2},p^{(1)}_2,p^{(2)}_2,i_{2,5,2},n_{3,3},p^{(1)}_3,p^{(2)}_3,i_{2,5,3},t}$ \qquad \qquad \qquad \qquad \qquad \qquad\qquad \qquad \qquad \qquad \qquad \qquad \qquad \qquad \\
$\qquad \qquad \qquad \qquad= \frac{1}{\sqrt{3}}
\begin{bmatrix} {\bf{w}}^1_{q_1,q_2,n_1,n_2,n_{3,1},p^{(1)}_1,p^{(2)}_1,i_{2,5,1},t}  \ \ {\bf{w}}^2_{q_1,q_2,n_1,n_2,n_{3,2},p^{(1)}_2,p^{(2)}_2,i_{2,5,2},t} \ \
{\bf{w}}^3_{q_1,q_2,n_1,n_2,n_{3,3},p^{(1)}_3,p^{(2)}_3,i_{2,5,3},t}
\end{bmatrix}$} }\\
\hline
$v=4$ & {\linespread{2}\selectfont\makecell{$ {\bf{W}}^{(4)}_{q_1,q_2,n_1,n_2,n_{3,1},p^{(1)}_1,p^{(2)}_1,i_{2,5,1},n_{3,2},p^{(1)}_2,p^{(2)}_2,i_{2,5,2},n_{3,3},p^{(1)}_3,p^{(2)}_3,i_{2,5,3},n_{3,4},p^{(1)}_4,p^{(2)}_4,i_{2,5,4},t}$ \qquad \qquad \qquad \qquad \qquad \qquad \qquad \qquad \qquad \qquad\\
$= \frac{1}{2}
\begin{bmatrix} {\bf{w}}^1_{q_1,q_2,n_1,n_2,n_{3,1},p^{(1)}_1,p^{(2)}_1,i_{2,5,1},t}  \ \ {\bf{w}}^2_{q_1,q_2,n_1,n_2,n_{3,2},p^{(1)}_2,p^{(2)}_2,i_{2,5,2},t}  \ \cdots \
{\bf{w}}^4_{q_1,q_2,n_1,n_2,n_{3,4},p^{(1)}_4,p^{(2)}_4,i_{2,5,4},t}
\end{bmatrix}$} }\\
\hline
\multicolumn{2}{|Sc|}{\linespread{2}\selectfont  
\makecell{  Precoding vector: ${\bf{w}}^l_{q_1,q_2,n_1,n_2,n_{3,l},p^{(1)}_l,p^{(2)}_l,i_{2,5,l},t}
=\frac{1}{\sqrt{N_1N_2\gamma_{t,l}}}
\begin{bmatrix}\sum^{L-1}_{i=0}{\bf{v}}_{m_1^{(i)}m_2^{(i)}}p^{(1)}_{l,0}\sum^{M_v-1}_{f=0}y^{(f)}_{t,l}p^{(2)}_{l,i,f}\varphi_{l,i,f}\\
\sum^{L-1}_{i=0}{\bf{v}}_{m_1^{(i)}m_2^{(i)}}p^{(1)}_{l,1}\sum^{M_v-1}_{f=0}y^{(f)}_{t,l}p^{(2)}_{l,i+L,f}\varphi_{l,i+L,f}
\end{bmatrix},\;\;l=1,2,3,4,$ \\
$\gamma_{t,l}=\sum^{2L-1}_{i=0}(P_{l,\left\lfloor \frac{i}{L}\right\rfloor}^{(1)})^2\begin{vmatrix}\sum^{M_v-1}_{f=0}y^{(f)}_{t,l}p^{(2)}_{l,i,f}\varphi_{l,i,f}\end{vmatrix}^2$.\\
  }
}\\ 
\hline
\end{tabular}  
\end{table*}

\subsection{R16 Enhanced Type II Regular Codebook}\label{enhanceregularII}
The Type II Regular Codebook proposed in Release 15 reduces the feedback overhead for the UE by applying spatial domain compression, which represents the space domain of $N_1N_2$ dimensions through a linear combination of $L$ spatial bases. However, for every subband, the UE is required to report the combination coefficients for $L$ beams to generate a subband-level PMI matrix. In Release 16, the Enhanced Type II Codebook further considers frequency-domain compression by representing the frequency domain of $N_3$ dimensions through a linear combination of $M_v$ spectral bases. The spatial bases can be referred to as beams, whereas the spectral bases can be referred to as taps (or delay taps). The reason compression can be applied in both the spatial and frequency domains is that the wireless channel in a MIMO system, as well as its associated precoding matrices, exhibit sparsity in both the angular domain and the delay domain naturally, particularly in mmWave bands.

The expression of the Enhanced Type II Regular Codebook in the protocol is presented in Table \ref{tabesII}. As can be seen, the supported layer number set is increased compared to the Type II Regular Codebook, with \(v\in\{1,2,3,4\}\). Moreover, this codebook combines the spatial bases ${\bf{v}}_{{m_1},m_2}$ from (\ref{vmm}) with the newly introduced spectral basis ${{\bf{y}}_f}$, i.e., 
\begin{equation}
 {{\bf{y}}_f} = {\left[ {y_{0,l}^{(f)},y_{1,l}^{(f)},\ldots,y_{{N_3} - 1,l}^{(f)}} \right]^T},   
\end{equation}
where $f=0,1,\ldots,M_v-1$ is the index of the selected tap, $M_v$ is the total number of selected taps in the delay domain (which is similar to the parameter $L$ in the angular domain), and 
\begin{equation}\label{b45}
y_{t,l}^{(f)} = {e^{j\frac{{2\pi tn_{3,l}^{(f)}}}{{{N_3}}}}},\;\;n_{3,l}^{(f)} \in \{ 0,1,\ldots,{N_3} - 1\},
\end{equation}
where $t=0,1,\ldots,N_3-1$ represents the index of the reported precoding matrix in the frequency domain. $n_{3,l}^{(f)}$ represents the index of the $f$-th selected tap in the delay domain. For ease of exposition, we refer to $t$ as the frequency index.\footnote{Although unusual, in the protocol codebook, $t$ is used to represent frequency index for precoding matrices on various subbands, while $f$ denotes the time index of the main taps for precoding matrices in the delay domain.} Although not explicitly stated in the protocol, for ease of understanding the codebook, we rewrite the spectral basis  as 
\begin{equation}\label{b46}
{{\bf{y}}_f} = {\Big[ {1,{e^{j\frac{{2\pi n_{3,l}^{(f)}}}{{{N_3}}}}},\ldots,{e^{j\frac{{2\pi ({N_3} - 1)n_{3,l}^{(f)}}}{{{N_3}}}}}} \Big]^T},    
\end{equation}
which admits a DFT vector with dimension of $N_3$. Thus, there are a total of $N_3$ orthogonal DFT vectors with $n_{3,l}^{(f)}\in \{0,1,\ldots,N_3-1\}$ from which $M_v$ vectors are selected in the $l$-th layer, indexed by $f=0,1,\ldots,M_v-1$, to represent the precoding matrices across various subbands.

The parameter $R \in \{1,2\}$, configured by the higher-layer parameter \emph{numberOfPMI-SubbandsPerCQI-Subband,} denotes the number of precoding matrices per subband. It determines the total number of precoding matrices $N_3$ across the entire BWP, as indicated by the PMI, based on the configured subbands in \emph{csi-ReportingBand}. As illustrated in Fig. \ref{pic_13}, the number of configured subbands is determined by the BWP size and associated subband size given by Table \ref{tabCSS} in the Appendix \ref{reporting}. As there are $R$ precoding matrices reported in each subband, the number of configured subbands can also be expressed as

\begin{figure}[t] 
    \centering 
    \includegraphics[width=0.45\textwidth]{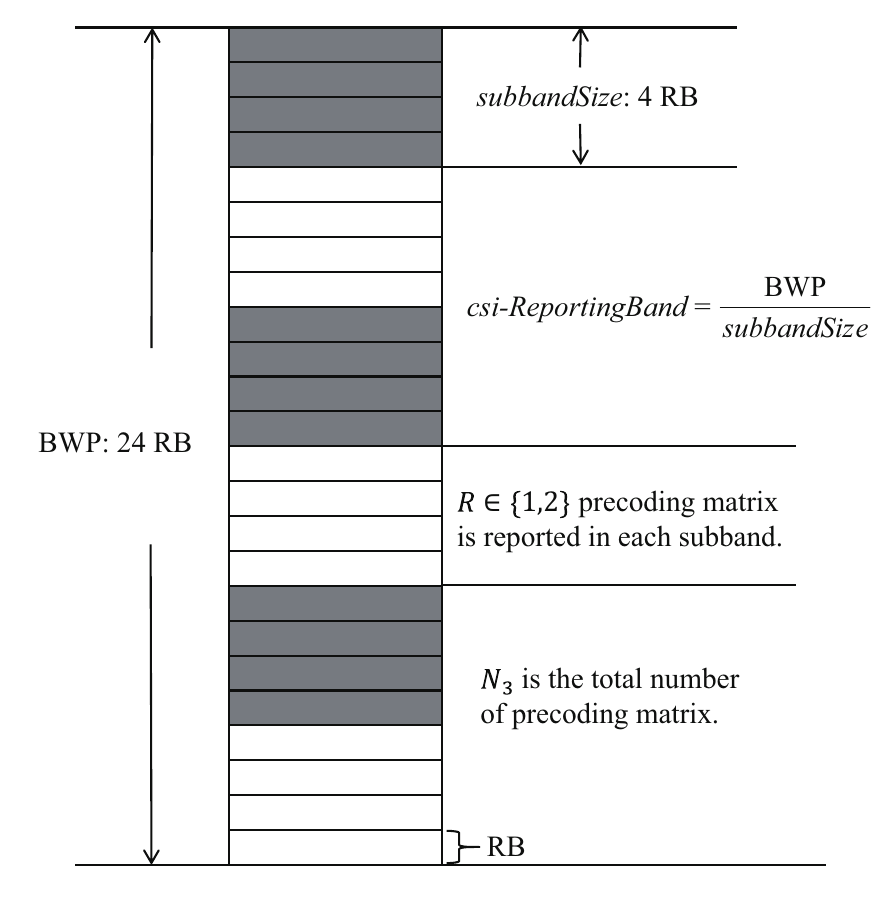} 
    \caption{Reporting granularity of the precoding matrix in the frequency domain for Enhanced Type II Regular Codebook.} 
    \label{pic_13} 
    \vspace{-5pt}
\end{figure}
\begin{figure*}[t] 
    \centering 
    \includegraphics[width=0.95\textwidth]{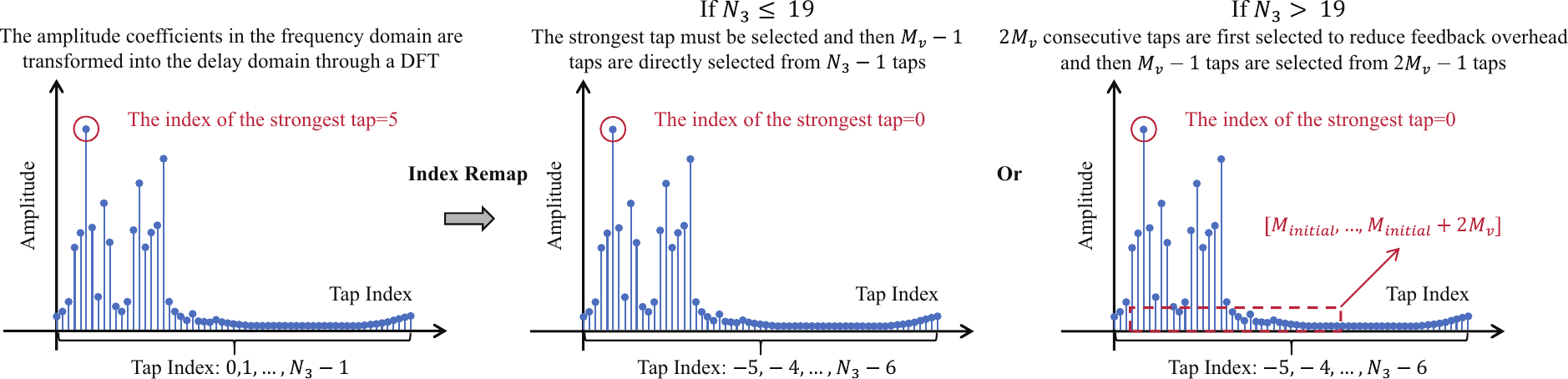} 
    \caption{Tap indication for wideband precoding matrices in Enhanced Type II Codebook.} 
    \label{pic_14} 
    \vspace{-5pt}
\end{figure*}
%The parameter $R \in \{1,2\}$ configured with the higher-layer parameter \emph{numberOfPMI-SubbandsPerCQI-Subband} \red{determines} the total number of precoding matrices $N_3$ indicated by the PMI as a function of the number of configured subbands in \emph{csi-ReportingBand}. As illustrated in Fig. \ref{pic_13}, the number of configured subbands in determined by the BWP size and associated subband size given by Table \ref{tabCSS}. As there are $R$ precoding matrices reported in each subband, the number of configured subbands can be expressed as

\begin{equation}
csi\text{-}ReportingBand = \frac{{{N_3}}}{R}.
\end{equation}
According to Table \ref{tabCSS}, it can be inferred that the range of possible values for $csi\text{-}ReportingBand$ is $\{3, 4, \ldots, 18\}$ and for $N_3$ is $\{3, 4, \ldots, 36\}$. The number of selected delay taps \( M_v \) is less than the number of configured subbands, which further reduces the number of PMI reports in the frequency domain \( N_3 \). In the protocol, \( M_v \) is configured by calculating
\begin{equation}\label{c66}
 M_v  = \left\lceil {{p_v }\frac{{{N_3}}}{R}} \right\rceil, \;\;\;{p_v} \in \left\{ {\frac{1}{4},\frac{1}{8}} \right\},
\end{equation}
where $\left\lceil \cdot \right\rceil$ is a ceiling operator and ${p_v}$ is a compression factor. For the common $100$ MHz bandwidth with a $15$ kHz subcarrier spacing, the BWP typically comprises $273$ RBs. According to Table \ref{tabCSS}, the subband size can be $16$, resulting in $18$ subbands. Consequently, $M_v$ can be either $3$ or $5$. The \( M_v \) selected taps among $N_3$ available taps by UE should be reported to gNB, which are indicated by the indices ${i_{1,5}} \in \left\{ {0,1, \ldots ,2{M_v } - 1} \right\}$ (for $N_3>19$) and 
\begin{equation*}
{i_{1,6,l}} \in \left\{ {\begin{aligned}
&{\left\{ {0,1, \ldots ,C\left(N_3-1,M_v-1  \right) - 1}\right\}},&{{N_3} \le 19},\\
&{\left\{ {0,1, \ldots ,C\left(2M_v-1,M_v-1  \right) - 1} \right\}},&{{N_3} > 19},
\end{aligned}} \right.
\end{equation*}
for $l=1,2,\ldots,v$ where $C(x,y)$ is defined in (\ref{cxy}). As shown in Fig. \ref{pic_14}, the UE first determines the beamforming coefficients across the delay domain and identifies the tap with the strongest amplitude. The UE then remaps the tap indices so that the index of the strongest tap is $0$. To compress the representation of the precoding matrix, it is necessary to select \( M_v \) taps from \( N_3 \) available taps. Since the tap with index zero is always selected, it is only necessary to choose \( M_v - 1 \) taps from the remaining \( N_3 - 1 \) taps. The selection result can be represented by a combinatorial mode index \( {i_{1,6,l}} \). When \( N_3 \) is relatively large, i.e., ${N_3} > 19$, a two-level indication method is adopted to further reduce overhead. First, \( {i_{1,5}} \) indicates a group of \( 2M_v \) taps that includes the index zero tap, and then \( {i_{1,6,l}} \) is used to indicate \( M_v - 1 \) taps from within that group. In particular, \( {i_{1,5}} \) indicates the initial of a consecutive group of \( 2M_v \) taps, i.e., 
\begin{equation}
  M_{\textrm{initial}}\in\left\{-2M_v+1,-2M_v+2,\ldots,0\right\},
\end{equation}
and the relationship between ${i_{1,5}}$ and $M_{\textrm{initial}}$ is given by
\begin{equation}
i_{1,5}=\left\{\begin{matrix}M_{\textrm{initial}},&M_{\textrm{initial}}=0,\ \\M_{\textrm{initial}}+2M_v,&M_{\textrm{initial}}<0.\\\end{matrix}\right.    
\end{equation}
Let $n_{3,l}^{(f_l^\ast)}$ denote the index of the strongest tap. Then, the reported tap indices $n_{3,l}=\left[n_{3,l}^{(0)},\ldots,n_{3,l}^{(M_v-1)}\right]$ are remapped with respect to $n_{3,l}^{(f_l^\ast)}$ as 
\begin{equation}
 n_{3,l}^{(f)}=\left(n_{3,l}^{(f)}-n_{3,l}^{(f_l^\ast)}\right) \mathrm{mod} \; N_3,
\end{equation}
where $n_{3,l}^{(f_l^\ast)}=0$ after remapping. The index $f$ in (\ref{b46}) is remapped with respect to $f_l^\ast$ as
\begin{equation}
    f=\left(f-f_l^\ast\right)\; mod\; M_v,
\end{equation}
where $f_l^\ast=0$ after remapping. The indication for the $M_v$ selected taps is performed after index remapping. Consequently, we have $n_{3,l}^{(0)}=0$ and the tap indices $n_{3,l}^{(1)},\ldots,n_{3,l}^{(M_v-1)}$ are found from \( i_{1,6,l} \) for \( N_3 \leq 19 \), and from \( i_{1,6,l} \) and \( M_{\textrm{initial}} \) for \( N_3 > 19 \), using the following Algorithm 3.

\renewcommand{\algorithmicrequire}{\textbf{Input:}}
\renewcommand{\algorithmicensure}{\textbf{Output:}}
\begin{algorithm}
\caption{Mapping of \( i_{1,6,l} \) and \( M_{\textrm{initial}} \) to indices of selected taps $n_{3,l}^{(1)},\ldots,n_{3,l}^{(M_v-1)}$}
\label{ciawf}       %
\begin{algorithmic}[1]
\Require  $M_v$, $ M_{\textrm{initial}}$, $N_3$
\State \textbf{Initialize:  $s_{0}=0$ }
\For {$f=1,\ldots,M_v-1$}
\State Find the largest $x^\ast\in\left\{M_v-1-f,\ldots,N_3-1-f\right\}$ such that $i_{1,6,l}\ -s_{f-1}\geq C\left(x^\ast,M_v-f\right)$.
\State $e_f=C\left(x^\ast,M_v-f\right)$
\State $s_f=s_{f-1}+e_f$
\If{$N_3 < 20$}
\State $n_{3,l}^{(f)}=N_3-1-x^\ast$
\Else 
\State $n_l^{(f)}=2M_v-1-x^\ast$
\If{${n_l^{(f)}\le M_{\textrm{initial}}+2M_v-1}$}
\State $n_{3,l}^{(f)}=n_l^{(f)}$
\Else 
\State $n_{3,l}^{(f)}=n_l^{(f)}+(N_3-2M_v)$
\EndIf
\EndIf
\EndFor
\Ensure $n_{3,l}^{(1)},\ldots,n_{3,l}^{(M_v-1)}$
\end{algorithmic}
\end{algorithm}

The indices of $i_{2,4,l}$, $i_{2,5,l}$ and $i_{1,7,l}$ indicate amplitude coefficients, phase coefficients and bitmap after remapping, respectively. Specifically, the wideband amplitude indicator is given by 
\begin{equation}
i_{2,3,l}=\left[\begin{matrix}k_{l,0}^{\left(1\right)},k_{l,1}^{\left(1\right)}\\\end{matrix}\right], \;\;k_{l,p}^{(1)}\in\left\{1,\ldots,15\right\},
\end{equation}
where $k_{l,0}^{\left(1\right)}$ and $k_{l,1}^{\left(1\right)}$ represent indicators for the first and second polarization directions, respectively. The subband amplitude indicator compressed in the delay domain is given by
\begin{equation}\label{c72}
    i_{2,4,l}=\left[{\bf{k}}_{l,0}^{(2)},\ldots ,{\bf{k}}_{l,M_v-1}^{(2)}\right],
\end{equation}
where $M_v$ coefficient vectors are included and each represents the amplitude indicators for $2L$ beams on both polarization directions, i.e.,
\begin{equation}
{\bf{k}}_{l,f}^{(2)}=\left[k_{l,0,f}^{(2)},\ldots, k_{l,2L-1,f}^{(2)}\right], \;\;k_{l,i,f}^{(2)}\in\left\{0,\ldots,7\right\}.
\end{equation}
The mapping from $k_{l,p}^{(1)}$ to the amplitude coefficient $p_{l,p}^{(1)}$ is given in Table \ref{tabmapkuan} and the mapping from $k_{l,i,f}^{(2)}$ to the amplitude coefficient $p_{l,i,f}^{(2)}$ is given in Table \ref{tabmapzhai}. The wideband amplitude coefficients are represented by 
\begin{equation}
p_l^{\left( 1 \right)} = \left[ {\begin{aligned}
{p_{l,0}^{\left( 1 \right)}},{p_{l,1}^{\left( 1 \right)}}
\end{aligned}} \right],   
\end{equation}
and the subband ones mapping from (\ref{c72}) can be written as
\begin{equation}
\begin{aligned}
{\bf{p}}_l^{(2)} &= \left[ {{\bf{p}}_{l,0}^{(2)}, \ldots ,{\bf{p}}_{l,{M_v } - 1}^{(2)}} \right],\\
{\bf{p}}_{l,f}^{(2)} &= \left[ {p_{l,0,f}^{(2)}, \ldots ,p_{l,2L - 1,f}^{(2)}} \right].
\end{aligned}
\end{equation}
The phase coefficient indicator is given by 
\begin{equation}
 {i_{2,5,l}} = \left[ {{{\bf{c}}_{l,0}} \ldots {{\bf{c}}_{l,{M_v } - 1}}} \right], \; {{\bf{c}}_{l,f}} = \left[ {{c_{l,0,f}} \ldots {c_{l,2L - 1,f}}} \right],  
\end{equation}
where the phase coefficients are expressed as 
\begin{equation}
\begin{aligned}
\varphi_{l,i,f}=\ e^{j\frac{2\pi c_{l,i,f}}{16}},\;\;{c_{l,i,f}} \in \left\{ {0, \ldots ,15} \right\}.
\end{aligned}
\end{equation}

\begin{table}[t]   
\begin{center} 
\centering  
\captionsetup{
  labelsep=newline, % 让标题另起一行
}
\renewcommand{\arraystretch}{1.5}  %每行的高度为1.5倍
\caption{Parameter Configurations for $L$, $\beta$,
and $p_v$}  \label{tabp}
\begin{tabular}{|c|c|c|c|c|}   
\hline  
\multirow{2}{*}{} & \multirow{2}{*}{}& \multicolumn{2}{c|}{$p_v$} & \multirow{2}{*}{}  \\ 
\cline{3-4}   
\multirow{-2}{*}{paramCombination-r16}& \multirow{-2}{*}{$L$} 
 & \textbf{$v \in \{1,2\}$} &\textbf{$v \in \{3,4\}$} & \multirow{-2}{*}{$\beta$}\\
\hline  1&2&$1/4$&$1/8$&$1/4$\\
\hline  2&2&$1/4$&$1/8$&$1/2$\\
\hline  3&4&$1/4$&$1/8$&$1/4$\\
\hline  4&4&$1/4$&$1/8$&$1/2$\\
\hline  5&4&$1/4$&$1/4$&$3/4$\\
\hline  6&4&$1/2$&$1/4$&$1/2$\\
\hline  7&6&$1/4$&$-$&$1/2$\\
\hline  8&6&$1/4$&$-$&$3/4$ \\
\hline
\end{tabular}   
\end{center}   
\vspace{-10pt}
\end{table}

To limit the overhead of reporting, the protocol defines a parameter $ K_0 $ as
\begin{equation}
K_0=\left\lceil\beta2LM_1\right\rceil ,   
\end{equation}
where $M_1$ is given by $M_v$ in (\ref{c66}) with $v=1$, and $\beta$ is provided in Table \ref{tabp}. The number of feedback entries for the weights across all taps and all beams on layer $l$, denoted by $K_l^{NZ}$, must not exceed $K_0$. To this end, a bitmap whose nonzero bits identify which coefficients in $i_{2,4,l}$ and $i_{2,5,l}$ are reported, which is indicated by 
\begin{equation}
i_{1,7,l}=\left[{\bf{k}}_{l,0}^{\left(3\right)},\ldots, {\bf{k}}_{l,M_v-1}^{\left(3\right)}\right],    
\end{equation}
where 
\begin{equation}
 {\bf{k}}_{l,f}^{\left(3\right)}=\left[k_{l,0,f}^{\left(3\right)}\ldots k_{l,2L-1,f}^{\left(3\right)}\right],   k_{l,i,f}^{(3)}\in\left\{0,1\right\},
\end{equation}
for $l=1,\ldots,v$, such that
\begin{equation}
K_l^{NZ}=\sum_{i=0}^{2L-1}\sum_{f=0}^{M_v-1}k_{l,i,f}^{(3)}\le K_0    
\end{equation}
is the number of nonzero coefficients. The total number of nonzero coefficients satisfies
\begin{equation}
K^{NZ}=\sum_{l=1}^{v}K_l^{NZ}\le2K_0.    
\end{equation}
Let $i_l^\ast\in\left\{0,1,\ldots,2L-1\right\}$ be the index that identifies the strongest coefficient of the strongest tap on the layer $l$. This index is indicated by 
\begin{equation}
{i_{1,8,l}} = \left\{ {\begin{aligned}
&{\sum\nolimits_{i = 0}^{i_1^ * } {k_{1,i,0}^{(3)}}  - 1,\;\; v = 1, }\\
&\;\;\;\;{i_l^ * ,\;\;\;\;\;\;\;\;1 < v \le 4,}
\end{aligned}} \right.
\end{equation}
for $l=1,\ldots,v$. As shown in Fig. \ref{pic_17}, there are two different methods to indicate the strongest beam, tailored respectively for single-stream $v = 1$ and multi-stream $1 < v \le 4$ scenarios. One method directly indicates the index in the $2L$ blue beams, while the other method indicates the index in the green beams using a bitmap. The rationale behind this design is more influenced by standardization requirements than by scientific perspectives. Let $p^\ast=\left\lfloor\frac{i_l^\ast}{L}\right\rfloor$ be the index that identifies the strongest polarization direction index, and the amplitude and phase coefficient indicators are reported as follows:
\begin{itemize}
    \item $k_{l,p^\ast}^{(1)}=15, k_{l,i_l^\ast,0}^{(2)}=7, k_{l,i_l^\ast,0}^{(3)}=1$ and $c_{l,i_l^\ast,0}=0$.
    \item The indicators $k_{l,p^\ast}^{(1)}$, $k_{l,i_l^\ast,0}^{(2)}$, and $c_{l,i_l^\ast,0}$ are not reported.
    \item The indicator $k_{l,(p^\ast+1)\ mod\ 2}^{(1)}$ is reported.
    \item The $K^{NZ}-v$ indicators $k_{l,i,f}^{(2)}$ for which $k_{l,i,f}^{(3)}=1$, $i\neq i_l^\ast,\ f\neq0$ are reported. The $K^{NZ}-v$ indicators $c_{l,i,f}$ for which $k_{l,i,f}^{(3)}=1$, $i\neq i_l^\ast,\ f\neq 0$ are reported. 
    \item The remaining $2L\cdot M_v\cdot v-K^{NZ}$ indicators $k_{l,i,f}^{(2)}$ are not reported. The remaining $2L\cdot M_v\cdot v-K^{NZ}$ indicators $c_{l,i,f}$ are not reported.
\end{itemize}
\begin{figure}[t]
	\centering
	\includegraphics[width=0.48\textwidth]{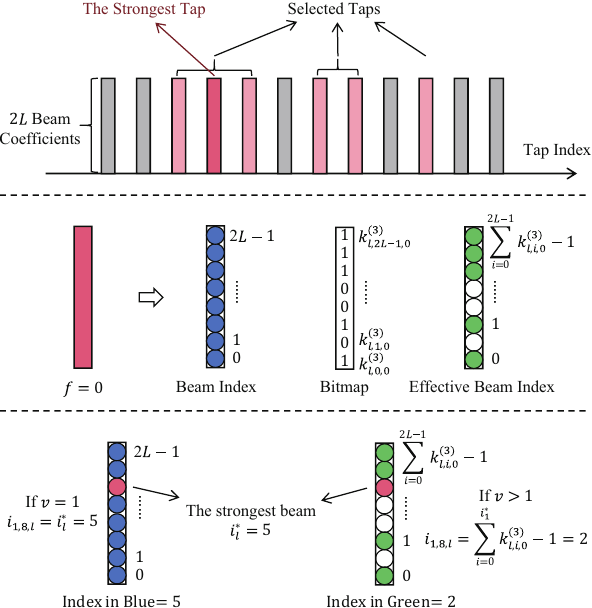}
	\caption{Indication for the strongest beam on the strongest tap by $i_{1,8,l}$ in Enhanced Type II Codebook.}\label{pic_17}
    \vspace{-5pt}
\end{figure}
The emerging parameters are interpreted in Table \ref{paraEtypeii}. The PMI value corresponds to the codebook indices of $i_1$ and $i_2$, which can be expressed by 
\begin{align}\label{a85}
&{i_1} = \Big[ {{i_{1,1}}\;\;{i_{1,2}}\;\;{i_{1,5}}\;\;\{ {i_{1,6,l}}\} _{l = 1}^v\;\;\{ {i_{1,7,l}}\} _{l = 1}^v\;\;\{ {i_{1,8,l}}\} _{l = 1}^v} \Big],\notag\\
&{i_2} = \Big[ \{ {{i_{2,3,l}}\} _{l = 1}^v\;\;\{ {i_{2,4,l}}\} _{l = 1}^v\;\;\{ {i_{2,5,l}}\} _{l = 1}^v} \Big].
\end{align}

The corresponding R16 Enhanced Type II Port-Selection Codebook is delegated to Appendix \ref{r16ps}.

\begin{table}[t]   
\begin{center} 
\centering  
\captionsetup{
  labelsep=newline, % 让标题另起一行
}
\renewcommand{\arraystretch}{2}  %每行的高度为1.5倍
\caption{Mapping of Elements of $i_{2,3,l}$: $k_{l,p}^{(1)}$ to $p_{l,p}^{(1)}$}  \label{tabmapkuan}
\begin{tabular}{|Sc|Sc|Sc|Sc|}   
\hline  
\rowcolor{gray!20}
$k_{l,p}^{(1)}$ &$p_{l,p}^{(1)}$&$k_{l,p}^{(1)}$ &$p_{l,p}^{(1)}$ \\
\hline  
0 &Reserved&8&$( \frac{1}{128})^{1/4}$\\
\hline  1&$\frac{1}{\sqrt{128}}$&9&$\frac{1}{\sqrt{8}}$\\
\hline  2&$( \frac{1}{8192})^{1/4}$&10&$( \frac{1}{32})^{1/4}$\\
\hline  3&$\frac{1}{8}$&11&$\frac{1}{2}$\\
\hline  4&$( \frac{1}{2048})^{1/4}$&12&$( \frac{1}{8})^{1/4}$\\
\hline  5&$\frac{1}{2\sqrt{8}}$&13&$\frac{1}{\sqrt{2}}$\\
\hline  6&$( \frac{1}{512})^{1/4}$&14&$( \frac{1}{2})^{1/4}$\\
\hline  7&$\frac{1}{4}$&15&$1$\\
\hline
\end{tabular}   
\end{center} 
\vspace{-5pt}
\end{table}

\begin{table}[t]   
\begin{center} 
\centering  
\captionsetup{
  labelsep=newline, % 让标题另起一行
}
\renewcommand{\arraystretch}{2}  %每行的高度为1.5倍
\caption{Mapping of Elements of $i_{2,4,l}$: $k_{l,i,f}^{(2)}$ to $p_{l,i,f}^{(2)}$}  \label{tabmapzhai}

\begin{tabular}{|Sc|Sc|Sc|Sc|}   
\hline  
\rowcolor{gray!20}
$k_{l,p}^{(1)}$ &$p_{l,p}^{(1)}$&$k_{l,p}^{(1)}$ &$p_{l,p}^{(1)}$ \\
\hline  0&$\frac{1}{8\sqrt{2}}$&4&$\frac{1}{2\sqrt{2}}$\\
\hline  1&$\frac{1}{8}$&5&$\frac{1}{2}$\\
\hline  2&$\frac{1}{4\sqrt{2}}$&6&$\frac{1}{\sqrt{2}}$\\
\hline  3&$\frac{1}{4}$&7&$1$\\
\hline
\end{tabular}   
\end{center} 
\vspace{-12pt}
\end{table}

\begin{table}[t]   
\begin{center} 
\centering  
\captionsetup{
  labelsep=newline, % 让标题另起一行
}
\renewcommand{\arraystretch}{1.5}  %每行的高度为1.5倍
\caption{New Parameters in Enhanced Type II Codebook.}  
\label{paraEtypeii} 
\begin{tabular}{|c|m{6cm}|}   
\hline   \textbf{Parameters} &\textbf{Interpretation} \\  
\hline   $i_{1,1}$ &  Indicates the index of the selected orthogonal beam group among the oversampled beam groups. \\
\hline   $i_{1,2}$ &  Indicates the index of the combination of $L$ beams within the selected beam group. \\
\hline   $i_{1,5}$ & Indicates the initial of a consecutive group of $2M_v$ taps for the two-level indication when $N_3 >19$.\\
\hline   $i_{1,6,l}$ & Indicates $M_v-1$ taps from $N_3-1$ taps (or from $2M_v$ taps when $N_3>19$) on layer \(l\).\\
\hline   $i_{1,7,l}$ & Indicates the bitmap of nonzero reported coefficients in $i_{2,4,l}$ and $i_{2,5,l}$ on layer \(l\).\\
\hline   $i_{1,8,l}$ & Indicates the index of the strongest beam on the strongest tap on layer \(l\).\\
\hline   $i_{2,3,l}$ &  Indicates the wideband amplitude coefficients of the selected beams on layer \(l\). \\
\hline   $i_{2,4,l}$ & Represents the subband amplitude coefficients  of the selected beams on layer \(l\). \\
\hline   $i_{2,5,l}$ & Represents the subband phase coefficients that are compressed in the delay domain of the selected beams on layer \(l\). \\
\hline   ${\bf{v}}_{{m_1},m_2}$ & The beam basis on direction of $({m_1},m_2).$ \\
\hline   ${\bf{y}}_{f}$ & The $f$-th spectral basis in delay domain.  \\
\hline   ${N_3}$ & The total number of available taps.\\
\hline   ${\overline \beta  _l}$ & The power scaling coefficient on layer \(l\).\\
\hline   $f$ & Represents the index of the selected tap.\\
\hline   $t$ & Represents the index of the reported precoding matrix in the frequency domain.\\
\hline   $n_{3,l}^{(f)}$ &  The index of the $f$-th selected tap in delay domain. \\
\hline $M_v$ & Represents the total number of the selected taps.\\
\hline $M_{\textrm{initial}}$ & Represents the initial index for the consecutive group of $2M_v$ taps for the two-level indication when $N_3 >19$.\\
\hline $K_0$ & Indicates the report limitation on each layer.\\
\hline $K_l^{NZ}$ & Represents the number of reported coefficients in $i_{2,4,l}$ and $i_{2,5,l}$ on layer \(l\).\\
\hline $K^{NZ}$ & Represents the number of reported coefficients in $i_{2,4,l}$ and $i_{2,5,l}$ on all layers.\\
\hline $f_l^\ast$ & Denotes the index of the strongest tap among $\{0,1,\ldots,M_v-1\}$ selected taps.\\
\hline $n_{3,l}^{(f_l^\ast)}$ &  Denotes the index of the strongest tap among $\{0,1,\ldots,N_3-1\}$ all taps.\\
\hline $i_l^\ast$ & Denotes the index of the strongest coefficient of the strongest tap on layer \(l\).\\
\hline $p^\ast$ & The index of the strongest polarization direction.\\
\hline
\end{tabular}   
\end{center}   
\vspace{-12pt}
\end{table}

\subsection{R17 Further Enhanced Type II Port-Selection Codebook}
In Release 17, the protocol was updated with the Further Enhanced Port-Selection Type II Codebook, which is a modification of the Enhanced Port-Selection Type II Codebook. Despite a substantial amount of new descriptions given to this codebook in the protocol, its essence remains highly similar to the Enhanced Port-Selection Type II Codebook. Firstly, the UE still compresses the wideband precoding matrices in both the spatial and frequency domains, with the spatial domain involving port selection, and the frequency domain based on the selection over DFT bases, i.e., selection over taps in the delay domain. The difference lies in the selection principle for both spatial and spectral bases as follows:
\begin{itemize}
    \item Instead of selecting consecutive $L$ ports, the UE can now freely indicate any $L$ ports. 
    \item The number of taps selected in the frequency domain is the same for both single-stream and multi-stream transmission, with $M_v$ being unified to $M$. 
     \item The number of selected taps $M$ has been reduced by $M \in \{1,2\}$ and the two-level indication method is canceled.
\end{itemize}
The fundamental rationale for reducing feedback overhead lies in the consistency of path angles and delays between uplink and downlink channels. This inherent reciprocity allows the base station to enhance the reported downlink precoding matrix by channel estimated from uplink sounding, thereby minimizing the need for extensive downlink CSI feedback. Besides, the protocol also defines some new parameters, such as the total number of beams reported under dual polarization by $K_1=2L$ and the spatial compression ratio factor by $\alpha$. Consequently, the protocol for this codebook has been rearticulated to accommodate these new parameters.

The expression of the Further Enhanced Type II Port-Selection Regular Codebook in the protocol is presented in Table \ref{tabfesp}. The supported number of CSI-RS ports is given by ${P_{{\text{CSI-RS}}}} \in \left\{ {4,8,12,16,24,32} \right\}$ as configured by the higher layer parameter \emph{nrofPorts}. The number of selected beams $L$ is no longer directly configured by the higher layer parameter \emph{numberOfBeams}, but is instead derived through the parameter $\alpha$ via 
\begin{equation}
    L=K_1/2,\;\;K_1=\alpha P_{\text{CSI-RS}},
\end{equation}
where the value $\alpha$ is determined by the higher layer parameter \emph{paramCombination-r17} provided in Table \ref{tabp17}. The $K_1$ selected ports from $P_{\text{CSI-RS}}$ ports are identified by 
\begin{equation}
\begin{aligned}
{\bf{m}} &= \left[ {{m^{(0)}},\ldots,{m^{(L - 1)}}} \right],\\
{m^{(i)}} &\in \left\{ {0,1,\ldots,\frac{{{P_{\text{CSI-RS}}}}}{2} - 1} \right\},
\end{aligned}
\end{equation}
which are indicated by the index $i_{1,2}$, where
\begin{equation}
  i_{1,2}\in\left\{0,1,\ldots,\left(\begin{matrix}P_{\text{CSI-RS}}/2\\L\\\end{matrix}\right)-1\right\}.
\end{equation}
If $\alpha=1$, we have $m^{(i)}=i$ for $i=1,\ldots,L-1$ and $i_{1,2}$ is not reported. Otherwise, the elements of $\bf{m}$ are found from $i_{1,2}$ using $C(x,y)$ defined in (\ref{cxy}) and the following Algorithm 4.

\renewcommand{\algorithmicrequire}{\textbf{Input:}}
\renewcommand{\algorithmicensure}{\textbf{Output:}}
\begin{algorithm}
\caption{Mapping of $i_{1,2}$ to  ${{m^{(0)}},\ldots,{m^{(L - 1)}}} $}
\label{ciawf}       %
\begin{algorithmic}[1]
\Require  $L$, $P_{\text{CSI-RS}}$
\State \textbf{Initialize:  $s_{-1}=0$ }
\For {$i = 0,1,\ldots, L-1$}
\State Find the largest $x^\ast \in \{L-1-i,\ldots,P_{\text{CSI-RS}}/2-1-i\}$ such that $i_{1,2} - {s_{k - 1}} \ge C\left( {{x^*},4 - k} \right)$.
\State ${e_i} = C\left( {{x^*},L - i} \right)$
\State ${s_i} = {s_{i - 1}} + {e_i}$
\State $ m^{(i)}= P_{\text{CSI-RS}}/2-1- {x^*}$
\EndFor
\Ensure ${{m^{(0)}},\ldots,{m^{(L - 1)}}}$
\end{algorithmic}
\end{algorithm}

\begin{table*}[t]  
\centering   
\captionsetup{
  labelsep=newline, % 让标题另起一行
}
\renewcommand{\arraystretch}{2}  
\caption{R17 Further Enhanced Type II Port Selection Codebook (TS 38.214 Table 5.2.2.2.7-3)}  \label{tabfesp}
\begin{tabular}{|Sc|Sc|}   
\hline  
\rowcolor{gray!20}
\textbf{Layers}  &  \\
\hline   
$v=1$ & ${\bf{w}}^{(1)}_{m,n_3,p^{(1)}_1,p^{(2)}_1,i_{2,5,1},t}
={\bf{w}}^1_{m,n_3,p^{(1)}_1,p^{(2)}_1,i_{2,5,1},t}$  \\
\hline
% $v=2,3,4$& ... \\
$v=2$ & $ {\bf{W}}^{(2)}_{m,n_3,p^{(1)}_1,p^{(2)}_1,i_{2,5,1},p^{(1)}_2,p^{(2)}_2,i_{2,5,2},t}
= \frac{1}{\sqrt{2}}
\begin{bmatrix} {\bf{w}}^1_{m,n_3,p^{(1)}_1,p^{(2)}_1,i_{2,5,1},t} \ \ {\bf{w}}^2_{m,n_3,p^{(1)}_2,p^{(2)}_2,i_{2,5,2},t}
\end{bmatrix}$ \\
\hline
$v=3$ & {\linespread{2}\selectfont\makecell{$ {\bf{W}}^{(3)}_{m,n_3,p^{(1)}_1,p^{(2)}_1,i_{2,5,1},p^{(1)}_2,p^{(2)}_2,i_{2,5,2},p^{(1)}_3,p^{(2)}_3,i_{2,5,3},t}$ \qquad \qquad \qquad \qquad \qquad \qquad\qquad \qquad \qquad \qquad \qquad \qquad \qquad \qquad \\
$\qquad \qquad \qquad \qquad= \frac{1}{\sqrt{3}}
\begin{bmatrix} {\bf{w}}^1_{m,n_3,p^{(1)}_1,p^{(2)}_1,i_{2,5,1},t}  \ \ {\bf{w}}^2_{m,n_3,p^{(1)}_2,p^{(2)}_2,i_{2,5,2},t} \ \
{\bf{w}}^3_{m,n_3,p^{(1)}_3,p^{(2)}_3,i_{2,5,3},t}
\end{bmatrix}$} }\\
\hline
$v=4$ & {\linespread{2}\selectfont\makecell{$ {\bf{W}}^{(4)}_{m,n_3,p^{(1)}_1,p^{(2)}_1,i_{2,5,1},p^{(1)}_2,p^{(2)}_2,i_{2,5,2},p^{(1)}_3,p^{(2)}_3,i_{2,5,3},p^{(1)}_4,p^{(2)}_4,i_{2,5,4},t}$ \qquad \qquad \qquad \qquad \qquad \qquad \qquad \qquad \qquad \qquad\\
$= \frac{1}{2}
\begin{bmatrix} {\bf{w}}^1_{m,n_3,p^{(1)}_1,p^{(2)}_1,i_{2,5,1},t}  \ \ {\bf{w}}^2_{m,n_3,p^{(1)}_2,p^{(2)}_2,i_{2,5,2},t}  \ \ {\bf{w}}^3_{m,n_3,p^{(1)}_3,p^{(2)}_3,i_{2,5,3},t} \ \ 
{\bf{w}}^4_{m,n_3,p^{(1)}_4,p^{(2)}_4,i_{2,5,4},t}
\end{bmatrix}$} }\\
\hline
\multicolumn{2}{|Sc|}{\linespread{2}\selectfont  
\makecell{Precoding vector: ${\bf{w}}^l_{m,n_3,p^{(1)}_l,p^{(2)}_l,i_{2,5,l},t}
=\frac{1}{\sqrt{\gamma_{t,l}}}
\begin{bmatrix}\sum^{L-1}_{i=0}{\bf{v}}_{m^{(i)}}p^{(1)}_{l,0}\sum^{M-1}_{f=0}y^{(f)}_{t}p^{(2)}_{l,i,f}\varphi_{l,i,f}\\
\sum^{L-1}_{i=0}{\bf{v}}_{m^{(i)}}p^{(1)}_{l,1}\sum^{M-1}_{f=0}y^{(f)}_{t}p^{(2)}_{l,i+L,f}\varphi_{l,i+L,f}
\end{bmatrix}, \;\;l=1,2,3,4,$ \\
$\gamma_{t,l}=\sum^{2L-1}_{i=0}(p_{l,\left\lfloor \frac{i}{L}\right\rfloor }^{(1)})^2\begin{vmatrix}\sum^{M-1}_{f=0}y^{(f)}_{t}p^{(2)}_{l,i,f}\varphi_{l,i,f}\end{vmatrix}^2.$\\
 }
}\\ 
\hline
\end{tabular}  
\end{table*} 

\begin{table}[htb]   
\begin{center} 
\centering  
\captionsetup{
  labelsep=newline, % 让标题另起一行
}
\renewcommand{\arraystretch}{2}  %每行的高度为1.5倍
\caption{Parameter Configurations for $\alpha$, $M$ and $\beta$}  \label{tabp17}
\begin{tabular}{|Sc|Sc|Sc|Sc|}   
\hline  
paraCombination-r17 &$M$&$\alpha$ &$\beta$ \\ 
\hline  1&1&$3/4$&$1/2$\\
\hline  2&1&1&$1/2$\\
\hline  3&1&1&$3/4$\\
\hline  4&1&1&1\\
\hline  5&2&$1/2$&$1/2$\\
\hline  6&2&$3/4$&$1/2$\\
\hline  7&2&1&$1/2$\\
\hline  8&2&1&$3/4$\\
\hline
\end{tabular}   
\end{center}   
\vspace{-10pt}
\end{table}

In addition to selecting $L$ beams from the $P_{\text{CSI-RS}}$ beams in the spatial domain, it is also necessary to choose $M$ taps from the $N_3$ taps in the frequency domain. Given that the channel and its associated precoding matrices are sparse in the delay domain and are concentrated near the LoS tap, the UE can leverage this prior information to compute the precoding matrices only for a subset of the delay taps, thereby reducing the DFT computing overhead. Consequently, the protocol restricts the range of tap selection: If $N_3$ exceeds a threshold $N$, the taps can only be selected from the set $\{0, 1, \ldots, N-1\}$. As the value of $M$ is limited to either $1$ or $2$, the indication method for the taps becomes much simpler:
\begin{itemize}
    \item If $M=1$, or $M=2$ and $N=2$, $i_{1,6}$ is not reported.
    \item If $M=2$ and $N=4$, the nonzero offset between $n_3^{(0)}$and $n_3^{(1)}$ is reported with $i_{1,6}$ assuming that $n_3^{(0)}$ (reference of the strongest tap for the offset) is $0$. 
\end{itemize}
Let $n_3^{(f)}$ be the index of the $f$-th tap with $f\in \{0,\ldots,M-1\}$ such that $n_3^{(f)}$ increases with $f$. Based on the above rules, we have 
\begin{equation}
n_3^{(f)} \in \left\{ {\begin{array}{*{20}{c}}
{\;\;\;\;\;\qquad\quad\;\;\;\;\;\{ 0\} ,\;\;\;\;\qquad\quad M = 1,}\\
{\{ 0,1,\ldots,\min (N,{N_3}) - 1\} ,\;\;M = 2.}
\end{array}} \right. 
\end{equation}
The amplitude coefficient indicators $i_{2,3,l}$ and $i_{2,4,l}$, for $l=1,\ldots,v$ are given by
\begin{equation}
i_{2,3,l}=\left[\begin{matrix}k_{l,0}^{\left(1\right)},k_{l,1}^{\left(1\right)}\\\end{matrix}\right], \;\;k_{l,p}^{(1)}\in\left\{1,\ldots,15\right\},
\end{equation}
where $k_{l,0}^{\left(1\right)}$ and $k_{l,1}^{\left(1\right)}$ represent indicators for the first and second polarization directions, respectively. The subband amplitude indicator in the delay domain is 
\begin{equation}\label{a72}
    i_{2,4,l}=\left[{\bf{k}}_{l,0}^{(2)},\ldots ,{\bf{k}}_{l,M-1}^{(2)}\right],
\end{equation}
where $M$ coefficient vectors are included and each vector represents the amplitude indicators for $2L$ beams on both polarization directions, i.e.,
\begin{equation}
{\bf{k}}_{l,f}^{(2)}=\left[k_{l,0,f}^{(2)},\ldots, k_{l,K_1-1,f}^{(2)}\right], \;\;k_{l,i,f}^{(2)}\in\left\{0,\ldots,7\right\}.
\end{equation}
The mapping from $k_{l,p}^{(1)}$ to the amplitude coefficient $p_{l,p}^{(1)}$ is given in Table \ref{tabmapkuan} and the mapping from $k_{l,i,f}^{(2)}$ to the amplitude coefficient $p_{l,i,f}^{(2)}$ is given in Table \ref{tabmapzhai}. The wideband amplitude coefficients are represented by 
\begin{equation}
p_l^{\left( 1 \right)} = \left[ {\begin{aligned}
{p_{l,0}^{\left( 1 \right)}},{p_{l,1}^{\left( 1 \right)}}
\end{aligned}} \right],   
\end{equation}
and the subband ones mapping from (\ref{a72}) can be written as
\begin{equation}
\begin{aligned}
{\bf{p}}_l^{(2)} &= \left[ {{\bf{p}}_{l,0}^{(2)}, \ldots ,{\bf{p}}_{l,{M } - 1}^{(2)}} \right],\\
{\bf{p}}_{l,f}^{(2)} &= \left[ {p_{l,0,f}^{(2)}, \ldots ,p_{l,K_1 - 1,f}^{(2)}} \right].
\end{aligned}
\end{equation}
The phase coefficient indicator is given by 
\begin{equation}
 {i_{2,5,l}} = \left[ {{{\bf{c}}_{l,0}} \ldots {{\bf{c}}_{l,{M } - 1}}} \right], \; {{\bf{c}}_{l,f}} = \left[ {{c_{l,0,f}} \ldots {c_{l,K_1 - 1,f}}} \right],  
\end{equation}
where the phase coefficients are expressed as 
\begin{equation}
\begin{aligned}
\varphi_{l,i,f}=\ e^{j\frac{2\pi c_{l,i,f}}{16}},\;\;{c_{l,i,f}} \in \left\{ {0, \ldots ,15} \right\}.
\end{aligned}
\end{equation}

To limit the overhead of reporting, the number of feedback entries for the weights across all taps and all beams on layer $l$, denoted by $K_l^{NZ}$, must not exceed $K_0$ where
\begin{equation}
K_0=\left\lceil\beta K_1M\right\rceil ,   
\end{equation}
where $M$ and $\beta$ are determined by the higher layer parameter \emph{paramCombination-r17} presented in Table \ref{tabp17}. To this end, a bitmap whose nonzero bits identify which coefficients in $i_{2,4,l}$ and $i_{2,5,l}$ are reported is indicated by 
\begin{equation}
i_{1,7,l}=\left[{\bf{k}}_{l,0}^{\left(3\right)},\ldots, {\bf{k}}_{l,M-1}^{\left(3\right)}\right],    
\end{equation}
where 
\begin{equation}
 {\bf{k}}_{l,f}^{\left(3\right)}=\left[k_{l,0,f}^{\left(3\right)}\ldots k_{l,K_1-1,f}^{\left(3\right)}\right], \;\;  k_{l,i,f}^{(3)}\in\left\{0,1\right\},
\end{equation}
for $l=1,\ldots,v$, such that
\begin{equation}
\begin{aligned}
    K_l^{NZ}&=\sum_{i=0}^{K_1-1}\sum_{f=0}^{M-1}k_{l,i,f}^{(3)}\le K_0,    \\
    K^{NZ}&=\sum_{l=1}^{v}K_l^{NZ}\le2K_0. 
\end{aligned}
\end{equation}
Let $i_l^\ast\in\left\{0,1,\ldots,2L-1\right\}$ be the index that identifies the strongest coefficient of the strongest tap on the layer $l$. The index that identifies the strongest beam coefficient of all taps on the layer $l$ is indicated by
\begin{equation}
{i_{1,8,l}} = {K_1}f_l^ \ast  + i_l^ \ast,\;\; {i_{1,8,l}} \in \{ 0,1, \ldots ,{K_1}M - 1\},
\end{equation}
for $l=1,\ldots,v$. Let $p^\ast=\left\lfloor\frac{i_l^\ast}{L}\right\rfloor$ be the index that identifies the strongest polarization direction and the amplitude and phase coefficient indicators are reported as follows:
\begin{itemize}
    \item $k_{l,p^\ast}^{(1)}=15, k_{l,i_l^\ast,0}^{(2)}=7, k_{l,i_l^\ast,0}^{(3)}=1$ and $c_{l,i_l^\ast,0}=0$.
    \item The indicators $k_{l,p^\ast}^{(1)}$, $k_{l,i_l^\ast,0}^{(2)}$, and $c_{l,i_l^\ast,0}$ are not reported.
    \item The indicator $k_{l,(p^\ast+1)\ mod\ 2}^{(1)}$ is reported.
    \item The $K^{NZ}-v$ indicators $k_{l,i,f}^{(2)}$ for which $k_{l,i,f}^{(3)}=1$, $i\neq i_l^\ast,\ f\neq f^\ast$ are reported. The $K^{NZ}-v$ indicators $c_{l,i,f}$ for which $k_{l,i,f}^{(3)}=1$, $i\neq i_l^\ast,\ f\neq f^\ast$ are reported. 
    \item The remaining $K_1Mv-K^{NZ}$ indicators $k_{l,i,f}^{(2)}$ are not reported. The remaining $K_1Mv-K^{NZ}$ indicators $c_{l,i,f}$ are not reported.
\end{itemize}
The PMI value corresponds to the codebook indices of $i_1$ and $i_2$, which can be expressed by 
\begin{align}\label{a104}
&{i_1} = \Big[ {{i_{1,2}}\;\; i_{1,6}\;\;\{ {i_{1,7,l}}\} _{l = 1}^v\;\;\{ {i_{1,8,l}}\} _{l = 1}^v} \Big],\notag\\
&{i_2} = \Big[ \{{{i_{2,3,l}}\} _{l = 1}^v\;\;\{ {i_{2,4,l}}\} _{l = 1}^v\;\;\{ {i_{2,5,l}}\} _{l = 1}^v} \Big].
\end{align}
Comparing the indices of (\ref{a104}) in the Further Enhanced Type II Port-Selection Codebook to those of (\ref{a86}) in the Enhanced Type II Port-Selection Codebook, some changes can be interpreted as follows. The $i_{1,1}$ has been replaced by $i_{1,2}$, because this codebook no longer indicates the initial index for the consecutive $L$ beams, but instead indicates the combination index for any $L$ beams from all ports. The $i_{1,5}$ has been removed because a two-level indication for selected taps is no longer needed. The $\{ {i_{1,6,l}}\} _{l = 1}^v$ has been unified into ${i_{1,6,l}}$, because the number of taps selected in the frequency domain is the same for both single-stream and multi-stream transmission.

\begin{table*}[t]  
\centering   
\captionsetup{
  labelsep=newline, % 让标题另起一行
}
\renewcommand{\arraystretch}{2}  
\caption{R18 Enhanced Type II Codebook for Predicted PMI (TS 38.214 Table 5.2.2.2.10-2)}  \label{tab1}
\begin{tabular}{|Sc|Sc|}   
\hline  
\rowcolor{gray!20}
\textbf{Layers}  &  \\
\hline   
$v=1$ & ${\bf{w}}^{(1)}_{q_1,q_2,n_1,n_2,n_{3,1},n_{4,1},p^{(1)}_1,p^{(2)}_1,\varphi_1,t,\iota}
={\bf{w}}^1_{q_1,q_2,n_1,n_2,n_{3,1},n_{4,1},p^{(1)}_1,p^{(2)}_1,\varphi_1,t,\iota}$  \\
\hline
% $v=2,3,4$& ... \\
$v=2$ & 
{\linespread{2}\selectfont\makecell{${\bf{W}}^{(2)}_{q_1,q_2,n_1,n_2,n_{3,1},n_{4,1},p^{(1)}_1,p^{(2)}_1,\varphi_1,n_{3,2},n_{4,2},p^{(1)}_2,p^{(2)}_2,\varphi_2,t,\iota}$\qquad \qquad\qquad \qquad \qquad \qquad \qquad\qquad\qquad \qquad \qquad \qquad\qquad\\
\qquad \qquad \qquad \qquad\qquad$= \frac{1}{\sqrt{2}}
\begin{bmatrix} {\bf{w}}^1_{q_1,q_2,n_1,n_2,n_{3,1},n_{4,1},p^{(1)}_1,p^{(2)}_1,\varphi_1,t,\iota} \ \ {\bf{w}}^2_{q_1,q_2,n_1,n_2,n_{3,2},n_{4,2},p^{(1)}_2,p^{(2)}_2,\varphi_2,t,\iota}
\end{bmatrix}$} }\\
\hline
$v=3$ & {\linespread{2}\selectfont\makecell{$ {\bf{W}}^{(3)}_{q_1,q_2,n_1,n_2,n_{3,1},n_{4,1},p^{(1)}_1,p^{(2)}_1,\varphi_1,n_{3,2},n_{4,2},p^{(1)}_2,p^{(2)}_2,\varphi_2,n_{3,3},n_{4,3},p^{(1)}_3,p^{(2)}_3,\varphi_3,t,\iota}$ \qquad \qquad \qquad\qquad \qquad \qquad \qquad \qquad \qquad \qquad \\
$= \frac{1}{\sqrt{3}}
\begin{bmatrix} {\bf{w}}^1_{q_1,q_2,n_1,n_2,n_{3,1},n_{4,1},p^{(1)}_1,p^{(2)}_1,\varphi_1,t,\iota} \ \ {\bf{w}}^2_{q_1,q_2,n_1,n_2,n_{3,2},n_{4,2},p^{(1)}_2,p^{(2)}_2,\varphi_2,t,\iota} \ \
{\bf{w}}^3_{q_1,q_2,n_1,n_2,n_{3,3},n_{4,3},p^{(1)}_3,p^{(2)}_3,\varphi_3,t,\iota}
\end{bmatrix}$} } \\
\hline
$v=4$ & {\linespread{2}\selectfont\makecell{$ {\bf{W}}^{(4)}_{q_1,q_2,n_1,n_2,n_{3,1},n_{4,1},p^{(1)}_1,p^{(2)}_1,\varphi_1,n_{3,2},n_{4,2},p^{(1)}_2,p^{(2)}_2,\varphi_2,n_{3,3},n_{4,3},p^{(1)}_3,p^{(2)}_3,\varphi_3,n_{3,4},n_{4,4},p^{(1)}_4,p^{(2)}_4,\varphi_4,t,\iota}$  \qquad \qquad \qquad \qquad \qquad \qquad \qquad\\
$= \frac{1}{2}
\begin{bmatrix} {\bf{w}}^1_{q_1,q_2,n_1,n_2,n_{3,1},n_{4,1},p^{(1)}_1,p^{(2)}_1,\varphi_1,t,\iota} \ \ {\bf{w}}^2_{q_1,q_2,n_1,n_2,n_{3,2},n_{4,2},p^{(1)}_2,p^{(2)}_2,\varphi_2,t,\iota} \ \cdots \;\;
{\bf{w}}^4_{q_1,q_2,n_1,n_2,n_{3,4},n_{4,1},p^{(1)}_4,p^{(2)}_4,\varphi_4,t,\iota}
\end{bmatrix}$} } \\
\hline
\multicolumn{2}{|Sc|}{\linespread{2}\selectfont  
\makecell{ Precoding vector: $
{\bf{w}}^l_{q_1,q_2,n_1,n_2,n_{3,l},n_{4,l},p^{(1)}_l,p^{(2)}_l,\varphi_l,t,\iota}
=\frac{1}{\sqrt{N_1N_2\gamma_{t,\iota,l}}}
\begin{bmatrix}\sum^{L-1}_{i=0}{\bf{v}}_{m_1^{(i)}m_2^{(i)}}p^{(1)}_{l,0}\sum^{M_v-1}_{f=0}y^{(f)}_{t,l}\sum^{Q-1}_{\tau=0}z^{(\tau)}_{\iota,l}p^{(2)}_{l,i,f,\tau}\varphi_{l,i,f,\tau}\\
\sum^{L-1}_{i=0}{\bf{v}}_{m_1^{(i)}m_2^{(i)}}p^{(1)}_{l,1}\sum^{M_v-1}_{f=0}y^{(f)}_{t,l}\sum^{Q-1}_{\tau=0}z^{(\tau)}_{\iota,l}p^{(2)}_{l,i+L,f,\tau}\varphi_{l,i+L,f,\tau}
\end{bmatrix},$ \\
$\gamma_{t,\iota,l}=\sum^{2L-1}_{i=0}(p_{l,|\frac{i}{L}|}^{(1)})^2\begin{vmatrix}\sum^{M_v-1}_{f=0}y^{(f)}_{t,l}\sum^{Q-1}_{\tau=0}z^{(\tau)}_{\iota,l}p^{(2)}_{l,i,f,\tau}\varphi_{l,i,f,\tau}\end{vmatrix}^2 ,\;\;l=1,2,3,4$.\\
  }
}\\ 
\hline
\end{tabular}  
\end{table*} 
\subsection{R18 Enhanced Type II Codebook for Predicted PMI}
The Enhanced (resp. Further Enhanced) Type II Codebook proposed in Release 16 (resp. 17) reduces the feedback overhead for the UE by applying compression in both the spatial and frequency domains, which reduces the space domain of $N_1N_2$ dimensions through a linear combination of $L$ spatial bases and the frequency domain of $N_3$ dimensions through a linear combination of $M_v$ (resp. $M$) spectral bases. However, for every slot interval configured for reporting, the UE has to report the combination coefficients to generate an interval-level PMI matrix. In Release 18, the Enhanced Type II Codebook for Predicted PMI further considers temporal compression by representing the time domain of $N_4$ dimensions through a linear combination of $Q$ temporal bases. The spatial and spectral bases can be referred to as beams and taps (or delay taps), whereas the temporal bases can be referred to as shifts (or Doppler shifts). Fig. \ref{R18tech} shows the techniques used for feedback compression adopted in the R18 Codebook, where the adoption of those in historical codebooks is also included.
\begin{figure}[t]
	\centering
	\includegraphics[width=0.5\textwidth]{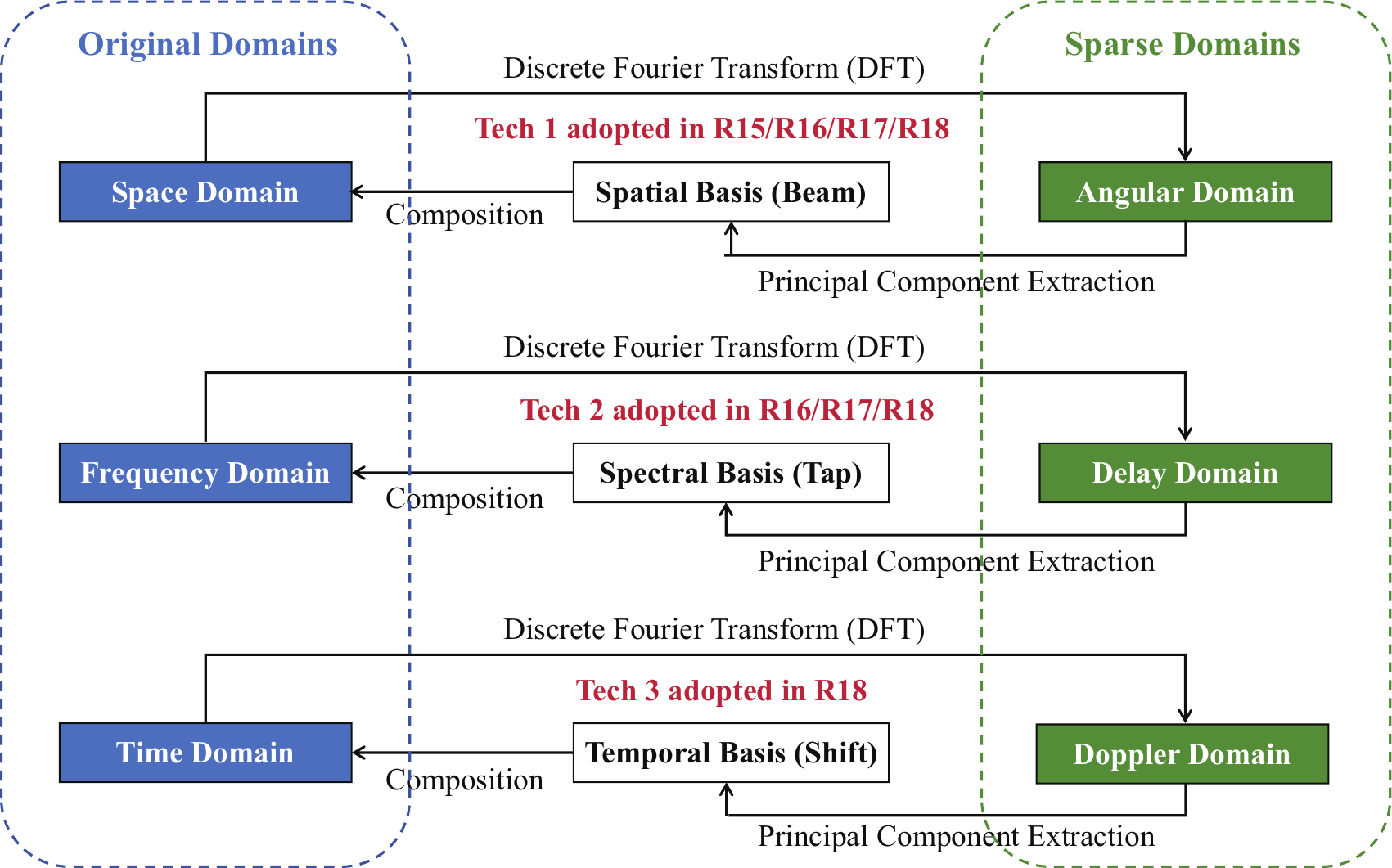}
	\caption{Techniques used in R18 (5G-A) codebooks.}\label{R18tech}
\end{figure}

The expression of the Enhanced Type II Regular Codebook for Predicted PMI in the protocol is presented in Table \ref{tabpredic}. The newly introduced temporal basis ${{\bf{z}}_{\tau}}$ can be written as
\begin{equation}
{{\bf{z}}_{\tau}} = {\left[ {z_{0,l}^{({\tau})},z_{1,l}^{({\tau})},\ldots,z_{{N_4} - 1,l}^{({\tau})}} \right]^T},   
\end{equation}
where $\tau=0,\ldots,Q-1$ is the index of the selected shift and $Q$ is the total number of selected shifts in the Doppler domain (which is similar to the parameter $L$ in the angular domain and parameter $M$ in the delay domain).  Currently, the protocol freezes the value of $Q$ at $2$, so the range of $\tau$ is $0$ and $1$. The element in the temporal basis ${{\bf{z}}_{\tau}}$ is given by 
\begin{equation}\label{b87}
z_{\iota,l}^{(\tau)}=e^{j\frac{2\pi\cdot\iota\cdot n_{4,l}^{(\tau)}}{N_4}},
\end{equation}
where $\iota=0,1,\ldots N_4-1$ represents the index of the reported precoding matrix in the time domain. For ease of exposition, we refer to $\tau$ as the time index. Although not explicitly stated in the protocol, for ease of understanding the codebook, we have rewritten the temporal basis as 
\begin{equation}\label{b88}
{{\bf{z}}_{\tau}} = {\Big[ {1,e^{j\frac{2\pi n_{4,l}^{(\tau)}}{N_4}},\ldots,{e^{j\frac{{2\pi ({N_4} - 1)n_{4,l}^{(\tau)}}}{{{N_4}}}}}} \Big]^T},    
\end{equation}
which admits a DFT vector with dimension of $N_4$. Thus, there are a total of $N_4$ orthogonal DFT vectors with $n_{4,l}^{(\tau)} \in \{0,1,\ldots,N_4-1\}$ from which $Q=2$ vectors are selected in the $l$-th layer, indexed by $\tau=0,1$, to represent the precoding matrices across various slot intervals. 

The value of $R\in{1,2}$ is configured with the higher-layer parameter \emph{numberOfPMI-SubbandsPerCQI-Subband-Doppler-r18}, where $R$ and the corresponding value of $N_3$ have been defined in Sec. \ref{enhanceregularII} for Enhanced Type II Regular Codebook. The UE shall report the RI value $\upsilon$ according to the configured higher layer parameter \emph{typeII-Doppler-RI-Restriction-r18}. The bitmap parameter \emph{typeII-Doppler-RI-Restriction-r18} forms the bit sequence $r_3,r_2,r_1,r_0$ where $r_0$ is the LSB and $r_3$ is the MSB. When $r_i$ is zero, $i\in\left\{0,1,\ldots,3\right\}$, PMI and RI reporting are not allowed to correspond to any precoder associated with $v=i+1$ layers. The value of $N_4\in \{1,2,4,8\}$ is configured by the higher layer parameter \emph{N4}, such that the PMI indicates $N_3$ precoder matrices for each of the $N_4$ consecutive slot intervals, each with a duration of $d$ slots. If $N_4=1$, the precoder matrices for 1-4 layers are obtained from the PMI codebook as in Table \ref{tabesII}, i.e., the Enhanced Type II Codebook for Predicted PMI degrades to the Enhanced Type II Codebook. 

The indication methods for the beam and tap, i.e., $v_{m_1^{\left(i\right)},m_2^{(i)}}$ and ${{\bf{y}}_f}$, are in line with the methods for Enhanced Type II Regular Codebook, which will not be elaborated here. The indication for $Q=2$ shifts can be expressed as 
\begin{equation}
n_{4,l}=\left[n_{4,l}^{\left(0\right)}\ n_{4,l}^{\left(1\right)}\right], \;\;n_{4,l}^{(\tau)}\in{0,1,\ldots,N_4-1},
\end{equation}
with the indices $\tau\in{0,1}$ assigned such that $n_{4,l}^{(\tau)}$ increases with $\tau$. The parameter $n_{4,l}^{(\tau)}$ represents the index of the $\tau$-th selected shift among all $N_4-1$ shifts, which is indicated by
\begin{equation}
 i_{1,10,l}\in\left\{0,1,\ldots,N_4-2\right\},
\end{equation}
where $n_{4,l}^{(0)}$ is set as a reference point after remapping and $i_{1,10,l}$ is the nonzero offset between $n_{4,l}^{(0)}$ and $n_{4,l}^{(1)}$. The offset value $1$ mapped to the index value $0$. Thus, we have 
\begin{equation}
    n_{4,l}^{(0)} = 0, \;\;n_{4,l}^{(1)} =  i_{1,10,l}+1.
\end{equation}
The wideband amplitude indicator is given by
\begin{equation}
i_{2,3,l}=\left[\begin{matrix}k_{l,0}^{\left(1\right)},k_{l,1}^{\left(1\right)}\\\end{matrix}\right], \;\;k_{l,p}^{(1)}\in\left\{1,\ldots,15\right\},
\end{equation}
where $k_{l,0}^{\left(1\right)}$ and $k_{l,1}^{\left(1\right)}$ represent indicators for the first and second polarization directions, respectively. The subband amplitude indicator compressed in the delay domain is given by
\begin{equation}
    i_{2,4,l}=\left[i_{2,4,l,0},i_{2,4,l,1}\right],
\end{equation}
which contains the indicators for coefficients on the two selected shifts, i.e.,
\begin{equation}\label{a113}
    i_{2,4,l,\tau}=\left[{\bf{k}}_{l,0,\tau}^{(2)},\ldots ,{\bf{k}}_{l,M_v-1,\tau}^{(2)}\right],
\end{equation}
where $M_v$ coefficient vectors are included and each represents the amplitude indicators for $2L$ beams on both polarization directions, i.e.,
\begin{equation}
{\bf{k}}_{l,f,\tau}^{(2)}=\left[k_{l,0,f,\tau}^{(2)},\ldots, k_{l,2L-1,f,\tau}^{(2)}\right], \;\;k_{l,i,f,\tau}^{(2)}\in\left\{0,\ldots,7\right\}.
\end{equation}
The mapping from $k_{l,p}^{(1)}$ to the amplitude coefficient $p_{l,p}^{(1)}$ is given in Table \ref{tabmapkuan} and the mapping from $k_{l,i,f}^{(2)}$ to the amplitude coefficient $p_{l,i,f}^{(2)}$ is given in Table \ref{tabmapzhai}, which is identical to that for Enhanced Type II Regular Codebook. The wideband amplitude coefficients are represented by 
\begin{equation}
p_l^{\left( 1 \right)} = \left[ {\begin{aligned}
{p_{l,0}^{\left( 1 \right)}},{p_{l,1}^{\left( 1 \right)}}
\end{aligned}} \right],   
\end{equation}
and the subband ones mapping from (\ref{a113}) can be written as
\begin{equation}
\begin{aligned}
{\bf{p}}_{l,\tau}^{(2)} &= \left[ {{\bf{p}}_{l,0,\tau}^{(2)}, \ldots ,{\bf{p}}_{l,{M_v } - 1,\tau}^{(2)}} \right],\\
{\bf{p}}_{l,f,\tau}^{(2)} &= \left[ {p_{l,0,f,\tau}^{(2)}, \ldots ,p_{l,2L - 1,f,\tau}^{(2)}} \right].
\end{aligned}
\end{equation}
The phase coefficient indicator is given by
\begin{equation}
  i_{2,5,l}=[i_{2,5,l,0}\ i_{2,5,l,1}]  
\end{equation}
in which
\begin{equation}
\begin{aligned}
 {i_{2,5,l,\tau}} &= \left[ {{{\bf{c}}_{l,0,\tau}} \ldots {{\bf{c}}_{l,{M_v } - 1,\tau}}} \right], \\ {{\bf{c}}_{l,f,\tau}} &= \left[ {{c_{l,0,f,\tau}} \ldots {c_{l,2L - 1,f,\tau}}} \right],      
\end{aligned}
\end{equation}
where the phase coefficients are expressed as 
\begin{equation}
\begin{aligned}
\varphi_{l,i,f,\tau}=\ e^{j\frac{2\pi c_{l,i,f,\tau}}{16}},\;\;{c_{l,i,f,\tau}} \in \left\{ {0, \ldots ,15} \right\}.
\end{aligned}
\end{equation}

\begin{table}[htb]   
\begin{center} 
\centering  
\captionsetup{
  labelsep=newline, % 让标题另起一行
}
\renewcommand{\arraystretch}{1.5}  %每行的高度为1.5倍
\caption{Codebook Parameter Configurations for $L$, $\beta$ and $p_v$}  \label{tabpredic}
\begin{tabular}{|c|c|c|c|c|}   
\hline  
\multirow{2}{*}{} & \multirow{2}{*}{}& \multicolumn{2}{c|}{$p_v$} & \multirow{2}{*}{}  \\ 
\cline{3-4}   
\multirow{-2}{*}{\makecell{paramCombination-\\Doppler-r18}}& \multirow{-2}{*}{$L$} 
 & \textbf{$v \in \{1,2\}$} &\textbf{$v \in \{3,4\}$} & \multirow{-2}{*}{$\beta$}\\
\hline  1&2&$1/8$&$1/16$&$1/4$\\
\hline  2&2&$1/4$&$1/8$&$1/2$\\
\hline  3&4&$1/4$&$1/8$&$1/4$\\
\hline  4&4&$1/4$&$1/4$&$1/4$\\
\hline  5&4&$1/4$&$1/4$&$1/2$\\
\hline  6&4&$1/4$&$1/4$&$3/4$\\
\hline  7&4&$1/2$&$1/4$&$1/2$\\
\hline  8&6&$1/4$&$-$&$1/2$ \\
\hline  9&6&$1/4$&$-$&$3/4$ \\
\hline
\end{tabular}   
\end{center}   
\vspace{-10pt}
\end{table}

Similarly, to control the reporting overhead, the number of feedback entries for the weights across all taps, beams, and slot intervals on layer $l$, represented as \( K_l^{NZ} \), should not exceed
\begin{equation}
K_0=\left\lceil2\beta LM_1Q\right\rceil ,   
\end{equation}
in which  \( M_1 \) is configured by calculating
\begin{equation}\label{a66}
 M_v  = \left\lceil {{p_v }\frac{{{N_3}}}{R}} \right\rceil, \;\;\;{p_v} \in \left\{ {\frac{1}{4},\frac{1}{8}} \right\},
\end{equation}
with $v=1$, where ${p_v}$ is a compression factor. The parameters $\beta$, $L$, $p_v$ are determined by the higher layer parameter \emph{paramCombination-Doppler-r18} presented in Table \ref{tabpredic}. To this end, a bitmap whose nonzero bits identify which coefficients in $i_{2,4,l}$ and $i_{2,5,l}$ are reported is indicated by 
\begin{equation}
\begin{aligned}
 i_{1,7,l}&=[i_{1,7,l,0}, i_{1,7,l,1}],\\
 i_{1,7,l,\tau}&=\left[{\bf{k}}_{l,0,\tau}^{\left(3\right)},\ldots, {\bf{k}}_{l,M_v-1,\tau}^{\left(3\right)}\right],   
\end{aligned} 
\end{equation}
where 
\begin{equation}
 {\bf{k}}_{l,f,\tau}^{\left(3\right)}=\left[k_{l,0,f,\tau}^{\left(3\right)},\ldots, k_{l,2L-1,f,\tau}^{\left(3\right)}\right], \;\;  k_{l,i,f}^{(3)}\in\left\{0,1\right\},
\end{equation}
for $l=1,\ldots,v$, such that

\begin{table}[t]   
\begin{center} 
\centering  
\captionsetup{
  labelsep=newline, % 让标题另起一行
}
\renewcommand{\arraystretch}{1.5}  %每行的高度为1.5倍
\caption{New Parameters in Type II Codebook for Predicted PMI.}  
\label{paraEtypeiiP} 
\begin{tabular}{|c|m{6cm}|}   
\hline   \textbf{Parameters} &\textbf{Interpretation} \\  
\hline   $i_{1,10,l}$ &  Indicates the index of the selected shifts among all $N_4-1$ shifts. \\
\hline   ${\bf{v}}_{{m_1},m_2}$ & The beam basis on direction of $({m_1},m_2).$ \\
\hline   ${\bf{y}}_{f}$ & The $f$-th spectral basis in delay domain.  \\
\hline   ${\bf{z}}_{\tau}$ & The $\tau$-th temporal basis in the Doppler domain.  \\
\hline   ${N_3}$ & The total number of available taps.\\
\hline   $N_4$ & The number of consecutive slot intervals. \\
\hline   $f$ & Represents the index of the selected tap.\\
\hline   $t$ & Represents the index of the reported precoding matrix in the frequency domain.\\
\hline   $\tau$ &   Represents the index of the selected shifts. \\
\hline   $\iota$ & Represents the index of the reported precoding matrix in the time domain.\\
\hline $M_v$ & Represents the total number of the selected taps.\\
\hline $Q$ & Represents the total number of the selected shifts.\\
\hline
\end{tabular}   
\end{center}   
\vspace{-12pt}
\end{table}

\begin{equation}
\begin{aligned}
    K_l^{NZ}&=\sum_{\tau=0}^{Q-1}\sum_{i=0}^{2L-1}\sum_{f=0}^{M_v-1}k_{l,i,f,\tau}^{(3)}\le K_0,    \\
    K^{NZ}&=\sum_{l=1}^{v}K_l^{NZ}\le2K_0. 
\end{aligned}
\end{equation}
Let $\tau_l^\ast\in{0,1}$, $f_l^\ast\in\left\{0,1,\ldots,M_v-1\right\}$ and $i_l^\ast\in\left\{0,1,\ldots,2L-1\right\}$ be the indices which identify the strongest coefficient of layer $l$. The codebook indices of $f$ and $n_{3,l}$ are remapped with respect to
\begin{equation}
\begin{aligned}
   f&=\left(f-f_l^\ast\right)\;\mathrm{mod}\ M_v, \\
   n_{3,l}^{(f)}&=\left(n_{3,l}^{(f)}-n_{3,l}^{(f_l^\ast)}\right)\;\mathrm{mod}\ N_3, 
\end{aligned}
\end{equation}
where $f_l^\ast=0$ and $n_{3,l}^{(f_l^\ast)}=0$ after remapping. The index that identifies the strongest beam coefficient of all taps and all shifts on the layer $l$ is indicated by
\begin{equation}
{i_{1,8,l}} = \left\{ {\begin{aligned}
&{\sum\nolimits_{I=0}^{{2L\tau_1^\ast+i}_1^\ast}\kappa_{1,I,0}^{(3)}-1,\;\; v = 1, }\\
&\;\;\;\;{{2L\tau_l^\ast+i}_l^\ast ,\;\;\;\;\;\;1 < v \le 4,}
\end{aligned}} \right.
\end{equation}
for $l=1,\ldots,v$. The meaning of this classification indication method is the same as that revealed in Fig. \ref{pic_17}, except that this codebook requires indicating the strongest beam among multiple shifts of the strongest tap. Let $p^\ast=\left\lfloor\frac{i_l^\ast}{L}\right\rfloor$ be the index of the strongest polarization direction, and the amplitude and phase coefficient indicators are reported as follows:
\begin{itemize}
    \item $k_{l,p^\ast}^{(1)}=15, k_{l,i_l^\ast,0,\tau_l^\ast}^{(2)}=7, k_{l,i_l^\ast,0,\tau_l^\ast}^{(3)}=1$ and $c_{l,i_l^\ast,0,\tau_l^\ast}^{(3)}=0$, where $k_{l,p^\ast}^{(1)}$, $k_{l,i_l^\ast,0,\tau_l^\ast}^{(2)}$, and $c_{l,i_l^\ast,0,\tau_l^\ast}$ are not reported.
    \item The indicator $k_{l,(p^\ast+1)\ mod\ 2}^{(1)}$ is reported.
    \item The $K^{NZ}-v$ indicators $k_{l,i,f,\tau}^{(2)}$ for which $k_{l,i,f,\tau}^{(3)}=1$, $\tau \neq \tau_l^\ast$, $i\neq i_l^\ast$, $f\neq 0$ are reported. The $K^{NZ}-v$ indicators $c_{l,i,f,\tau}$ for which $k_{l,i,f,\tau}^{(3)}=1$, $\tau \neq \tau_l^\ast$, $i\neq i_l^\ast$, $f\neq 0$  are reported. 
    \item The remaining $2LM_vQv-K^{NZ}$ indicators $k_{l,i,f,\tau}^{(2)}$ are not reported. The remaining $2LM_vQv-K^{NZ}$ indicators $c_{l,i,f,\tau}$ are not reported.
\end{itemize}
The PMI value corresponds to the codebook indices of $i_1$ and $i_2$, which can be expressed by 
\begin{equation}\label{a127}
\begin{aligned}
&{i_1} = \Big[ i_{1,1}\;\;{i_{1,2}}\;\;i_{1,5}\;\; \{ {i_{1,6,l}}\} _{l = 1}^v\;\;\{ {i_{1,7,l}}\} _{l = 1}^v \\
&\qquad\qquad\qquad\qquad \qquad \{ {i_{1,8,l}}\} _{l = 1}^v \;\;\{ {i_{1,10,l}}\} _{l = 1}^v\Big],\\
&{i_2} = \Big[ \{{{i_{2,3,l}}\} _{l = 1}^v\;\;\{ {i_{2,4,l}}\} _{l = 1}^v\;\;\{ {i_{2,5,l}}\} _{l = 1}^v} \Big].
\end{aligned}
\end{equation}
Comparing the indices of (\ref{a127}) in the Enhanced Type II Codebook for Predicted PMI to that of (\ref{a85}) in the Enhanced Type II Regular Codebook, an additional indicator $\{ {i_{1,10,l}}\} _{l = 1}^v$ has been introduced to indicate the selected shifts. While the content indicated by the other indicators remains unchanged, their composition structure might be slightly modified. This completes the tutorial on R18 Type II Codebook, and the emerging parameters are interpreted in Table \ref{paraEtypeiiP}.

\section{Discussion of 5G Codebooks}\label{comparisonandapp}
In the previous section, we delved into various codebook frameworks and their parameter components. In this section, we will distill simplified models for each codebook, analyze their performance disparities, evaluate feedback overhead, and summarize their application scenarios.

\subsection{Compact Model}

 Each version of the codebook includes intricate mathematical models within the protocol. If we set aside the modeling of certain secondary factors, such as the scaling, normalization, quantization, bitmap, and other elements, we can utilize a compact model to represent these codebooks. This simplified expression allows us to better understand the key features and structural makeup of the codebooks.

To define the bases for the space domain, frequency domain, and time domain, we respectively model the \emph{full bases} and \emph{effective bases} as follows:

\subsubsection{Bases in the Space Domain}
Considering the beam oversampling and dual polarization, the representation of bases in the space domain is more complex compared to that in the frequency and time domains. The oversampled DFT matrix ${{\bf{V}}_s}\in \mathbb{C}^{{{N_1}{N_2}} \times {N_1}{O_1}{N_2}{O_2}}$ can be expressed as
\begin{equation}
{{\bf{V}}_s} = \left[{{\bf{v}}_{0,0}},{{\bf{v}}_{0,1}}, \cdots ,{{\bf{v}}_{{N_1}{O_1} - 1,{N_2}{O_2} - 1}}\right],    
\end{equation}
in which each beam ${\bf{v}}_{l,m}$ is a DFT vector and has been defined in (\ref{beamv}). Once the orthogonal beam group is determined by $i_{1,1}=[q_1,q_2]$, the associated DFT matrix ${\bar{\bf{V}} _s} \in \mathbb{C}^{{N_1}{N_2} \times {N_1}{N_2}}$ is given by
\begin{equation}
{\bar{\bf{V}} _s} = \left[ {{{\bar{\bf{v}} }_{0,0}},{{\bar{\bf{v}} }_{0,1}}, \cdots {{\bar{\bf{v}} }_{{N_1} - 1,{N_2} - 1}}} \right],    
\end{equation}
where
\begin{equation}
{\bar{\bf{v}} _{i,j}} = {{\bf{v}}_{i{O_1} + {q_1},j{O_2} + {q_2}}},  
\end{equation}
with $i=0,1,\ldots,N_1-1$ and $j=0,1,\ldots,N_2-1$. The DFT matrix ${\bar{\bf{V}} _s}$ provides $N_1N_2$ orthogonal beams, and $L$ beams should be selected for each precoding vector. The full bases refer to the matrix that represents the complete set of orthogonal spatial basis vectors for the codebook. Due to the presence of dual polarization, the matrix of full bases in regular codebooks can be written as
\begin{equation}
  {{\bf{W}}_s} = \left[ {\begin{array}{*{20}{c}}
{{{\bar{\bf{V}} }_s}}&{\bf{0}}\\
{\bf{0}}&{{{\bar{\bf{V}} }_s}}
\end{array}} \right]\in \mathbb{C}^{2{N_1}{N_2} \times 2{N_1}{N_2}},
\end{equation}
where the dimension of $ {{\bf{W}}_s}$ can also be written as ${P_{{\text{CSI-RS}}}} \times {P_{{\text{CSI-RS}}}}$ since we have ${P_{{\text{CSI-RS}}}}=2N_1N_2$.
The matrix of effective bases represents the beam set composed of the selected beams in the PMI report, i.e., 
\begin{equation}
{{{\bf{\hat W}}}_s} = \left[ {\begin{array}{*{20}{c}}
{{{{\bf{\hat V}}}_s}}&{\bf{0}}\\
{\bf{0}}&{{{{\bf{\hat V}}}_s}}
\end{array}} \right]\in \mathbb{C}^{2{N_1N_2} \times 2L},
\end{equation}
in which
\begin{equation}
{{{\bf{\hat V}}}_s} = \left[ {{{\bf{v}}_{m_1^{(0)},m_2^{(0)}}} \cdots {{\bf{v}}_{m_1^{(L - 1)},m_2^{(L - 1)}}}} \right],
\end{equation}
where ${\bf{v}}_{m_1^{(i)},m_2^{(i)}}$ represent the $i$-th selected beam from the full matrix. The index of $(m_1^{(i)},m_2^{(i)})$ is related to $i_{1,1}$ and $i_{1,2}$.

For the port-selection codebooks, the bases are no longer composed of DFT vectors but are instead composed of standard basis vectors. Consequently, the matrix of full bases in port-selection codebooks can be expressed as\footnote{To unify the compact model of the codebooks in the subsequent discussion, we represent the spatial bases of both regular and port-selection codebooks using the same matrix notation. However, it is important for the reader to understand that the spatial bases are different for each case.}
\begin{equation}\label{b115}
{{\bf{W}}_s} = \left[ {\begin{array}{*{20}{c}}
{{{\bf{I}}_{{P_{{\text{CSI-RS}}}}/2}}}&{\bf{0}}\\
{\bf{0}}&{{{\bf{I}}_{{P_{{\text{CSI-RS}}}}/2}}}
\end{array}} \right]\in {\mathbb{C}}^{{{P_{{\text{CSI-RS}}}}} \times {{P_{{\text{CSI-RS}}}}}},
\end{equation}
where matrix ${{{\bf{I}}_{{P_{{\text{CSI-RS}}}}/2}}}$ is an identity matrix of dimension ${{P_{{\text{CSI-RS}}}}/2}$. The matrix of effective bases in port-selection codebooks can be written as 
\begin{equation}\label{b116}
{{\bf{\hat W}}_s} = \left[ {\begin{array}{*{20}{c}}
{{{{\bf{\hat I}}}_{{P_{{\text{CSI-RS}}}/2} \times L}}}&{\bf{0}}\\
{\bf{0}}&{{{{\bf{\hat I}}}_{{P_{{\text{CSI-RS}}}/2} \times L}}}
\end{array}} \right]\in {\mathbb{C}}^{{{P_{{\text{CSI-RS}}}}} \times {2L}},
\end{equation}
where ${{{{\bf{\hat I}}}_{{P_{{\text{CSI-RS}}}} \times L}}}$ is a truncated identity matrix that selects $L$ columns from a complete identity matrix.

\subsubsection{Bases in the Frequency Domain}
The matrix of full bases in the frequency domain can be expressed by 
\begin{equation}\label{a115}
{{\bf{W}}_f} = \left[ {{{{\bf{\bar y}}}_0},{{{\bf{\bar y}}}_1},\ldots,{{{\bf{\bar y}}}_{{N_3-1}}}} \right]\in \mathbb{C}^{N_3\times N_3},
\end{equation}
in which 
\begin{equation}
 {{{\bf{\bar y}}}_n} = {[1,{e^{j\frac{{2\pi n}}{{{N_3}}}}},\ldots,{e^{j\frac{{2\pi ({N_3} - 1)n}}{{{N_3}}}}}]^T}  , 
\end{equation}
for $n = 0,1,\ldots,{N_3} - 1$. The matrix of effective bases in the frequency domain can be written as
\begin{equation}\label{a117}
{{{\bf{\hat W}}}_f} = \left[ {{{\bf{y}}_0},{{\bf{y}}_1},\ldots,{{\bf{y}}_{{M_v} - 1}}} \right]\in \mathbb{C}^{N_3\times M_v},
\end{equation}
where each delay tap ${{\bf{y}}_n}$ has been defined in (\ref{b46}). The selected taps in (\ref{a117}) have the following relation to the taps in (\ref{a115}), i.e., 
\begin{equation}
{{\bf{y}}_n} = {{{\bf{\bar y}}}_{n_{3,l}^{(n)}}}, \;\; n=0,1,\ldots,M_v-1, 
\end{equation}
where the values of $\{{n_{3,l}^{(n)}}\}_{n=0}^{M_v-1}$ are determined by $i_{1,6,l}$.

\subsubsection{Bases in the Temporal Domain}
The  matrix of full bases  in the temporal domain can be expressed by 
\begin{equation}\label{a119}
{{\bf{W}}_t} = \left[ {{{{\bf{\bar z}}}_0},{{{\bf{\bar z}}}_1},\ldots,{{{\bf{\bar z}}}_{{N_4-1}}}} \right]\in \mathbb{C}^{N_4\times N_4},
\end{equation}
in which 
\begin{equation}
 {{{\bf{\bar z}}}_n} = {[1,{e^{j\frac{{2\pi n}}{{{N_4}}}}},\ldots,{e^{j\frac{{2\pi ({N_4} - 1)n}}{{{N_4}}}}}]^T}  , 
\end{equation}
for $n = 0,1,\ldots,{N_4} - 1$. The matrix of effective bases in the temporal domain can be written as
\begin{equation}\label{a121}
{{{\bf{\hat W}}}_t} = \left[ {{{\bf{z}}_0},{{\bf{z}}_1},\ldots,{{\bf{z}}_{Q - 1}}} \right]\in \mathbb{C}^{N_4\times Q},
\end{equation}
where each shift ${{\bf{z}}_n}$ has been defined in (\ref{b88}). The selected shifts in (\ref{a121}) from (\ref{a119}) have the following relation, i.e., 
\begin{equation}
{{\bf{z}}_n} = {{{\bf{\bar z}}}_{n_{4,l}^{(n)}}}, \;\; n=0,1,\ldots,Q-1, 
\end{equation}
where the values of $\{{n_{4,l}^{(n)}}\}_{n=0}^{Q-1}$ are determined by $i_{1,10,l}$.

\subsubsection{Compact Model for R15 Type I}
The R15 Type I Codebook selects only a single beam, with both polarizations choosing the same beam, and there is a phase difference between the polarizations. Therefore, its compact model $ {\bf{w}} \in \mathbb{C}^{{P_{{\text{CSI-RS}}}}\times 1}$ can be expressed as
\begin{equation}
 {\bf{w}} = \left[ {\begin{array}{*{20}{c}}
{{{\bar {\bf{V}} }_s}}&{\bf{0}}\\
{\bf{0}}&{{{\bar {\bf{V}} }_s}}
\end{array}} \right]\underbrace {\left[ {\begin{array}{*{20}{c}}
{{{\bf{e}}_{l,m}}}&{\bf{0}}\\
{\bf{0}}&{{{\bf{e}}_{l,m}}}
\end{array}} \right]\left[ {\begin{array}{*{20}{c}}
1\\
{{\varphi _n}}
\end{array}} \right]}_{{{\bf{w}}_{{\rm{PMI}}}}} = {{\bf{W}}_s}{{\bf{w}}_{{\rm{PMI}}}},   
\end{equation}
where the vector ${{\bf{e}}_{l,m}}$ has only one element equal to $1$, with all other elements being $0$. The position of the element $1$ is determined by the indices $l$ and $m$. The information fed back by the UE determines the parameters in the matrix ${{\bf{w}}_{{\rm{PMI}}}}$.
\subsubsection{Compact Model for R15 Type II}
The R15 Type II Codebook selects \( L \) beams, with both polarizations utilizing the same set of beams. However, the beam combination coefficients for different polarizations are distinct. Different data streams also employ the same beams but have different combination coefficients. Each subband independently reports its own precoding matrix. Consequently, the precoding vector $ {\bf{w}} \in \mathbb{C}^{{P_{{\text{CSI-RS}}}}\times 1}$ for a single data stream within a single subband can be expressed as
\begin{align}
{\bf{w}} & \!=\! \left[ {\begin{array}{*{20}{c}}
{{{{\bf{\hat V}}}_s}}&{\bf{0}}\\
{\bf{0}}&{{{{\bf{\hat V}}}_s}}
\end{array}} \right]\underbrace {\left[ {\begin{array}{*{20}{c}}
{p_0^{(1)}}&{}&{\bf{0}}\\
{}& \ddots &{}\\
{\bf{0}}&{}&{{p_{2L - 1}^{(1)}}}
\end{array}} \right]\left[ {\begin{array}{*{20}{c}}
{p_0^{(2)}{\varphi _0}}\\
 \vdots \\
{p_{2L - 1}^{(2)}{\varphi _{2L - 1}}}
\end{array}} \right]}_{{{\bf{w}}_c}} \notag\\
& \mathop  = \limits^{(a)} \mathop {{{{\bf{\hat W}}}_s}}\limits_{{P_{{\text{CSI-RS}}}} \times 2L} \times \mathop {{{\bf{w}}_c}}\limits_{2L \times 1} \mathop  = \limits^{(b)} \mathop {{{\bf{W}}_s}}\limits_{{P_{{\text{CSI-RS}}}} \times {P_{{\text{CSI-RS}}}}} \times \mathop {{{\bf{w}}_{{\text{PMI}}}}}\limits_{{P_{{\text{CSI-RS}}}} \times 1} ,
\end{align}  
in which model $(a)$ utilizes the expression of effective bases, whereas model $(b)$ employs the expression of full bases. The vector \( {\bf{w}}_c \) represents the weights of \( 2L \) beams, which are composed of wideband amplitudes, subband amplitudes, and phases. It is important to note that in model $(a)$, \( {\bf{w}}_c \) does not convey all the information of the PMI because it does not account for the beam selection process. In contrast, in model $(b)$, \( {\bf{w}}_{\rm{PMI}} \) can represent the complete information of the PMI. \( {\bf{w}}_{\rm{PMI}} \) is a sparse vector with a sparsity of \( 2L \), where its support set indicates the beam selection process, and its nonzero values represent the weights of the \( 2L \) beams.

\subsubsection{Compact Model for R16 Type II}
The R16 Type II Codebook no longer follows the approach where each subband independently reports its own precoding matrix. Instead, it jointly compresses and reports the precoding matrices for \( N_3 \) subbands. Firstly, this codebook selects \( L \) beams, with both polarizations utilizing the same set of beams. Additionally, different subbands have different beam combination coefficients, but these coefficients are compressed onto \( M_v \) spectral bases. Therefore, the codebook also needs to select the dominant \( M_v \) taps out of \( N_3 \) taps in the delay domain to constitute the precoding matrix. Consequently, the precoding matrix for all subbands \( {\bf{W}} \in \mathbb{C}^{{P_{{\text{CSI-RS}}}}\times N_3} \) for a single data stream can be expressed as
\begin{align}
{\bf{W}} &= \left[ {\begin{array}{*{20}{c}}
{{{{\bf{\hat V}}}_s}}&{\bf{0}}\\
{\bf{0}}&{{{{\bf{\hat V}}}_s}}
\end{array}} \right]\left[ {\begin{array}{*{20}{c}}
{p_0^{(1)}}&{}&{}&{}&{}&{\bf{0}}\\
{}& \ddots &{}&{}&{}&{}\\
{}&{}&{p_0^{(1)}}&{}&{}&{}\\
{}&{}&{}&{p_1^{(1)}}&{}&{}\\
{}&{}&{}&{}& \ddots &{}\\
{\bf{0}}&{}&{}&{}&{}&{p_1^{(1)}}
\end{array}} \right] \notag\\
& \; \times \left[ {\begin{array}{*{20}{c}}
{p_{0,0}^{(2)}{\varphi _{0,0}}}\\
 \vdots \\
{p_{2L - 1,0}^{(2)}{\varphi _{2L - 1,0}}}
\end{array}\begin{array}{*{20}{c}}
 \cdots \\
 \vdots \\
 \cdots 
\end{array}\begin{array}{*{20}{c}}
{p_{0,{M_v} - 1}^{(2)}{\varphi _{0,{M_v} - 1}}}\\
 \vdots \\
{p_{2L - 1,{M_v} - 1}^{(2)}{\varphi _{2L - 1,{M_v} - 1}}}
\end{array}} \right]\notag\\
&\qquad \qquad  \qquad \qquad \quad \; \times \left[ {\begin{array}{*{20}{c}}
{y_0^{(0)}}\\
 \vdots \\
{y_0^{({M_v} - 1)}}
\end{array}\begin{array}{*{20}{c}}
 \cdots \\
 \vdots \\
 \cdots 
\end{array}\begin{array}{*{20}{c}}
{y_{{N_3} - 1}^{(0)}}\\
 \vdots \\
{y_{{N_3} - 1}^{({M_v} - 1)}}
\end{array}} \right] \notag\\
&\mathop  = \limits^{(a)} \mathop {{{{\bf{\hat W}}}_s}}\limits_{{P_{{\text{CSI-RS}}}} \times 2L} \times \mathop {{{\bf{W}}_c}}\limits_{2L \times {M_v}} \times \mathop {{\bf{\hat W}}_f^T}\limits_{{M_v} \times {N_3}} \notag\\
&\mathop  = \limits^{(b)} \mathop {{{\bf{W}}_s}}\limits_{{P_{{\text{CSI-RS}}}} \times {P_{{\text{CSI-RS}}}}} \times \mathop {{{\bf{W}}_{{\rm{PMI}}}}}\limits_{{P_{{\text{CSI-RS}}}} \times {N_3}} \times \mathop {{\bf{W}}_f^T}\limits_{{N_3} \times {N_3}} ,
\end{align}
where model $(a)$ utilizes the expression of effective bases and model $(b)$ employs the expression of full bases. The definition of ${y_t^{(f)}}$ is the same as that given in (\ref{b45}) by omitting the index of layer $l$. Unlike the R15 Type II Codebook, where each of the $2L$ beams has an independent wideband amplitude, in the R16 Type II Codebook, the $L$ beams under each polarization share the same wideband amplitude. Through the $2L \times M_v$ parameters in ${{{\bf{W}}_c}}$, the codebook utilizes spectral bases to obtain the beam combination coefficients for $N_3$ subbands. ${\bf{W}}_{\rm{PMI}}$ is a sparse matrix, where only $2L$ rows and $M_v$ columns have non-zero values. Its support set reflects the selection of beams and taps, with the non-zero values representing the beam combination coefficients.

\subsubsection{Compact Model for R17 Type II}
The R17 Type II Codebook employs the same space-frequency compression technique as the R16 Type II Codebook. However, the port-selection codebook now supports a more flexible port selection, where the continuous $L$ ports from dual polarization are replaced by freely selecting $L$ ports. Additionally, the number of beam selections is indicated by $K_1$, and the number of tap selections is indicated by $M$.  Consequently, the precoding matrix for all subbands \( {\bf{W}} \in \mathbb{C}^{{P_{{\text{CSI-RS}}}}\times N_3} \) for a single data stream can be expressed as
\begin{align}
{\bf{W}} &= \left[ {\begin{array}{*{20}{c}}
{{{{\bf{\hat V}}}_s}}&{\bf{0}}\\
{\bf{0}}&{{{{\bf{\hat V}}}_s}}
\end{array}} \right]\left[ {\begin{array}{*{20}{c}}
{p_0^{(1)}}&{}&{}&{}&{}&{\bf{0}}\\
{}& \ddots &{}&{}&{}&{}\\
{}&{}&{p_0^{(1)}}&{}&{}&{}\\
{}&{}&{}&{p_1^{(1)}}&{}&{}\\
{}&{}&{}&{}& \ddots &{}\\
{\bf{0}}&{}&{}&{}&{}&{p_1^{(1)}}
\end{array}} \right] \notag\\
& \; \times \left[ {\begin{array}{*{20}{c}}
{p_{0,0}^{(2)}{\varphi _{0,0}}}\\
 \vdots \\
{p_{K_1 - 1,0}^{(2)}{\varphi _{K_1 - 1,0}}}
\end{array}\begin{array}{*{20}{c}}
 \cdots \\
 \vdots \\
 \cdots 
\end{array}\begin{array}{*{20}{c}}
{p_{0,{M} - 1}^{(2)}{\varphi _{0,{M} - 1}}}\\
 \vdots \\
{p_{K_1 - 1,{M} - 1}^{(2)}{\varphi _{K_1 - 1,{M} - 1}}}
\end{array}} \right]\notag\\
&\qquad \qquad  \qquad \qquad \quad \; \times \left[ {\begin{array}{*{20}{c}}
{y_0^{(0)}}\\
 \vdots \\
{y_0^{({M} - 1)}}
\end{array}\begin{array}{*{20}{c}}
 \cdots \\
 \vdots \\
 \cdots 
\end{array}\begin{array}{*{20}{c}}
{y_{{N_3} - 1}^{(0)}}\\
 \vdots \\
{y_{{N_3} - 1}^{({M} - 1)}}
\end{array}} \right] \notag\\
&\mathop  = \limits^{(a)} \mathop {{{{\bf{\hat W}}}_s}}\limits_{{P_{{\text{CSI-RS}}}} \times K_1} \times \mathop {{{\bf{W}}_c}}\limits_{K_1 \times {M}} \times \mathop {{\bf{\hat W}}_f^T}\limits_{{M} \times {N_3}} \notag\\
&\mathop  = \limits^{(b)} \mathop {{{\bf{W}}_s}}\limits_{{P_{{\text{CSI-RS}}}} \times {P_{{\text{CSI-RS}}}}} \times \mathop {{{\bf{W}}_{{\rm{PMI}}}}}\limits_{{P_{{\text{CSI-RS}}}} \times {N_3}} \times \mathop {{\bf{W}}_f^T}\limits_{{N_3} \times {N_3}} ,
\end{align}
where model $(a)$ utilizes the expression of effective bases and model $(b)$ employs the expression of full bases. Since Release 17 only enhances the port-selection codebook, the spatial domain bases of the R17 Codebook Type II specifically refer to the standard vector basis, as defined in (\ref{b115}) and (\ref{b116}).

\begin{table*}[t]
{\renewcommand{\arraystretch}{2} 
\centering
\caption{Comprehensive Evaluation of Codebooks Across Different Versions}
\label{mastructure}
\resizebox{\textwidth}{!}{%
\begin{tabular}{|c|c|c|c|c|c|c|}
\hline
\rowcolor[HTML]{EFEFEF} 
\textbf{Codebook Type} & \textbf{Spectral Compression} & \textbf{Port Selection} & \textbf{Mobility} & \textbf{Feedback Overhead} & \textbf{Beamforming Performance} & \textbf{Codebook Performance} \\ \hline
R15 Type I & No & Unsupported & Unsupported & Low & Poor & Fair \\ \hline
R15 Type II & No & Supported & Unsupported & High & Excellent & Good \\ \hline
R16 Type II & Yes & Supported & Unsupported & Medium & Fair & Good \\ \hline
R17 Type II & Yes & Supported & Unsupported & Medium & Good & Excellent \\ \hline
R18 Type II & Yes & Unsupported & Supported & Medium & Good & Excellent \\ \hline
\end{tabular}}
}
\end{table*}

\subsubsection{Compact Model for R18 Type II}
The R18 Type II Codebook employs three types of bases, i.e., spatial bases, spectral bases, and temporal bases. The codebook projects the weights of \( P_{\text{CSI-RS}} \) ports onto \( 2L \) beams, the weights of \( N_3 \) subbands onto \( M_v \) taps, and the weights of \( N_4 \) slot intervals onto \( Q \) shifts. The R18 Type II Codebook jointly feeds back the precoding matrices of the space, frequency, and time domains on a tensor. Therefore, the dimensions of the precoding tensor should be \( P_{\text{CSI-RS}} \times N_3 \times Q \). To represent this using a matrix, we flatten the spatial and frequency domains into the row dimension. To this end, we vectorize the R16 Codebook by
\begin{equation}
 {\rm{Vec}}({{{\bf{\hat W}}}_s}{{\bf{W}}_c}{\bf{\hat W}}_f^T) = \left( {{{{\bf{\hat W}}}_f} \otimes {{{\bf{\hat W}}}_s}} \right){\rm{Vec}}({{\bf{W}}_c}),   
\end{equation}
where  $\otimes$ is the Kronecker product operator and
\begin{equation}
{\rm{Vec}}({{\bf{W}}_c}) = \left[ {\begin{array}{*{20}{c}}
{p_0^{(1)}p_{0,0}^{(2)}{\varphi _{0,0}}}\\
 \vdots \\
{p_1^{(1)}p_{2L - 1,0}^{(2)}{\varphi _{2L - 1,0}}}\\
 \vdots \\
{p_1^{(1)}p_{2L - 1,{M_v} - 1}^{(2)}{\varphi _{2L - 1,{M_v} - 1}}}
\end{array}} \right].
\end{equation}
Then, the precoding matrix for all slot intervals \( {\bf{W}} \in \mathbb{C}^{{P_{{\text{CSI-RS}}}N_3}\times Q} \) for a single data stream can be expressed as
\begin{align}
{\bf{W}} &= \left( {\left[ {\begin{array}{*{20}{c}}
{y_0^{(0)}}\\
 \vdots \\
{y_{{N_3} - 1}^{(0)}}
\end{array}\begin{array}{*{20}{c}}
 \cdots \\
 \vdots \\
 \cdots 
\end{array}\begin{array}{*{20}{c}}
{y_0^{({M_v} - 1)}}\\
 \vdots \\
{y_{{N_3} - 1}^{({M_v} - 1)}}
\end{array}} \right] \otimes \left[ {\begin{array}{*{20}{c}}
{{{{\bf{\hat V}}}_s}}&{\bf{0}}\\
{\bf{0}}&{{{{\bf{\hat V}}}_s}}
\end{array}} \right]} \right) \times \notag\\
&\qquad \left[ {\begin{array}{*{20}{c}}
{p_{0,0}^{(1)}p_{0,0,0}^{(2)}{\varphi _{0,0,0}}}\\
 \vdots \\
{p_{1,0}^{(1)}p_{2L - 1,0,0}^{(2)}{\varphi _{2L - 1,0,0}}}\\
 \vdots \\
{p_{1,0}^{(1)}p_{2L - 1,{M_v} - 1,0}^{(2)}{\varphi _{2L - 1,{M_v} - 1,0}}}
\end{array}\begin{array}{*{20}{c}}
 \cdots \\
 \vdots \\
 \cdots \\
 \vdots \\
 \cdots 
\end{array}} \right.\notag\\
& \qquad \;\;\left. {\begin{array}{*{20}{c}}
 \cdots \\
 \vdots \\
 \cdots \\
 \vdots \\
 \cdots 
\end{array}\begin{array}{*{20}{c}}
{p_{0,Q - 1}^{(1)}p_{0,0,Q - 1}^{(2)}{\varphi _{0,0,Q - 1}}}\\
 \vdots \\
{p_{1,Q - 1}^{(1)}p_{2L - 1,0,Q - 1}^{(2)}{\varphi _{2L - 1,0,Q - 1}}}\\
 \vdots \\
{p_{1,Q - 1}^{(1)}p_{2L - 1,{M_v} - 1,Q - 1}^{(2)}{\varphi _{2L - 1,{M_v} - 1,Q - 1}}}
\end{array}} \right]\notag\\
& \qquad \qquad  \qquad \qquad \qquad  \; \times \left[ {\begin{array}{*{20}{c}}
{z_0^{(0)}}\\
 \vdots \\
{z_0^{(Q - 1)}}
\end{array}\begin{array}{*{20}{c}}
 \cdots \\
 \vdots \\
 \cdots 
\end{array}\begin{array}{*{20}{c}}
{z_{{N_4} - 1}^{(0)}}\\
 \vdots \\
{z_{{N_4} - 1}^{(Q - 1)}}
\end{array}} \right]\notag\\
&\mathop  = \limits^{(a)} \mathop {\left( {{{{\bf{\hat W}}}_f} \otimes {{{\bf{\hat W}}}_s}} \right)}\limits_{{P_{{\text{CSI-RS}}}}{N_3} \times 2L{M_v}}  \times \mathop {{{\bf{W}}_c}}\limits_{2L{M_v} \times Q}  \times \mathop {{\bf{\hat W}}_t^T}\limits_{Q \times {N_4}} \notag\\
&\mathop  = \limits^{(b)} \mathop {\left( {{{\bf{W}}_f} \otimes {{\bf{W}}_s}} \right)}\limits_{{P_{{\text{CSI-RS}}}}{N_3} \times {P_{{\text{CSI-RS}}}}{N_3}} \times \mathop {{{\bf{W}}_{{\rm{PMI}}}}}\limits_{{P_{{\text{CSI-RS}}}}{N_3} \times {N_4}} \times \mathop {{\bf{W}}_t^T}\limits_{{N_4} \times {N_4}},
\end{align}
where model $(a)$ utilizes the expression of effective bases and model $(b)$ employs the expression of full bases. The definition of ${z_\iota^{(\tau)}}$ is the same as that given in (\ref{b87}) by omitting the index of layer $l$. Through the $2LM_v \times Q$ parameters in ${{{\bf{W}}_c}}$, the codebook leverages spectral and temporal bases to obtain the beam combination coefficients for $N_3$ subbands and $N_4$ slot intervals. ${\bf{W}}_{\rm{PMI}}$ is a sparse matrix, where only $2LM_v$ rows and $Q$ columns have non-zero values. Its support set reflects the selection of beams, taps, and shifts, with the non-zero values representing the beam combination coefficients. 

Since this codebook jointly reports the precoding matrices for all three domains, it can be simplified using a tensor representation. Specifically, let \(\mathcal{W} \in \mathbb{C}^{P_{\text{CSI-RS}} \times N_3 \times N_4}\) represent the precoding matrix for all subbands and all slot intervals for a single data stream. Then, the structure of the R18 Type II Codebook can be expressed using Tucker decomposition, i.e., 
\begin{align}\label{tucker}
\mathcal{W}\; &\mathop  = \limits^{(a)} \mathop {{{\mathcal{W}}_c}}\limits_{2L \times {M_v} \times Q} { \times _1}\mathop {{{{\bf{\hat W}}}_s}}\limits_{{P_{{\text{CSI-RS}}}} \times 2L} { \times _2}\mathop {{{{\bf{\hat W}}}_f}}\limits_{{N_3} \times {M_v}} { \times _3}\mathop {{{{\bf{\hat W}}}_t}}\limits_{{N_4} \times Q} \\
&\mathop  = \limits^{(b)} \mathop {{{\mathcal{W}}_{\rm{PMI}}}}\limits_{{P_{{\text{CSI-RS}}}} \times {N_3} \times {N_4}} { \times _1}\mathop {{{\bf{W}}_s}}\limits_{{P_{{\text{CSI-RS}}}} \times {P_{{\text{CSI-RS}}}}} { \times _2}\mathop {{{\bf{W}}_f}}\limits_{{N_3} \times {N_3}} { \times _3}\mathop {{{\bf{W}}_t}}\limits_{{N_4} \times {N_4}},\notag 
\end{align}
where $\times _n$ is the mode-$n$ product operator.  \(\mathcal{W}_{c}\) is the core tensor of the Tucker decomposition, which encompasses all the beam weights in the codebook. \(\mathcal{W}_{\rm{PMI}}\) is a sparse tensor, where only \(2L\) rows, \(M_v\) columns, and \(Q\) tubes contain non-zero values, representing the beam combination coefficients.

\subsection{Codebook Performance}
The performance of a codebook is not equivalent to the performance of beamforming. The codebook performance refers to the trade-off between the accuracy of the reported precoding matrices and the feedback overhead. Regarding beamforming performance, the smaller the error between the reported precoding matrices and the ideal ones, the higher the beamforming performance. Thus, if the ideal precoding matrices were directly reported, the beamforming performance would be maximized. However, reporting the ideal ones results in significant feedback overhead, making the performance of this unstructured codebook practically poor.

The Table \ref{mastructure} compares codebooks across different 3GPP releases, illustrating a progressive enhancement in performance from R15 to R18. This evolution is driven by advancements such as the introduction of spectral compression in R16 to reduce feedback overhead, optimization of port selection in R17 to improve beamforming accuracy, and supporting predicted PMI in R18 for mobile UE. While earlier versions prioritized basic functionality (R15 Type I) or high precision at the cost of overhead (R15 Type II), later releases strike a balance between efficiency and adaptability, ultimately achieving excellent codebook performance in R17 and R18 through refined compression techniques and scenario-specific trade-offs. These improvements reflect 3GPP’s focus on addressing diverse 5G-A deployment challenges, from static low-overhead use cases to high-speed mobility scenarios. 

\begin{figure}[t]
 	\centering 
 	\includegraphics[width=3.5in]{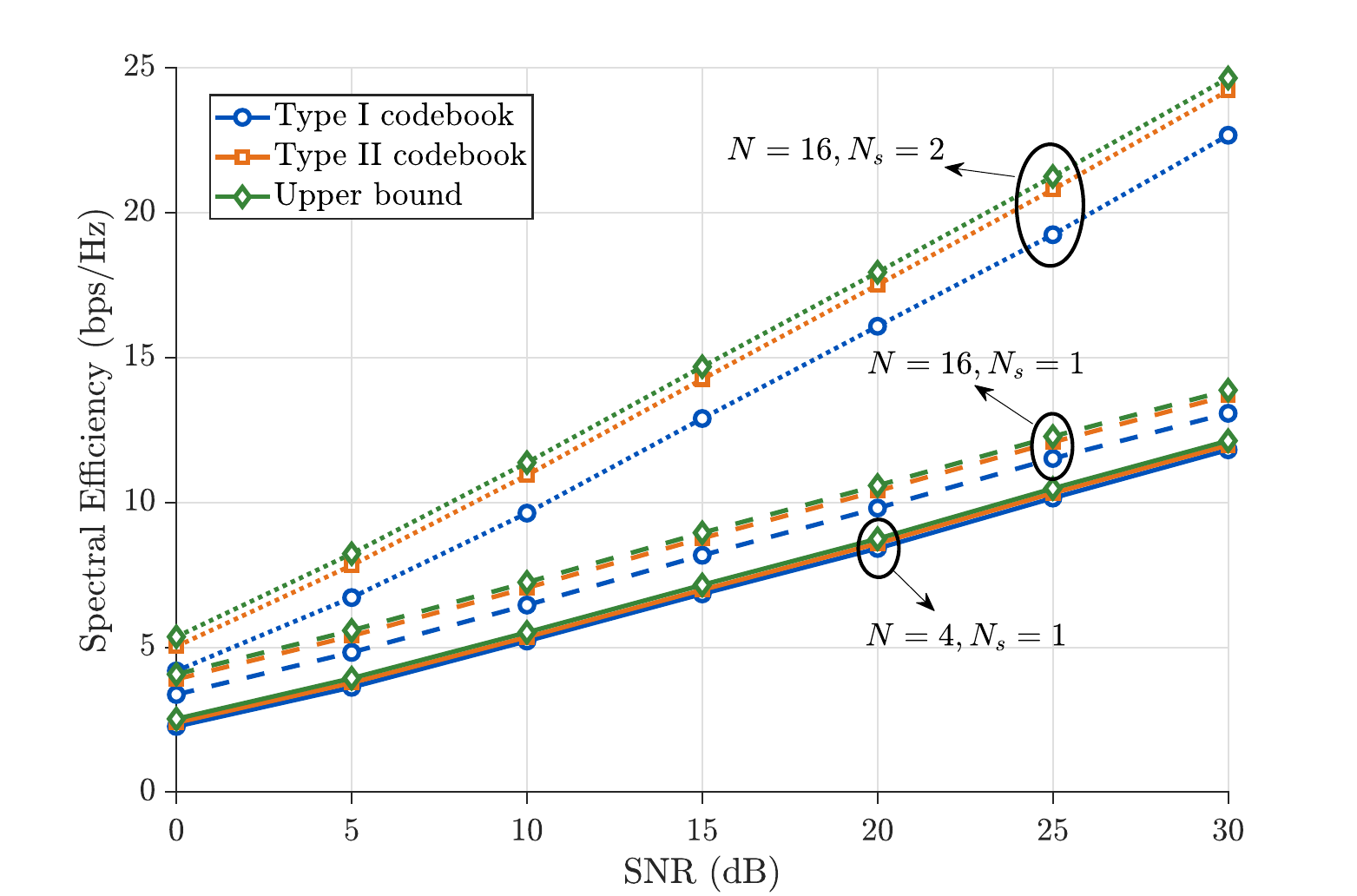}
 	\caption{The spectral efficiency versus SNR.}\label{type1and2} 
    \vspace{-10pt}
 \end{figure}

\begin{table*}[t]
\centering
{\renewcommand{\arraystretch}{2.2} 
\caption{Bit Overhead for $i_1$ Information Elements with $v=2$ and $N_3>19$}
\label{bit1}
\begin{tabular}{|c|c|c|c|c|c|c|}
\hline
\rowcolor[HTML]{EFEFEF} 
\cellcolor[HTML]{EFEFEF}\textbf{$i_1$ Elements} & \cellcolor[HTML]{EFEFEF} \textbf{R15 Regular} & \textbf{R15 Port-Selection} & \textbf{R16 Regular} & \textbf{R16 Port-Selection} & \textbf{R17 Port-Selection} & \textbf{R18 Regular} \\ \hline
$i_{1,1}$ & $\left\lceil {{{\log }_2}\left( {{O_1}{O_2}} \right)} \right\rceil $  & $\left\lceil {{{\log }_2}\left\lceil {\frac{{{P_\text{CSI-RS}}}}{{2d}}} \right\rceil } \right\rceil $ & $\left\lceil {{{\log }_2}\left( {{O_1}{O_2}} \right)} \right\rceil $  & $\left\lceil {{{\log }_2}\left\lceil {\frac{{{P_\text{CSI-RS}}}}{{2d}}} \right\rceil } \right\rceil $ & N/A & $\left\lceil {{{\log }_2}\left( {{O_1}{O_2}} \right)} \right\rceil $ \\ \hline
$i_{1,2}$ & $\left\lceil {{{\log }_2}\Big( {\begin{aligned}
&{{N_1}{N_2}}\\
&\;\;\;L
\end{aligned}} \Big)} \right\rceil$ &  N/A & $\left\lceil {{{\log }_2}\Big( {\begin{aligned}
&{{N_1}{N_2}}\\
&\;\;\;L
\end{aligned}} \Big)} \right\rceil$ &  N/A & $\left\lceil {{{\log }_2}\Big( {\begin{aligned}
&{{P_\text{CSI-RS}}/2}\\
&\;\;\;K_1/2
\end{aligned}} \Big)} \right\rceil$ & $\left\lceil {{{\log }_2}\Big( {\begin{aligned}
&{{N_1}{N_2}}\\
&\;\;\;L
\end{aligned}} \Big)} \right\rceil$ \\ \hline
$i_{1,3,l}$ & $\left\lceil {{{\log }_2}\left( {2L} \right)} \right\rceil $ & $\left\lceil {{{\log }_2}\left( {2L} \right)} \right\rceil $  &  N/A &  N/A & N/A  &  N/A \\ \hline
$i_{1,4,l}$ & $3\left( {2L - 1} \right)$ & $3\left( {2L - 1} \right)$ &  N/A &  N/A &  N/A &  N/A \\ \hline
$i_{1,5}$ &  N/A &  N/A & $\left\lceil\log_2(2M_v)\right\rceil$ & $\left\lceil\log_2(2M_v)\right\rceil$ &  N/A & $\left\lceil\log_2(2M_v)\right\rceil$  \\ \hline
$i_{1,6,l}$ &  N/A &  N/A & $\left\lceil {{{\log }_2}\Big( {\begin{aligned}
&2M_v-1\\
&\;M_v-1
\end{aligned}} \Big)} \right\rceil$ & $\left\lceil {{{\log }_2}\Big( {\begin{aligned}
&2M_v-1\\
&\;M_v-1
\end{aligned}} \Big)} \right\rceil$ &$\left\lceil\log_2(N-1)\right\rceil$  & $\left\lceil {{{\log }_2}\Big( {\begin{aligned}
&2M_v-1\\
&\;M_v-1
\end{aligned}} \Big)} \right\rceil$ \\ \hline
$i_{1,7,l}$ &  N/A & N/A &  $4LM_v$ & $4LM_v$ & $2K_1M$ & $4LM_vQ$ \\ \hline
$i_{1,8,l}$ &  N/A &  N/A &  $\left\lceil {{{\log }_2}\left( {2L} \right)} \right\rceil $ & $\left\lceil {{{\log }_2}\left( {2L} \right)} \right\rceil $ & $\left\lceil\log_2(K_1M)\right\rceil$ & $\left\lceil {{{\log }_2}\left( {2LQ} \right)} \right\rceil $ \\ \hline
$i_{1,10,l}$ &  N/A &  N/A & N/A &  N/A & N/A & $\left\lceil\log_2(N_4-1)\right\rceil$ \\ \hline
\end{tabular}}
\end{table*}

\begin{table*}
{\renewcommand{\arraystretch}{2.2} 
 \caption{Bit Overhead for $i_2$ Information Elements with $v=2$ and $N_3>19$}
\label{bit2}
\begin{tabular}{|c|c|c|c|c|c|c|}
\hline
\rowcolor[HTML]{EFEFEF} 
\cellcolor[HTML]{EFEFEF}\textbf{$i_2$ Elements} & \cellcolor[HTML]{EFEFEF} \textbf{R15 Regular} & \textbf{R15 Port-Selection (PS)} & \textbf{R16 Regular} & \textbf{R16 PS} & \textbf{R17 PS} & \textbf{R18 Regular} \\ \hline
$i_{2,1,l}$ & $\begin{aligned}&
\min \left( {{M_v},{K^{(2)}}} \right) \cdot {\log _2}{N_{{\text{PSK}}}} \\
& \qquad - {\log _2}{N_{{\text{PSK}}}}\\
 & \!+ 2 \left( {{M_v} \!-\! \min \left( {{M_v},{K^{(2)}}} \right)} \right)
\end{aligned}$ & $\begin{aligned}&
\min \left( {{M_v},{K^{(2)}}} \right) \cdot {\log _2}{N_{{\text{PSK}}}} \\
& \qquad - {\log _2}{N_{{\text{PSK}}}}\\
 & \!+ 2\left( {{M_v} \!-\! \min \left( {{M_v},{K^{(2)}}} \right)} \right)
\end{aligned}$ & N/A & N/A & N/A & N/A\\ \hline
$i_{2,2,l}$ & $\min \left( {{M_v},{K^{(2)}}} \right) - 1$ & $\min \left( {{M_v},{K^{(2)}}} \right) - 1$ & N/A & N/A & N/A & N/A\\ \hline
$i_{2,3,l}$ & $4$ & $4$ & $4$ & $4$ & $4$ & $4$\\ \hline
$i_{2,4,l}$ & $3(K^{NZ}-2)$ & $3(K^{NZ}-2)$  & $3(K^{NZ}-2)$ & $3(K^{NZ}-2)$  & $3(K^{NZ}-2)$  & $3(K^{NZ}-2)$\\ \hline
$i_{2,5,l}$ & $4(K^{NZ}-2)$ & $4(K^{NZ}-2)$  & $4(K^{NZ}-2)$ & $4(K^{NZ}-2)$  & $4(K^{NZ}-2)$ & $4(K^{NZ}-2)$\\ \hline
\end{tabular}%
}   
\end{table*}

Fig. \ref{type1and2} depicts the spectral efficiency of Type I and Type II codebooks for antenna configurations \( (N_1, N_2) = (4,1), (16,1) \), with a fixed oversampling factor \( (O_1, O_2) = (4,1) \). The analysis focuses on spatial compression for single-polarization, single-stream data to isolate per-carrier capacity differences. Under ideal feedback, the beamforming vector would be the dominant eigenvector of the channel covariance. The Type I codebook approximates this by feeding back the spatial basis with the maximum projected energy, minimizing overhead. The Type II codebook, however, employs a linear combination of bases. Our configuration uses \( L=4 \) beams. Wideband amplitude \( p^{(1)}_{l} \) is quantized to $3$ bits and subband amplitude reporting is disabled, with phase quantized to 3 bits, i.e., \( N_{\text{PSK}}=4 \). The Type II implementation selects the highest-energy beam group from four orthogonal sets, then uses orthogonal matching pursuit (OMP) to select four bases. Their weights, derived via least squares algorithm, are quantized in amplitude and phase. As can be seen from Fig. \ref{type1and2}, when the number of antennas is small, the performance gap between Type I and Type II is not significant. As the number of antennas increases, the gap becomes more pronounced. Furthermore, when the number of data streams increases, the gap is further widened.

\subsection{Feedback Comparison}
The feedback information in the codebook, i.e., PMI, primarily consists of two parts: $i_1$ and $i_2$. The feedback overhead can be represented by the bit overhead of the information elements within $i_1$ and $i_2$. Since each codebook typically supports multi-stream transmission, the feedback overhead varies across schemes with different data streams. Additionally, some codebooks, such as the R16 regular and port-selection codebooks, have different feedback mechanisms depending on the number of subbands $N_3 \leq 19$ and $N_3 > 19$. Therefore, to compare the feedback overhead of different versions of codebooks, we have considered the scenario with rank $v = 2$ and a subband number $N_3 > 19$. The bit overhead of the information elements for $i_1$ and $i_2$ is listed in Table \ref{bit1} and Table {\ref{bit2}}, respectively.

Fig. \ref{r15-18} compares the feedback overhead of the regular codebooks defined from Release 15 to Release 18. Note that Release 17 is excluded from this comparison as its enhancements are specific to the port selection codebook. The total feedback bitload was calculated for configurations with \(L = 1, 2, 3, 4\) under wideband and multi-time-interval reporting. The case of \(L=1\) effectively reduces the Type II codebook to a Type I codebook. The simulation parameters were set as follows: number of logical ports \(N_1 N_2 = 16\), oversampling factors \(O_1 O_2 = 4\), number of subcarriers \(N_3 = 18\), number of time intervals \(N_4 = 4\), time shifts \(Q = 2\), rank \(v = 2\), number of taps \(M_v = 5\), phase quantization bits \(N_{\text{PSK}} = 4\), maximum reported subband amplitudes per layer \(K^{(2)} = 6\), and total number of non-zero coefficients \(K^{NZ} = 20\).  The results indicate that the R15 codebook has the highest feedback overhead, which is more than ten times greater than that of the R16 codebook. The R18 codebook further reduces the feedback overhead compared to R16. As number of spatial bases $L$ increases, the feedback overhead of all three codebooks grows at the same rate.

 \begin{figure}[t]
 	\centering 
 	\includegraphics[width=3.5in]{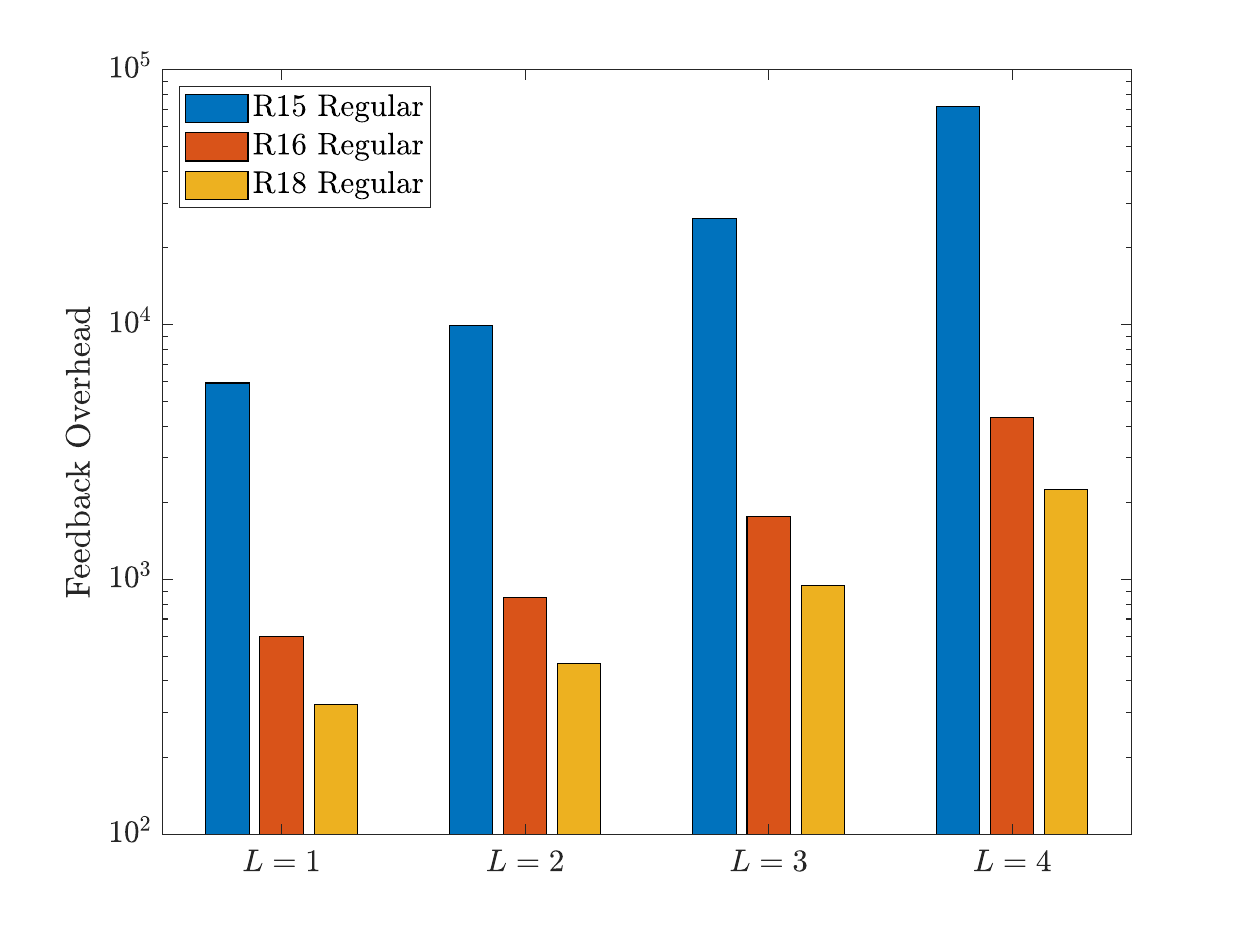}
 	\caption{The feedback overhead versus $L$.}\label{r15-18} 
    \vspace{-10pt}
 \end{figure}

\subsection{Applicable Scenarios}
The R15 Type I Codebook supports only single-stream and single-beam transmission, fulfilling the most fundamental functions of a codebook. It is suitable for scenarios with low complexity and low power consumption requirements. Consequently, all 5G mobile phones on the market must support this codebook. Building upon Type I, the R15 Type II Codebook enhances the precoding precision. It supports single-stream multi-beam combined transmission and higher-order MCS transmission modes. This makes it appropriate for scenarios that demand higher data rates and system capacity, such as enhanced mobile broadband (eMBB). Due to the complex structure, only about $20\%$ of 5G terminals on the market currently support this codebook. The R16 and R17 Type II Codebooks introduce frequency-domain compression techniques on top of the R15 Type II, reducing the feedback overhead. As a result, they are employed in scenarios that require high reliability and low latency, such as non-terrestrial networks (NTN) and intelligent transportation systems (ITS). The R18 Type II Codebook further incorporates temporal compression techniques. It predicts the channel for varying blocks, making it suitable for mobile user scenarios, such as high-speed trains and vehicles. This is crucial for supporting applications like eMBB and ultra-reliable low-Latency communication (URLLC). For versions R17 and above, it is likely that only flagship phones will support them, accounting for less than $1\%$ of 5G terminals on the market. The standardization of protocols typically precedes the market deployment. Therefore, more terminals will support these advanced codebooks, and the design of algorithms for these codebooks will become extremely important in the future.

\section{Future Directions and Other Scenarios}\label{futuredirection}
Beamforming, as a technique for multi-antenna transmission, has spawned a range of research topics, including channel estimation and codebook design. Within the 3GPP 5G protocol, the PMI codebook is primarily used for feedback of beamforming matrices in massive MIMO systems. In this section, we first outline research directions on beamforming codebooks that could influence future networks, yet are still far from the standardization process. Then, we summarize other scenarios beyond massive MIMO beamforming, where codebooks likely will be needed, and a new codebook design is needed due to their unique channel characteristics and transmission features.

\subsection{Research Directions for Massive MIMO Beamforming}

\subsubsection{Evolution of Codebook Bases}
The existing regular codebooks have been designed with spatial, spectral, and temporal bases to compress feedback on precoding matrices. However, these bases conform to either the DFT vectors (i.e., spectral and temporal bases) or the Kronecker product of DFT vectors (spatial bases). Such bases allow the precoding to be represented in the angular, delay, and Doppler domains and have been verified to possess sparse characteristics. Moving forward, we need to contemplate whether there are better bases that allow for a sparser representation of precoding vectors in some new domains. A classic example is in near-field communications, where each propagation path contributes with a curved or spherical wavefront, whereas DFT vectors characterize the response induced by plane waves when having uniform arrays. Therefore, the DFT spatial bases cannot sparsely represent the precoding matrix in the near-field, necessitating research into better bases for codebook construction. In the port-selection codebooks, the spatial bases are the standard basis vectors, i.e., one element is $1$, and the rest are $0$. Such bases are not as flexible as the bases of regular codebooks, which usually have multiple sets of oversampled DFT orthogonal bases. Thus, the design of oversampling bases in port-selection codebooks is also an important part of future research. Moreover, existing codebooks are typically designed for UPAs. If non-uniform arrays or other shapes of arrays, such as circular arrays, are considered, different bases should be provided for different topological structures.

\subsubsection{Innovative Frameworks for Codebook}
Future research and development on the codebook framework can progress in two main directions: the unified codebook framework and the artificial intelligence (AI)-enabled codebook framework. a) The former involves unifying the existing codebook structures. Currently, for single-panel codebooks, there are $15$ sets for Type I Codebook and $7$ sets for Type II Codebook, supporting a variety of precoding orders, spatial bases, array configurations, and layer requirements. Future codebooks should be more concise and flexible, necessitating a unified codebook framework whose reporting mechanisms can satisfy most or all of the functionalities of existing codebooks. For instance, this paper conceives a tensor structure for the codebook, as proposed in (\ref{tucker}), focusing on the reporting parameters and bases design on the Tucker core and the factor matrices. b) The second category aims to harness the AI technology for application in codebooks. The general approaches and underlying models in existing codebooks are deeply rooted in signal processing methodologies developed over the last few decades. Therefore, it is accurate to state that contemporary feedback designs are model-based. There is an opportunity for further enhancements using deep learning networks, which would make the feedback protocols data-aided \cite{deep1,deep2,deep3, dreifuerst2024ml, heathbeam}. The specific use of the AI codebook framework can be diverse, such as directly designing a standardized network for compressing feedback and reconstruction, similar to an autoencoder. Alternatively, a network can be standardized to learn and discover statistical structures that exist in the wireless channel and to generate the bases for the codebook, which will then become site-specific to capture the channel, hardware, mobility, and data traffic characteristics in a particular cell \cite{Heng2024a,Kwak2024a}.
Moreover, one can introduce trainable parameters into the model-based codebooks that can be specified using control signaling. The main purpose is to compensate for inaccuracies in the models underpinning the protocols, such as hardware impairments and channel characterizations, and to reach an enhanced solution that can be universally utilized in an operator's network.

\subsubsection{Hierarchical Codebook Structures}
The existing codebook structures in 5G systems provide direct feedback of the precoding matrix between the data streams and the CSI-RS ports. In high sparsity channel scenarios, such as those found in millimeter-wave or terahertz bands, hierarchical measurements can reduce pilot overhead through the use of multi-stage measurements and feedback, assisting the BS in determining the appropriate precoding matrix.  Academic research has explored the use of hierarchical beam training to swiftly identify a satisfactory DFT beam\cite{xiaocodebook,ningexh,ningwide}. Yet, there is no hierarchical codebook within the standards that employs multi-stage measurements and feedback to rapidly establish a satisfactory precoding matrix. The primary reason for this is that exhaustive beam training by the BS is a cell-level behavior; for instance, after one round of beam sweeping, all UEs can provide feedback simultaneously. In contrast, hierarchical beam training is a user-specific procedure, where different UEs must perform it independently. Moreover, while a hierarchical search can reduce the number of beam scans, it increases the number of feedback rounds. These, in turn, raise beam alignment latency and introduce additional scheduling overhead. Current 5G systems have implemented a two-stage measurement process, where wide beams are used to scan for SSB signals, and then CSI-RS signals are sent within the optimal SSB beam. However, the implementation of the measurement within the SSB beam is facilitated by the PEB, which is independent of the codebook. A hierarchical codebook, on the other hand, would allow for the definition of multi-stage precoding matrices, effectively creating one or more layers of virtual ports between the data streams and the CSI-RS ports. At each stage, the UE would feed back the precoding information across these different stages, with the process being defined by a hierarchical codebook. Notably, the precoding bases and the coefficient precision between different stages could differ.
\subsubsection{Analog Beamforming Codebooks}
Hybrid beamforming architectures are often adopted in current 5G BSs, where the number of antenna ports exceeds the number of digital ports, and the number of digital ports is greater than or equal to the number of CSI-RS ports. Within this framework, the linear mapping  (ignoring the non-linear distortion from power amplifiers) from digital ports to antenna ports is referred to as analog beamforming. Current codebooks are designed to feed back the precoding matrices from data streams to CSI-RS ports, which is essentially feedback for digital beamforming. Nonetheless, analog beamforming design plays a vital role in the system as well. In today's 5G protocol, BSs cover the entire service area by scanning through analog beams and select the best analog beam for subspace channel measurements. Thus, BSs require an analog beamforming codebook to ensure the capability of signal coverage. Since this action is internal to the BS, it has not been included in the standards. Future downlink channel estimation may shift from measuring the channel subspace via an analog beam to measuring the full space of the antenna ports. Under this new approach, the BS could transmit CSI-RS across different analog beams. If the UE is aware of the analog beamforming vectors and the results of channel measurements for various subspaces, it could combine this information to reconstruct the full channel profile. Consequently, the analog beamforming codebook must be standardized to allow the UE to understand the analog beamforming vectors employed by the BS with minimal overhead. 

\subsection{Other MIMO-Related Features}

The MIMO channel estimation and feedback mechanisms have been refined in every release of 5G. The basic goals are to deliver higher communication performance while reducing the feedback overhead, which can be achieved by exploiting inherent structures in the channels. The resources saved by these enhancements can either be used for data transmission or to support additional antenna ports.  Beyond these massive MIMO refinements, there are new emerging MIMO-related functionalities, frequency bands, and array geometries that will require radical changes to the existing protocols.

6G is expected to expand the supported frequency ranges to include the upper mid-band (e.g., 7-24 GHz) and terahertz band (above 100 GHz). The former is meant to provide higher capacity than 5G with similar coverage in urban scenarios, while the latter is envisioned for short-range links that require extreme capacity for wireless backhaul or access.
Future networks operating in the upper mid-band will need more physical antennas to maintain the same physical array aperture as in the lower mid-band (e.g., at 3.5 GHz). Since this could enable MU-MIMO transmission with even more data streams, it is desirable for the 3GPP standard to eventually support more CSI-RS antenna ports than currently. Future systems with 256, 512, or more ports have been proposed to be called gigantic MIMO (gMIMO) \cite{GiganticMIMO} to distinguish them from current massive MIMO systems with 16, 32, or 64 ports. These developments also opens the door for near-field communications and distributed beamforming, which will be discussed in detail below.

The research community has introduced many additional names for future MIMO-related features. 
Ultra-massive MIMO (UM-MIMO) is the common name for future MIMO technology in terahertz bands \cite{UM1,UM2,UM3}, which can squeeze in many thousands of elements in a small form factor with the primary goal of achieving beamforming gains that overcome pathloss issues.
By contrast, the terms extremely large aperture array (ELAA) \cite{emilten} and extremely large-scale MIMO (XL-MIMO) \cite{XL1,XL2} are used synonymously and refer to BSs with arrays that often are \emph{physically} larger than current BSs, but particularly are \emph{electrically} large compared to the wavelength. The primary reason is to achieve more spatial degrees of freedom by having at least an order-of-magnitude more antenna ports than current systems. 
Such systems will feature various near-field propagation effects that require new channel models and feedback mechanisms compared to 5G.
Since there is overlap between the different terms (and they are not always used consistently in the community), the remainder of this section will focus on the main technological features that require further research to support future standardization, but without using specific MIMO terms.

%{\color{red} Hi Prof. Bj\"ornson, we should also discuss UM-MIMO and XL-MIMO here. My understanding is that UM-MIMO emphasizes a large number of antenna elements, while XL-MIMO focuses on the large physical size of the antenna arrays. Perhaps you have more precise definitions to differentiate these terms.}

\subsubsection{Terahertz Beamforming}
The terahertz band, which spans frequencies from $0.1$ to $10$ terahertz, represents a new frontier in wireless communication technology, offering bandwidths that could enable data transmission rates of several terabits per second\cite{thz1,thz2,thz3}. This band of the electromagnetic spectrum is characterized by unique propagation features that separate it from lower frequency bands. A critical aspect is the necessity for using many antenna elements, since each element becomes tiny, leading to extremely directional beamforming transmission and reception.
The terahertz communication will be mostly limited to LoS transmission due to the significant molecular absorption, and weak reflecting/scattering capability experienced by terahertz waves\cite{ningmimo}. Furthermore, the high-frequency nature of terahertz waves results in a low-rank channel, indicating a limited number of significant propagation paths. The low-rank characteristics of terahertz channels imply that the PMI codebooks for these channels can be significantly different from those used in lower frequency bands. With fewer spatial and spectral bases to consider, i.e., beams and delay taps, this simplification also suggests that channel prediction or beamforming prediction in terahertz communications is more manageable compared to other beamforming scenarios. Moreover, the design of an analog beamforming codebook for terahertz beamforming is particularly challenging due to the impact of beam squint or beam split phenomena at large bandwidths\cite{ningcodebookwide,ningmaxmin,ningsplit}. When the transmitting array becomes very large compared to the wavelength, it is also possible to synthesize other kinds of wavefronts than what conventional beamforming generates, including Bessel and Airy beams \cite{Petrov2024}. If the network should switch between these wavefronts depending on the propagation scenario, then new feedback mechanisms are required for that purpose.

\subsubsection{Near-field Beamforming}
The existing beamforming codebooks are particularly designed for far-field channels, which can be described as a summation of plane waves arriving from different directions. If future systems use even larger arrays and/or higher carrier frequencies than currently, we can no longer rely on the far-field channel approximation because the Fraunhofer distance can become hundreds of meters. In such circumstances, each propagation path will instead contribute with a curved/spherical wavefront. This is called radiative near-field effect, and when communicating over such channels, the flashlight-like beamforming is replaced by lens-like beamfocusing, which is referred to as near-field beamforming\cite{Kosasih2025a}. This feature is attractive for future systems because it gives the BS a depth perception that can be utilized to focus different data streams at receivers located at different distances \cite{10273772}, even if those users are undistinguishable in the angular domain. Due to the near-field propagation features, the resulting sparsity will not be well represented in the DFT domain. A potential solution is to develop new codebooks based on polar-domain dictionaries \cite{10123941,10403506,9913211,Abdallah2025}, which can capture the channel variability inherent in a spherical wave. Future research needs to explore which are the most efficient refinements to capture near-field characteristics in CSI compression and feedback. Another complication in near-field communications is the presence of spatial non-stationary multi-path propagation \cite{XL2}, where the signals arriving from some scattering objects are only visible over parts of the array. Uncovering sparsity in such propagation environments will require very different bases than when the waves are fully visible over the array.

\subsubsection{Passive Beamforming}
Reconfigurable intelligent surfaces (RIS) are programmable metasurfaces composed of many passive elements with tunable electromagnetic properties\cite{9721205}. These properties can be adjusted during communications to perform passive beamforming, i.e., reflection of incident waves in desirable directions. The most compelling use case is to create virtual LoS paths, where the surface is used to perform beamforming around an obstacle to achieve a stronger received signal power than if the signal needs to pass through the obstacle \cite{IRS1,risbook2,risbook3}. An RIS can either be viewed as a noise-free wireless repeater that uses its physical size to enhance the propagation link instead of using noisy amplifiers, or it can be viewed as moving some of the antenna elements from a MIMO transmitter/receiver to an intermediate point.
Configuring an RIS is akin to conventional MIMO beamforming, in the sense that one needs to transmit reference signals, determine a suitable RIS configuration at the receiver, and then feed it back to the RIS over a control channel with limited resources. Some recent research on this topic is \cite{10600711, joseph2024TWC,Enqvist2025a} and the tools for beam training and feedback compression resemble those used in MIMO communications, but there are several key differences: a) A typical RIS has an order-of-magnitude more reflecting elements than a MIMO array has logical antenna ports; b) The use-cases for RIS might be more limited to spatially sparse environments; c) The RIS is unable of performing frequency-selective beamforming.  Hence, new beam training and feedback compression algorithms need to be designed \cite{9200578,10236999} and tested under realistic conditions. The RIS channel might feature near-field propagation, depending on the deployment scenario and carrier frequency.

% d) The RIS configuration might have to change more rapidly due to new scheduling decisions.
\subsubsection{Distributed MIMO Beamforming}
There is an ongoing trend of moving baseband processing from the BS sites to edge clouds that are shared between multiple sites. The principal motivations are to reduce the total compute resource requirements and use virtualized machines running on general-purpose hardware. Importantly, this development can also be the enabler for distributed MIMO, also known as cell-free massive MIMO \cite{Nayebi2017a,Ngo2017b,cellfreebook}, where neighboring BSs serve their users through coherent joint transmission. The academic literature promotes the use of uplink pilot signals for channel estimation (e.g., SRS), but a CSI codebook might still be needed, either for sharing CSI between BSs and the edge computer, or for CSI-RS-like channel estimation in practical networks. This raises the question: What kind of codebooks are needed? The simplest answer would be to adopt already standardized 3GPP protocols independently for each BS, but one would then ignore the channel correlation that exists in practice. The small-scale fading realizations will be practically uncorrelated between BSs, but the channels’ angular support, mobility characteristics, and frequency selectivity can be strongly correlated. This will be particularly important if multiple BS panels are deployed on the same or neighboring rooftops to increase the number of antennas and achieve a near-field-like depth perception \cite{kosasih2025}. The aforementioned site-specific enhancements will likely play an essential role to learn the CSI correlation between the BSs.

\subsubsection{Sensing Applications}
Future cellular networks have the potential to be turned into large-scale radar sensing systems, where transmission resources not needed for communications can be used to explore the propagation environment. This is often referred to as integrated sensing and communication (ISAC) \cite{Mishra2019a,Liu2022a
} and has been studied within 3GPP for several years \cite{10349862}. Sensing can make the network operation more efficient by tracking user movements and predicting future resource allocation needs. The environmental awareness achieved through sensing, regarding both static and moving objects, might also be useful for various applications running on top of the network, e.g., monitoring of roads and airspaces, intrusion detection in facilities, and real-time refinements of digital twins. A basic feature in a sensing system is to sweep around a beam in different directions and investigate how the signal is reflected by objects before reaching a receiver. One possible implementation is to let a BS send a radar pulse in a downlink slot and then let a different BS receive the ``echo''. Such bistatic sensing will require codebooks for the beam sweeping, and possibly different mechanisms for target detection and target tracking. The design of beamforming codebooks for ISAC is an ongoing research topic of practical importance \cite{10930389,hussain2025nearfieldisacsynergydualpurpose,10626219}.

\section{Conclusion}\label{conclusion}
This paper has presented a comprehensive and accessible tutorial on PMI codebooks, with the aim of helping researchers and engineers gain a clear understanding without needing to delve into complex protocol specifications. We began by reviewing the fundamental principles of codebook-based beamforming, followed by an in-depth discussion of the practical implementation of CSI-RS measurements and PMI feedback mechanisms. The evolution of 3GPP codebooks from Release 15 to Release 18 was then analyzed, covering Type I and Type II codebooks, including both regular and port-selection codebooks, along with relevant protocol implementation details.  Furthermore, we developed simplified models for each codebook, analyzed their performance disparities, evaluated their feedback overhead, and summarized their application scenarios. Finally, the tutorial highlighted promising future research directions and explored additional beamforming scenarios pertinent to the ongoing development of beamforming codebooks. It is our hope that this work will serve as a timely and valuable resource, thereby fostering more effective collaboration between academia and industry to advance this critical wireless technology.

\begin{appendices}    
\section{Details of Full CSI-based Beamforming}\label{appdA}
To enable researchers in the MIMO field to quickly reproduce classical beamforming schemes, we summarize several representative full CSI-based beamforming methods for both SU-MIMO and MU-MIMO systems.
\subsection{SU-MIMO}\label{BF-SU-MIMO}

Various beamforming techniques have been developed as summarized in Table \ref{SUMIMO}.

\begin{itemize}
    \item \emph{Singular Value Decomposition (SVD)}: This is the main method for SU-MIMO systems. The SVD decomposes the channel matrix into its singular values $\bm{\Sigma}$ and singular vectors $\mathbf{U}$ and $\mathbf{V}$. This decomposition facilitates optimal power allocation and effective interference management by transmitting along the channel’s eigenmodes. However, if CSI is only partially known by either BS or UE, then alternative methods can be considered.
    \item \emph{Maximum Ratio Transmission (MRT)}: This method maximizes the SNR by aligning the transmit beam with the channel’s dominant eigen-direction. However, it does not address the inter-stream interference inherent in MIMO systems, which makes it generally undesirable.
    \item \emph{Zero Forcing (ZF)}: By inverting the channel matrix, ZF effectively cancels inter-stream interference, thereby improving signal separation. When the condition number of the channel matrix is large, the elements of its inverse matrix can be quite large. This may result in the weights for certain antennas possibly exceeding the power amplifier's (PA) power constraints, requiring an overall reduction in transmission power.\footnote{For a ZF combiner, these large elements, when multiplied by the noise, can lead to noise amplification, thus reducing the SNR at the receiver.}
    \item \emph{Regularized Zero Forcing (RZF)}: This technique introduces a regularization factor $\xi$ to mitigate noise amplification. By tuning $\xi$, RZF can adapt to different operating regimes: it approaches MRT when $\xi\rightarrow\infty$, ZF when $\xi=0$, and MMSE when $\xi=\frac{\sigma^2}{P_t}$, where $P_t$ and $\sigma^2$ denote the transmit power and noise power, respectively.
A common way to select the regularization factor is $\xi=\frac{\sigma^2}{P_t}$, which is called the minimum mean square error (MMSE) beamforming and is listed separately in the table.
    \item \emph{Geometric Mean Decomposition (GMD)}: This method addresses the imbalance among spatial streams produced by SVD that negatively impacts the bit error ratio (BER). It provides a more balanced power-allocation approach to improve BER performance when multiple data streams are transmitted with a single codeword. Performing GMD on the channel obtains $\mathbf{H}=\mathbf{Q}\mathbf{R}\mathbf{P}^H$, where $\mathbf{R}$ is upper triangular with identical diagonal entries $\overline{\sigma}=\left(\prod_{i=1}^r\sigma_i\right)$, $r$ is the rank of $\mathbf{H}$, $\{\sigma_i\}_{i=1}^r$ denotes the singular values,
    and $\mathbf{Q}$ and $\mathbf{P}$ are unitary matrices.  
\end{itemize}

\begin{table}[]\centering
{\renewcommand{\arraystretch}{2} 
\caption{Beamforming Schemes for SU-MIMO}\label{SUMIMO}
\begin{tabular}{ccllc}
\hline
% \multicolumn{5}{|c|}{{\bf{Channel Capacity}}}                                                                                  \\ \hline
% \multicolumn{5}{|c|}{$ 	R = {\rm log_2}\left({\rm det}\left(\mathbf{I}_{N_r}+\frac{P_t}{\sigma_n^2}\mathbf{H}\mathbf{W}\mathbf{W}^{\rm H} \mathbf{H}^{\rm H}\right)\right)$}                                                                                  \\ \hline
\multicolumn{1}{|c|}{\bf{Schemes}} & \multicolumn{3}{c|}{\bf{Description}} & \multicolumn{1}{c|}{\begin{tabular}[c]{@{}c@{}}
\bf{Closed Form}\end{tabular}} \\ \hline
\multicolumn{1}{|c|}{SVD}     & \multicolumn{3}{c|}{ \begin{tabular}[c]{@{}c@{}}$\mathbf{H}=\mathbf{U}\bm{\Sigma}\mathbf{V}^H$, $\mathbf{W}=[\mathbf{V}]_{:,1:N_s}$\end{tabular}
}            & \multicolumn{1}{c|}{\checkmark}                        \\ \hline 
\multicolumn{1}{|c|}{MRT}      & \multicolumn{3}{c|}{$\mathbf{W}=\mathbf{H}^H$}            & \multicolumn{1}{c|}{\checkmark}                      \\  \hline
\multicolumn{1}{|c|}{ZF}      & \multicolumn{3}{c|}{$\mathbf{W}=\mathbf{H}^H\left(\mathbf{H}\mathbf{H}^H\right)^{-1}$}            & \multicolumn{1}{c|}{ \checkmark}                      \\ \hline
\multicolumn{1}{|c|}{RZF}     & \multicolumn{3}{c|}{  
$\mathbf{W}=\mathbf{H}^H\left(\mathbf{H}\mathbf{H}^H+\xi\mathbf{I}_{N_r}\right)^{-1}$}            & \multicolumn{1}{c|}{\checkmark }                      \\ \hline
\multicolumn{1}{|c|}{MMSE}    & \multicolumn{3}{c|}{$\mathbf{W}=\mathbf{H}^H\left(\mathbf{H}\mathbf{H}^H+\frac{\sigma^2}{P_t}\mathbf{I}_{N_r}\right)^{-1}$}            & \multicolumn{1}{c|}{ \checkmark}                      \\ \hline
\multicolumn{1}{|c|}{GMD}     & \multicolumn{3}{c|}{ \begin{tabular}[c]{@{}c@{}}     $\mathbf{H}=\mathbf{QR}\mathbf{P}^H$, $\mathbf{W}=[\mathbf{P}]_{:,1:N_s}$       \end{tabular}
}            & \multicolumn{1}{c|}{\checkmark}                        \\ \hline 
\end{tabular}}
\end{table}
\subsection{MU-MIMO}

Unlike SU-MIMO, MU-MIMO systems must explicitly address inter-user interference. Table \ref{MUMIMO} summarizes several commonly used full CSI-based beamforming methods for MU-MIMO. Notably, ZF, MMSE, and RZF beamforming techniques can be extended to MU-MIMO by stacking all users' channel matrices as
$\widetilde{\mathbf{H}}=[\mathbf{H}_1^H,\ldots,\mathbf{H}_K^H]^H$. Additional techniques are detailed below:

\begin{itemize}
\item  \emph{Eigen Zero-forcing (EZF)}: This approach combines SVD and ZF to maximize the signal strength while suppressing inter-user interference. For each user $k$, the dominant eigenvectors are extracted by applying the SVD, i.e.,
\begin{equation}
\mathbf{H}_k=\mathbf{U}_k\bm{\Sigma}_k\mathbf{V}^H_k,    
\end{equation}
where the right singular matrix can be given by
\begin{equation}
{{\bf{V}}_k} = \left[ {{{\bf{v}}_{1,k}},{{\bf{v}}_{2,k}},\ldots,{{\bf{v}}_{{N_t,k}}}} \right].    
\end{equation}
Let ${{{\bf{\hat V}}}_k} = \left[ {{{\bf{v}}_1},\ldots,{{\bf{v}}_{v_k}}} \right]$ denote the eigen-beamformer for user $k$ with $v_k$ representing the number of data streams allocated. To eliminate inter-user interference, the eigen-beamformers of all users are aggregated for ZF, i.e., 
\begin{equation}
{{\bf{V}}_{{\rm{eff}}}} = \left[ {{{{\bf{\hat V}}}_1},{{{\bf{\hat V}}}_2},\ldots,{{{\bf{\hat V}}}_K}} \right].    
\end{equation}
Assuming $v_k=v$, the EZF beamforming is given by
\begin{equation}
{\bf{W}} = {{\bf{V}}_{{\rm{eff}}}}{\left( {{\bf{V}}_{{\rm{eff}}}^H{{\bf{V}}_{{\rm{eff}}}} + \xi {{\bf{I}}_{Kv}}} \right)^{ - 1}},
\end{equation}
where $\xi$ is a regularization factor to balance noise.
    \item \emph{Block Diagonalization (BD)}: This method eliminates interference by projecting each user's signal onto the null space of other users' channels. For user $k$, compute the SVD of the aggregated interference channel, i.e.,
    \begin{equation}
     \mathbf{H}_{\overline{k}}=[\mathbf{H}_1^H,\cdots,\mathbf{H}_{k-1}^H,\mathbf{H}_{k+1}^H,\cdots,\mathbf{H}_K^H]^H.   
    \end{equation}
    Extract its null-space basis of $\mathbf{H}_{\overline{k}}$ by
    \begin{equation}
\mathbf{H}_{\overline{k}}=\mathbf{U}_{\overline{k}}\bm{\Sigma}_{\overline{k}}\left[\mathbf{V}_{\overline{k}}^{(1)},\mathbf{V}_{\overline{k}}^{(0)}\right]^H,
    \end{equation}
    and use $\mathbf{V}_{\overline{k}}^{(0)}$ as the pre-beamformer, which is the matrix containing the eigenvectors associated with the eigenvalue of $0$. As such, the effective channel $\overline{\mathbf{H}}_k = \mathbf{H}_k \mathbf{V}_{\overline{k}}^{(0)}$
  becomes interference-free, enabling independent single-user MIMO processing. Apply SVD to the effective channel for user $k$ as 
  \begin{equation}
\overline{\mathbf{H}}_k=\overline{\mathbf{U}}_k\overline{\bm{\Sigma}}_k\left[\overline{\mathbf{V}}^{(1)}_k,\overline{\mathbf{V}}^{(0)}_k\right]^H.   
  \end{equation}
  Then, the BD beamforming is given by
  \begin{equation}
{\bf{W}} = \left[ {{\bf{V}}_{\bar 1}^{(0)}\overline {\bf{V}} _1^{(1)}, \cdots ,{\bf{V}}_{\bar K}^{(0)}\overline {\bf{V}} _K^{(1)}} \right]. 
  \end{equation}
 This approach achieves perfect interference suppression but requires high computational complexity due to iterative null-space SVDs and is feasible only when $K \leq N_t$.
    \item \emph{WMMSE}: This iterative approach jointly optimizes the beamformer and combiner by introducing an auxiliary matrix that establishes an equivalence between the weighted sum-rate maximization problem and the WMMSE formulation \cite{wmmse,WMMSE1,DWMMSE1}. With the iterative number denoted by $\mathcal{I}$ and user priorities represented as $\left\{\chi_k\right\}_{k=1}^K$, the WMMSE beamforming algorithm is summarized in Algorithm \ref{WMMSE}.
    \item \emph{Dirty Paper Coding (DPC)}: As a nonlinear beamforming technique, DPC achieves optimal sum-rate performance in MU-MIMO systems by pre-canceling interference at the transmitter when perfect CSI is available. In practice, DPC is often implemented using Tomlinson-Harashima precoding (THP). Although DPC is theoretically capacity-achieving by eliminating multiuser interference, its high computational complexity and stringent CSI requirements limit its practical deployment.
\end{itemize}

\begin{table}[]\centering
{\renewcommand{\arraystretch}{2} 
\caption{Beamforming Schemes for MU-MIMO}\label{MUMIMO}
\begin{tabular}{ccllc}
\hline
% \multicolumn{5}{|c|}{{\bf{Spectral Efficiency}}}                                                                                  \\ \hline
% \multicolumn{5}{|c|}{\scriptsize $ \log_2\det\left(\mathbf{I}_{N_r}+\left(\sum_{i\neq k}\mathbf{H}_i\mathbf{W}_i\mathbf{W}_i^H\mathbf{H}_i^H+\sigma_k^2\mathbf{I}_{N_r}\right)^{-1}\mathbf{H}_k\mathbf{W}_k\mathbf{W}_k^H\mathbf{H}_k^H
% 		\right)$}                                                                                         \\ \hline
\multicolumn{1}{|c|}{\bf{Schemes}} & \multicolumn{3}{c|}{\bf{Description}} & \multicolumn{1}{c|}{\begin{tabular}[c]{@{}c@{}}
\bf{Closed Form}\end{tabular}} \\ \hline
\multicolumn{1}{|c|}{ZF}      & \multicolumn{3}{c|}{$\widetilde{\mathbf{H}}^H\left(\widetilde{\mathbf{H}}\widetilde{\mathbf{H}}^H\right)^{-1}$}            & \multicolumn{1}{c|}{\checkmark }                                       \\  \hline
\multicolumn{1}{|c|}{RZF}    & \multicolumn{3}{c|}{$\widetilde{\mathbf{H}}^H\left(\widetilde{\mathbf{H}}\widetilde{\mathbf{H}}^H+\xi\mathbf{I}_{KN_r}\right)^{-1}$}            & \multicolumn{1}{c|}{ \checkmark}    \\  \hline
\multicolumn{1}{|c|}{MMSE}      & \multicolumn{3}{c|}{$\widetilde{\mathbf{H}}^H\left(\widetilde{\mathbf{H}}\widetilde{\mathbf{H}}^H+\frac{\sigma ^2}{P_t}\mathbf{I}_{KN_r}\right)^{-1}$}            & \multicolumn{1}{c|}{ \checkmark}                      \\ \hline
\multicolumn{1}{|c|}{EZF}      & \multicolumn{3}{c|}{${{\bf{V}}_{{\rm{eff}}}}{\left( {{\bf{V}}_{{\rm{eff}}}^H{{\bf{V}}_{{\rm{eff}}}} + \xi {{\bf{I}}_{Kv}}} \right)^{ - 1}}$ }            & \multicolumn{1}{c|}{\checkmark }                          \\ \hline
\multicolumn{1}{|c|}{BD}     & \multicolumn{3}{c|}{\begin{tabular}[c]{@{}c@{}}
$\left[\mathbf{V}^{(0)}_{\overline{1}}\overline{\mathbf{V}}^{(1)}_1,\cdots,\mathbf{V}^{(0)}_{\overline{K}}\overline{\mathbf{V}}^{(1)}_K\right]$\end{tabular}}            & \multicolumn{1}{c|}{ \checkmark}     \\ \hline
\multicolumn{1}{|c|}{WMMSE}    & \multicolumn{3}{c|}{See Algorithm \ref{WMMSE}}            & \multicolumn{1}{c|}{\ding{55}}                     \\ \hline 
\multicolumn{1}{|c|}{DPC}     & \multicolumn{3}{c|}{ \begin{tabular}[c]{@{}c@{}}
Embedding  interference into    the \\ transmitted signal by nonlinear mapping\end{tabular}}            & \multicolumn{1}{c|}{\ding{55}}                       \\ \hline 
\end{tabular}}
\end{table}

\renewcommand{\algorithmicrequire}{\textbf{Input:}}
\renewcommand{\algorithmicensure}{\textbf{Output:}}
\begin{algorithm}
\caption{The procedure of WMMSE beamforming.}
\label{WMMSE}       %
\begin{algorithmic}[1]
\Require  Iterative number $\mathcal{I}$, transmit power $P_t$, user priority $\{\chi_k\}_{k=1}^K$,  noise power $\{\sigma_k^2\}_{k=1}^K$, and stacked channel $\widetilde{\mathbf{H}}$.
\State \textbf{Initialize:}  Random $\mathbf{W}$ with ${\rm Tr}\left(\mathbf{W}\mathbf{W}^{H}
		\right)= P_t$. 
\For {$i=1,\ldots,\mathcal{I}$}\\
\textbf{--------------- \emph{Update Combiner $\mathbf{C}_k,\forall k$} --------------}
\State $\gamma_{1,k}=\frac{\sigma^2_k}{P_t} {\rm Tr}\left(\mathbf{W}\mathbf{W}^{H}\right)$.
\State $\mathbf{C}_k=\left( \mathbf{H}_k\mathbf{W}\mathbf{W}^{H}\mathbf{H}_k^H+\gamma_{1,k}  \mathbf{I}_{N_r}
			\right)^{-1} \mathbf{H}_k\mathbf{W}_k$.\\
            \textbf{------------- \emph{Update Auxiliary Matrix $\mathbf{B}_k,\forall k$}-----------}
		\State ${\mathbf{B}}_k=\left(\mathbf{I}_D- \mathbf{W}_k^{H}\mathbf{H}_k^H\mathbf{C}_k\right)^{-1}.$\\ 
        \textbf{--------------- \emph{Update Beamformer $\mathbf{W}_k,\forall k$}---------------}\State $\gamma_2=\sum_{k}\frac{\chi_k \sigma_k^2}{P_t}{\rm Tr}\left( 	 \mathbf{C}_k \mathbf{B}_k\mathbf{C}^{H}_k\right)$.
        \State		$\mathbf{W}_k=\chi_k  (\sum_{k}
		\chi_k \mathbf{H}_k^H \mathbf{C}_k \mathbf{B}_k \mathbf{C}^{H}\mathbf{H}_k\newline {} \ \ \ \qquad \qquad  +  \gamma_2 \mathbf{I}_{N_t}
	)^{-1}\mathbf{H}^H_k\mathbf{C}_k\mathbf{B}_k$.
\EndFor
\State ${\bf{W}} = \left[ {{{\bf{W}}_1},{{\bf{W}}_2}, \cdots ,{{\bf{W}}_K}} \right].$
\State Normalize $\mathbf{W}\leftarrow \sqrt{\frac{P_t}{{\rm Tr}\left(\mathbf{W}\mathbf{W}^{H}\right)}}\mathbf{W}$.
\Ensure Beamformer $\mathbf{W}$ and combiner $\{\mathbf{C}_k\}_{k=1}^K$. 
\end{algorithmic}
\end{algorithm}

\subsection{Power Allocation}
While beamforming pre-processes transmitted signals by suppressing interference and matching channel characteristics to enhance transmission efficiency, power allocation dynamically adjusts transmit power across subchannels to maximize system capacity, energy efficiency, or user fairness.  In SU-MIMO, common power allocation schemes include equal power allocation, water-filling, and harmonic mean. In MU-MIMO, water-filling and quality of services (QoS)-aware dynamic allocation are typically employed. After beamforming is applied, the data streams are transmitted across mutually orthogonal (or quasi-orthogonal) subchannels, with \(\lambda_i^2\) representing SNR of the $i$-th subchannel. Then, the user's data rate can be expressed as
\begin{equation}
    R=\sum_{i=1}^{D} \log \left(1+P_i \lambda_i^2\right),
\end{equation}
where $P_i$ is the power allocated to the $i$-th subchannel.
\begin{itemize}
    \item \emph{Water-Filling}: Water-filling algorithm is a power allocation method in MIMO systems that maximizes data rate by distributing more power to stronger subchannels. The objective is formulated as
    \begin{equation}
     \begin{aligned}
&\mathop {\max }\limits_{{P_i}} \sum\limits_{i = 1}^D {\log } \left( {1 + {P_i}\lambda _i^2} \right),\\
&{\rm{s}}{\rm{.t}}{\rm{.}}\;\;{P_i} \ge 0,\;\;\sum\limits_{i = 1}^D {{P_i} =} {P_t}.
\end{aligned}   
    \end{equation}
By leveraging the Lagrange multiplier method, the optimal water-filling solution is given by  
\begin{equation}
P_i^* = \left( \mu - \frac{1}{\lambda_i^2} \right)^+,    
\end{equation}
where \( P_i^* \) is the optimal power allocated to the $i$-th subchannel,  \( \mu \) is the water level (determined by total power constraint \( \sum P_i^* = P_{t} \)), and \( (x)^+ = \max(0, x) \) ensures non-negative power allocation.  
 \item  \emph{Harmonic Mean}: The water-filling algorithm greedily allocates more power to stronger subchannels, which is theoretically optimal for maximizing channel capacity. However, in practical systems where multiple data streams are decoded jointly under a single codeword, an excessive power disparity between streams can degrade overall decoding performance. To strike a balance between capacity maximization and decoding robustness, an alternative approach is to maximize the harmonic mean of the subchannel SNRs, i.e., 
\begin{equation}
 \begin{aligned}
&\mathop {\max }\limits_{{P_i}}  {\frac{D}{{\sum\nolimits_{i = 1}^D {\frac{1}{{{P_i}\lambda _i^2}}} }}} \quad \\
&\;{\rm{s}}{\rm{.t}}{\rm{.}}\quad {P_i} \ge 0,\;\;\sum\limits_{i = 1}^D {{P_i} = } {P_t}.
\end{aligned}   
\end{equation}
By leveraging the Lagrange multiplier method, the optimal harmonic-mean solution is given by
\begin{equation}
{P_i} = \beta \frac{1}{\lambda },
\end{equation}
where \( \beta\) is the factor to satisfy the total power constraint.  
\item \emph{QoS-aware Dynamic Allocation}: QoS-aware dynamic power allocation is a method in MU-MIMO systems that ensures each user meets its specific quality-of-service requirements while efficiently utilizing available power resources. The objective is formulated as
    \begin{equation}
     \begin{aligned}
&\mathop {\min }\limits_{{P_i}} \sum\limits_{i = 1}^K {{P_i}} ,\\
&{\rm{s}}{\rm{.t}}{\rm{.}}\;\;{\gamma _i} \ge \gamma _i^{\rm{tgt}},\;\;{P_i} \ge 0,
\end{aligned}   
    \end{equation}
where \( K \) is the number of users, \( \gamma_i \) represents the received signal-to-interference-plus-noise ratio (SINR) for the $i$-th user, and \( \gamma_i^{\rm{tgt}} \) denotes the target SINR for QoS guarantee. By employing the Lagrange multiplier method, the optimal QoS-aware power allocation solution is  
\begin{equation}
P_i^* = \frac{\gamma_i^{\rm{tgt}}}{\lambda_i^2} \left( \sum_{j \neq i} P_j^* |h_{ij}|^2 + \sigma^2 \right),    
\end{equation}
where \( h_{ij} \) represents the interference channel coefficient between the $i$-th and $j$-th users, and \( \sigma^2 \) is the noise power. As the power distributions across data streams are mutually coupled, an iterative fixed-point algorithm should be employed until the solution converges. This approach provides a practical balance between system efficiency and user fairness, making it suitable for heterogeneous networks with diverse service demands.
\end{itemize}

\begin{table*}[t]  
\centering   
\captionsetup{
  labelsep=newline, % 让标题另起一行
}
\renewcommand{\arraystretch}{1.5}  %每行的高度为1.5倍
\caption{CSI-RS Positions Within a Slot (Part of TS 38.211 Table 7.4.1.5.3).}  \label{tabcdm}
\begin{tabular}{|c|c|c|c|c|c|c|c|}   
\hline  
\textbf{Row} &\textbf{Ports}  $X$& \textbf{Density $\rho$} & \textbf{CDM Type} & \textbf{$(\overline{k},\overline{l})$} & \textbf{CDM group  index $j$} & \textbf{$k'$} & \textbf{$l'$}\\
\hline   
1 & 1 &3& No CDM &$(k_0,l_0),(k_0+4,l_0),(k_0+8,l_0)$ &0,0,0 & 0 & 0 \\
\hline
2 & 1 &1,0.5 &No CDM &$(k_0,l_0)$ &0& 0 & 0 \\
\hline
3 & 2 &1,0.5 &FD-CDM2 &$(k_0,l_0)$ &0& 0,1 & 0 \\
\hline
4 & 4 &1& FD-CDM2 &$(k_0,l_0),(k_0+2,l_0)$ &0,1& 0,1 & 0 \\
\hline
5 & 4 &1 &FD-CDM2 &$(k_0,l_0),(k_0,l_0+1)$ &0,1 & 0,1 & 0 \\
\hline
6 & 8 &1& FD-CDM2 &$(k_0,l_0),(k_1,l_0),(k_2,l_0),(k_3,l_0)$ &0,1,2,3 & 0,1 & 0 \\
\hline
7 & 8 &1 &FD-CDM2 &$(k_0,l_0),(k_1,l_0),(k_0,l_0+1),(k_1,l_0+1)$ &0,1,2,3 & 0,1 & 0 \\
\hline
8 & 8 &1& CDM4 (FD2,TD2) &$(k_0,l_0),(k_1,l_0)$ &0,1 & 0,1 & 0,1 \\
\hline 
\end{tabular}   
\end{table*}

\section{Resource Allocation for CSI-RS}\label{resource}
The configuration of the CSI-RS in 5G NR systems involves a multilayered procedure that ensures precise placement of reference signals within the resource grid for effective channel estimation. The hierarchical approach to CSI-RS configuration can be delineated as follows:

(i)  Map CSI-RS symbols on REs within a single resource block (RB) via \emph{CSI-RS-ResourceMapping}.

(ii) Configure the RB locations in the frequency domain via \emph{CSI-RS-ResourceMapping.freqBand}.

(iii) Configure the periodicity and offset in time domain via  \emph{NZP-CSI-RS-Resource} and/or \emph{ZP-CSI-RS-Resource}{\footnote{CSI-RS can be allocated to different UEs, so when a UE occupies a PDSCH RB that contains multiple CSI-RSs, and these CSI-RSs may belong to different UEs, that is, they could either be its own CSI-RS or from other UEs. At this time, how does this UE identify these CSI-RS? The protocol defines the time-frequency resource sets for non-zero power CSI-RS (NZP-CSI-RS) and zero power CSI-RS (ZP-CSI-RS). For NZP-CSI-RS, this UE can perform CSI-RS measurements on the received signals. For the time-frequency resource sets of ZP-CSI-RS, this UE considers them to be the CSI-RS resources of other UEs and does not receive or process them.}}.

(iv) The comprehensive configuration, encompassing the aforementioned steps (i, ii, and iii), is then applied through the \emph{PDSCH-Config} and/or \emph{CSI-MeasConfig}. 

First, we explain what CSI-RS symbols are and how these symbols map to the REs on the resource grid. The CSI-RS symbols are generated by applying a pseudo-random sequence, which is then multiplied by orthogonal cover codes (OCCs) in the frequency and/or time domains, following appropriate power scaling. In the context of the same REs, multiple CSI-RS can be transmitted, each carrying a distinct OCC for the estimation of the channels related to multiple CSI-RS ports. One can think of a CSI-RS port as a single antenna, but it is typically a logical antenna port created by the BS's array and not a physical antenna.
This approach is referred to as code division multiplexing (CDM). Specifically, the pseudo-random sequence $r_l\left(m\right)$ is derived from a 31-length Gold sequence $c(m)$, and is expressed as 
\begin{equation}\label{gold-sequence} 
r_l\left(m\right)= \frac{1}{\sqrt{2}}\big(1-2c(2m)\big)+\frac{j}{\sqrt{2}}\big(1-2c(2m+1)\big).
\end{equation}
Then, the CSI-RS symbol corresponding to the $p$-th antenna port at the RE with frequency index $k$ and time index $l$ is given by
\begin{equation}\label{csi-rs-sequence} 
a^{\left(p\right)}_{k,l}=\beta \omega_f(k^{'}) \omega_t(l^{'})r_l\left(m\right),
\end{equation}
where $\beta$ denotes the power scaling coefficient. The terms $\omega_f(k^{'})$ and $\omega_t(l^{'})$ represent the OCC in the frequency and time domains, respectively. The frequency index $k$ and time index $l$ are defined as $k=nN_{SC}^{RB} + \overline{k} + k^{'}$ and $l = \overline{l}+l^{'}$, respectively. In these equations, $N_{SC}^{RB}$ denotes the number of subcarriers within an RB, and $n$ is the index of the RB. $\overline{k}$ and $\overline{l}$ denote the starting position of each OCC within an RB in the frequency and time domains, respectively. The set $(\overline{k}, \overline{l}, k^{'}, l^{'})$ is selected from a predefined table, which can refer to Table \ref{tabcdm}, that governs the mapping of CSI-RS symbols within a slot. Apart from the parameters $(\overline{k}, \overline{l}, k^{'}, l^{'})$, some important terms in Table \ref{tabcdm} are interpreted as below:
\begin{figure}[t]
	\centering
	\includegraphics[width=0.45\textwidth]{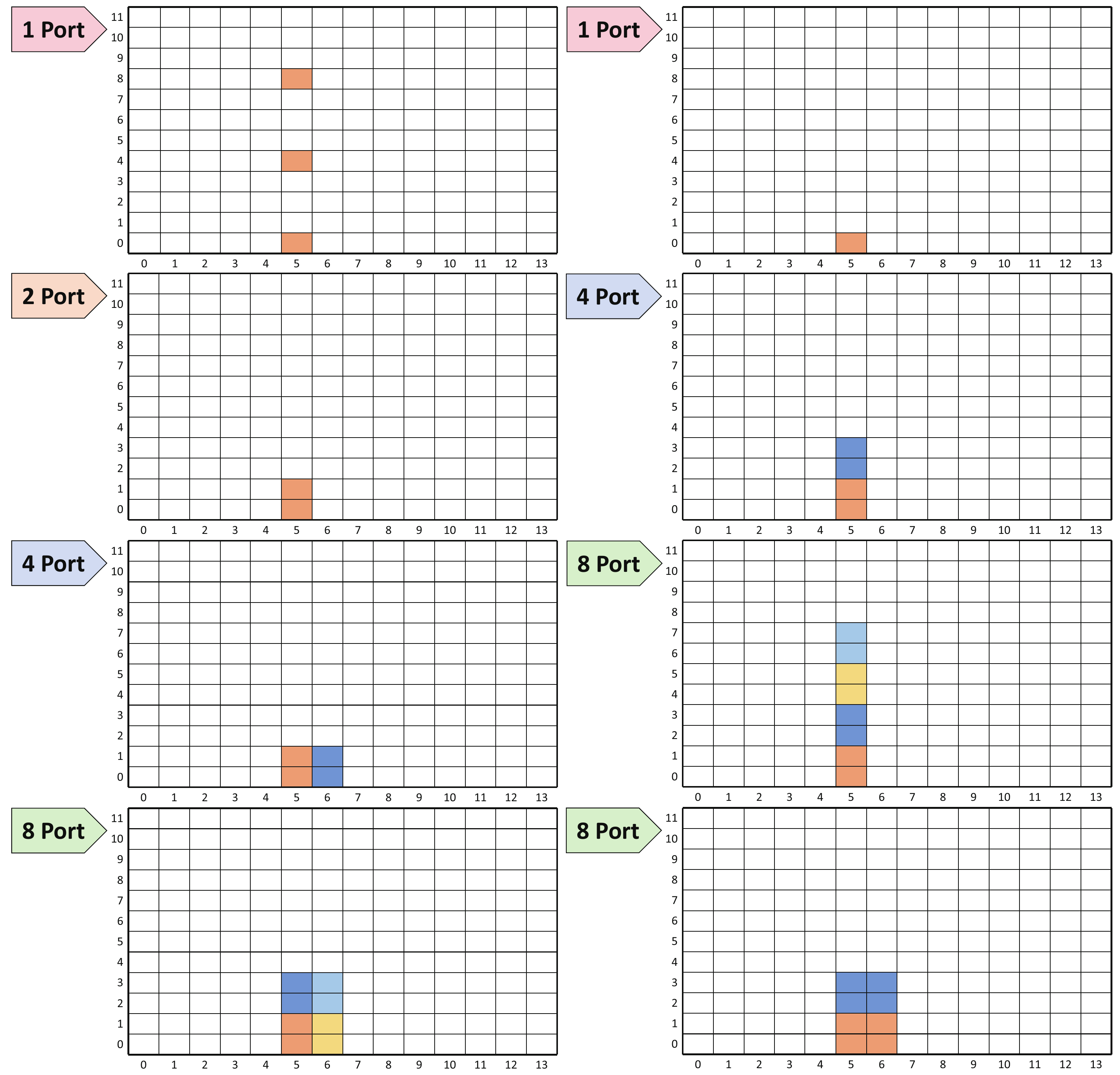}
	\caption{Schematic of the RE positions occupied by CSI-RS ports within a single RB in a slot.}\label{schematic}
	\label{RE positions}
\end{figure}

\begin{itemize}
    \item Ports: This denotes the number of antenna ports for CSI-RS, which is conveyed to the UE via the information element \emph{nrofPorts} and can be configured as {1, 2, 4, 8, 12, 16, 24, 32}. The configurations of 8, 12, 16, 24, and 32 ports are composed of combinations of 2 and 4 ports.

    \item Density: This refers to the density of CSI-RS, which is conveyed to the UE through the parameter $\rho$ within the information element \emph{CSI-RS-ResourceMapping} or \emph{CSI-RS-CellMobility}. The density can be set to 3, 1, or 0.5. A density of 3 indicates that there are three REs per RB; a density of 1 means that CSI-RS is present on every RB; a density of 0.5 signifies that CSI-RS is present on only one RB for every two RBs. 

    \item CDM Type: This indicates the multiplexing method for a set of CSI-RS, which is provided to the UE via the information element \emph{CSI-RS-ResourceMapping}. The configurations are as follows:
\end{itemize}
\begin{enumerate}
    \item No CDM: This signifies that there is no multiplexing of CSI-RS, i.e., one RE corresponds to one antenna port.
    \item FD-CDM2: This denotes that CSI-RS spans two consecutive REs in the frequency domain, using CDM with two superimposed OCC corresponding to two CSI-RS ports.
    \item CDM4 (FD2, TD2): This indicates that CSI-RS spans four REs, with two consecutive REs in the frequency domain and two consecutive REs in the time domain. CDM is employed using four OCCs corresponding to four CSI-RS ports.
\end{enumerate}

% \begin{table}[htb]   
% \begin{center}   
% \captionsetup{
%   labelsep=newline, % 让标题另起一行
% }
% \renewcommand{\arraystretch}{1.5}  %每行的高度为1.5倍
% \caption{ $\omega_f(k^{'})$ and $\omega_t(l^{'})$ for FD-CDM2.}\label{cdm2}
% \begin{tabular}{|c|c|c|}   
% \hline   \textbf{index} &$[w_f(0) \ \ w_f(1)]$& $w_t(0)$\\
% \hline   0 & [+1 \ \ +1] &1 \\
% \hline   1 & [+1 \ \ -1] &1 \\
% \hline 
% \end{tabular}   
% \end{center}   
% \vspace{-12pt}
% \end{table}

% \begin{table}[htb]   
% \begin{center}  
% \captionsetup{
%   labelsep=newline, % 让标题另起一行
% }
% \renewcommand{\arraystretch}{1.5}  %每行的高度为1.5倍
% \caption{ $\omega_f(k^{'})$ and $\omega_t(l^{'})$ for  CDM4 (FD2,TD2).} \label{cdm4}
% \begin{tabular}{|c|c|c|}   
% \hline   \textbf{index} &$[w_f(0) \ \ w_f(1)]$& $[w_t(0) \ \ w_t(1)]$\\  
% \hline   0 & [+1 \ \ +1] &[+1 \ \ +1] \\
% \hline   1 & [+1 \ \ -1] &[+1 \ \ +1] \\
% \hline   2 & [+1 \ \ +1] &[+1 \ \ -1]\\     
% \hline   3 & [+1 \ \ -1] &[+1 \ \ -1] \\
% \hline 
% \end{tabular}   
% \end{center}   
% \vspace{-12pt}
% \end{table}

Through Table \ref{tabcdm}, one can understand how CSI-RS symbols are mapped onto REs within a single RB, where different rows represent various mapping schemes. To facilitate understanding, we visually illustrate these mapping schemes in Fig. \ref{schematic}. It is worth noting that up to 3GPP Release 18, the protocol has defined 18 mapping schemes, of which only 8 are presented in Table \ref{tabcdm} and Fig. \ref{schematic}. Among these schemes, there are four types of CDM, namely, No CDM, FD-CDM2, CDM4 (FD2,TD2), and CDM8 (FD2,TD4). Different CDM types will result in different CSI-RS symbols in (\ref{csi-rs-sequence}).

% and Tables \ref{cdm2} and \ref{cdm4} detail the values of $\omega_f(k^{'})$ and $\omega_t(l^{'})$ in (\ref{csi-rs-sequence}) for FD-CDM2 and CDM4 (FD2,TD2), respectively.

\section{Reporting Configuration for CSI-RS}\label{reporting}
The CSI-RS reporting framework comprises two major components. The first component pertains to reporting configuration, which defines the parameters and resources for CSI-RS measurement and reporting. The second component relates to triggering/activation states, which are linked to a specific configuration and dictate when the UE should send CSI reports based on predefined conditions and events.

\begin{table*}[t]   
\begin{center}  
\captionsetup{
  labelsep=newline, % 让标题另起一行
}
\renewcommand{\arraystretch}{1.5}  %每行的高度为1.5倍
\caption{Reporting Granularity: Wideband vs Subband} \label{granularity}
\begin{tabular}{|c|c|>{\centering\arraybackslash}m{5.2cm}|}   
\hline   \textbf{Format Indicator} & \textbf{Wideband} & \textbf{Subband}\\  
\hline   CQI & A CQI is reported for the codeword for the entire BWP of a UE &One CQI for each codeword is reported for each subband in the CSI reporting band.  \\
\hline   PMI & A PMI is reported for the beamforming for the entire BWP of a UE&A wideband indicator for the entire BWP is reported via $i_1$ and the subband indication for each subband is report via $i_2$. \\
\hline 
\end{tabular}   
\vspace{10pt}

\caption{Triggering/Activation States of CSI Reporting}\label{trigger}
\begin{tabular}{|c|c|>{\centering\arraybackslash}m{5cm}|c|}   
\hline   \textbf{CSI-RS Configuration} & \textbf{Periodic CSI Reporting} & \textbf{Semi-Persistent CSI Reporting} & \textbf{Aperiodic CSI Reporting}\\  
\hline   Periodic CSI-RS & No dynamic  triggering/activation &For reporting on PUCCH, the UE receives an activation command; for reporting on PUSCH, the UE receives triggering on DCI & Triggered by DCI or subselection indication \\
\hline   Semi-Persistent CSI-RS & Not Supported &For reporting on PUCCH, the UE receives an activation command; for reporting on PUSCH, the UE receives triggering on DCI & Triggered by DCI or subselection indication.\\
\hline   Aperiodic CSI-RS & Not Supported &Not Supported & Triggered by DCI or subselection indication.\\     
\hline 
\end{tabular}   
\end{center}   
\vspace{-12pt}
\end{table*}

In this subsection, we mainly focus on the reporting configuration. It should be understood that gNB can only specify a \emph{CSI-ReportConfig} with the establishment of \emph{Resources}, \emph{ResourceSets}, and \emph{CSI-ResourceConfigs}. The significance of these information elements is elucidated below:

(i) \emph{Resource}: This parameter identifies the physical locations of the CSI-RS on the radio frame. There are three distinct types of resources in this category: NZP CSI-RS, ZP CSI-RS, and CSI interference measurement (CSI-IM) \emph{Resource}. The methodology to pinpoint the exact locations of these resources is complex and has been briefly introduced in Sec. \ref{resource}.

(ii) \emph{ResourceSet}: While a \emph{Resource} defines individual CSI-RS, in the context of 5G NR, all such individual resources must be grouped into a specific \emph{ResourceSet} to be operational. Even in scenarios where only a single resource is required, it must be associated with a \emph{ResourceSet}.  This approach streamlines the management and scheduling of resources, thereby optimizing resource utilization and simplifying the UE's reporting mechanism.

(iii) \emph{CSI-ResourceConfig}: This configuration determines the specifics of the reference signal transmission, such as NZP CSI-RS, ZP CSI-RS, CSI-IM \emph{Resource} and the transmission mode (periodic, aperiodic, or semi-persistent). While  \emph{Resource} and \emph{ResourceSet} outline the CSI resource structure, they do not trigger the transmission of these resources on their own. It is the role of \emph{CSI-ResourceConfig} to activate the transmission of the resources according to the defined schedule and conditions.

(iv) \emph{CSI-ReportConfig}: This configuration indicates which \emph{CSI-ResourceConfig} the UE should use for its measurements. It provides a mapping table that links the measurement type with the corresponding \emph{CSI-ReportConfig} ID. This configuration enables the UE to report accurate and relevant CSI data back to the BS, which is essential for optimizing network performance.

By integrating these elements into a cohesive framework, the 5G NR standard facilitates a flexible and robust mechanism for CSI acquisition and reporting, which is pivotal for the dynamic optimization of network resources and the enhancement of user experience. In particular, the three transmission modes mentioned in \emph{CSI-ResourceConfig} are clarified as below.
\begin{itemize}
    \item Periodic: The gNB configures the CSI-RS report period and the time-frequency resource location, and transmits according to the configured cycle. Periodic reporting is always carried out over the PUCCH. Thus, in the case of periodic reporting, the resource configuration also includes information about a periodically available PUCCH resource to be used for the reporting.
    \item Aperiodic: The UE is informed of each CSI-RS report by downlink control information (DCI) signaling, more specifically within a CSI-request field within the uplink scheduling grand format (DCI format 0-1). Since aperiodic reporting is always done on the scheduled PUSCH, an uplink scheduling grant is required. 
    
    \item Semi-persistent: The gNB configures the CSI-RS report period and time-frequency resource location. The medium access control control element (MAC CE) activates/deactivates the CSI-RS report and notifies the UE. Similar to periodic reporting, semi-persistent reporting can be performed on a periodically assigned PUCCH resource. Alternatively, it can also be transmitted over a semi-persistently allocated PUSCH, which is typically used for larger reporting payloads.
\end{itemize}

In the parameters for CSI reporting, the main elements include CQI, CRI, PMI, RI, and LI, which have been introduced in Sec. \ref{flow} (iii). Since different UEs may be required to report different parameters, a UE may be configured with a \emph{CSI-ReportConfig} with the higher layer parameter \emph{reportQuantity} to specify which parameters need to be reported. It is important to understand that not every CSI-RS measurement requires reporting. The \emph{reportQuantity} can be set to options such as `none', `cri-RI-PMI-CQI', `cri-RI-i1', `cri-RI-i1-CQI', `cri-RI-CQI', `cri-RSRP', and so on. Among these, an important concept is the reporting granularity, which indicates whether the report value represents the entire frequency band (i.e., wideband) or a specific segment of the band (i.e., subband). Reporting granularity is mainly applied to CQI and PMI. As shown in Table \ref{granularity}, the  reporting granularity determines whether the codeword\footnote{A codeword refers to a block of data that has undergone channel coding and modulation processing before transmission at the physical layer, with its size and structure depending on the feedback of CQI. The value of the CQI determines the modulation and coding scheme (MCS) that is most suitable for the current channel conditions. Based on the CQI feedback, gNB can select the appropriate modulation order and coding rate to generate the codeword.} and beamforming are the same across the entire band or differ across segments.

\begin{table}[t]   
\begin{center} 
\centering  
\captionsetup{
  labelsep=newline, % 让标题另起一行
}
\renewcommand{\arraystretch}{2}  %每行的高度为1.5倍
\caption{Configurable Subband Sizes}\label{tabCSS}
\begin{tabular}{|c|c|}   
\hline  
\rowcolor{gray!20}
\bf{Bandwidth Part (RBs)} &\bf{Subband Size (RBs)}  \\
\hline  24 - 72 &4 or 8\\
\hline  73 - 144&8 or 16\\
\hline  145 - 275&16 or 32\\
\hline
\end{tabular}   
\end{center}   
\vspace{-10pt}
\end{table}

 We already know that the CSI-RS configuration is categorized into three transmission modes: Periodic, Aperiodic, and Semi-persistent. Similarly, the CSI reporting also has three triggering/activation states: Periodic, Aperiodic, and Semi-persistent. The supported triggering/activation modes under different configurations are summarized in Table \ref{trigger}.  It should be noted that ``wideband" does not refer to the entire bandwidth of the BS, but rather to the bandwidth used by the UE, commonly known as the BWP. Theoretically, to achieve optimal performance in an OFDM system, each subcarrier requires independent beamforming, meaning that PMI feedback should be operated at every RE. However, to reduce feedback overhead, the smallest unit of feedback is the subband, which consists of several RBs. The 3GPP standard supports different sizes of subbands for different BWP configurations, as shown in Table \ref{tabCSS}.

\section{Port-Selection Codebooks in R15-16}

\begin{table*}[t]  
\centering   
\captionsetup{
  labelsep=newline, % 让标题另起一行
}
\renewcommand{\arraystretch}{2}  
\caption{R15 Type II Port-Selection Codebook (TS 38.214 Table 5.2.2.2.4-1)}  \label{tabps}
\begin{tabular}{|Sc|Sc|}   
\hline  
\rowcolor{gray!20}
\textbf{Layers}  &  \\
\hline   
$v=1$ & ${\bf{w}}^{(1)}_{i_{1,1},p^{(1)}_1,p^{(2)}_1,i_{2,1,1}}
={\bf{w}}^1_{i_{1,1},p^{(1)}_1,p^{(2)}_1,i_{2,1,1}}$  \\
\hline
$v=2$ & $ {\bf{W}}^{(2)}_{i_{1,1},p^{(1)}_1,p^{(2)}_1,i_{2,1,1},p^{(1)}_2,p^{(2)}_2,i_{2,1,2}}
= \frac{1}{\sqrt{2}}
\begin{bmatrix}{\bf{w}}^1_{i_{1,1},p^{(1)}_1,p^{(2)}_1,i_{2,1,1}} \ \ {\bf{w}}^2_{i_{1,1},p^{(1)}_2,p^{(2)}_2,i_{2,1,2}}
\end{bmatrix}$ \\
\hline
\multicolumn{2}{|Sc|}{{\linespread{2}\selectfont  
\makecell{ Precoding vector: ${\bf{w}}^l_{i_{1,1},p^{(1)}_l,p^{(2)}_l,i_{2,1,l}}
=\frac{1}{\sqrt{\sum^{2L-1}_{i=0}(p_{l,i}^{(1)}p_{l,i}^{(2)})^2}}
\begin{bmatrix}\sum^{L-1}_{i=0}{\bf{v}}_{i_{1,1}d+i}p^{(1)}_{l,i}p^{(2)}_{l,i}\varphi_{l,i}\\
\sum^{L-1}_{i=0}{\bf{v}}_{i_{1,1}d+i}p^{(1)}_{l,i+L}p^{(2)}_{l,i+L}\varphi_{l,i+L}
\end{bmatrix},\;\; l=1,2.$ }
}}\\ 
\hline
\end{tabular}  
\end{table*} 

\subsection{R15 Type II Port-Selection Codebook} \label{r15ps}
The Type II Port-Selection Codebook is derived from modifications to the Type II Regular Codebook, so understanding the design rationale behind this codebook first requires familiarity with Sec. \ref{regularII} presented in the main text. The Type II Port-Selection Codebook is intended for BSs employing a full-connected PEB, where the PEB utilizes the structural characteristics of the channel to transform the antenna-domain channel into a beam-domain channel, resulting in a sparse channel being measured by the CSI-RS ports. Moreover, since the PEB maps each logical antenna element, where each CSI-RS is transmitted to a physical antenna array, the UE does not need to be aware of the full physical antenna array structure. This mapping is transparent to the UE.

The new parameters are summarized in Table \ref{paratypeiips} and the expression of the Type II Port-Selection Codebook in the protocol is presented in Table \ref{tabps}. The Type II Regular Codebook selects $L$ beams on each layer to form a precoding vector, whereas the Type II Port-Selection Codebook selects $L$ ports on each layer to form a precoding vector. In fact, due to the presence of the PEB, selecting $L$ ports is tantamount to selecting $L$ external beams. Thus, the precoding vector of the Port-Selection Codebook combined with the PEB is equivalent to the precoding vector of the Regular Codebook. Based on Table \ref{tabps}, the precoding matrix can be written as
\begin{equation}
{\bf{w}}^l_{i_{1,1},p^{(1)}_l,p^{(2)}_l,i_{2,1,l}}
=\frac{1}{\sqrt{{\overline \beta  _l}}}
\begin{bmatrix}\sum^{L-1}_{i=0}a_{l,i}^1 \cdot {\bf{v}}_{i_{1,1}d+i}\\
\sum^{L-1}_{i=0}a_{l,i}^2 \cdot{\bf{v}}_{i_{1,1}d+i}
\end{bmatrix}, 
\end{equation}
where ${\overline \beta  _l} = \sum^{2L-1}_{i=0}(p_{l,i}^{(1)}p_{l,i}^{(2)})^2$. Compared to the regular one shown in (\ref{a46}), the main difference lies in the spatial basis ${\bf{v}}_{i_{1,1}d+i}$, which admits a form of
\begin{equation}\label{vd}
{\bf{v}}_{i_{1,1}d+i} = \left\{ {\begin{aligned}
&{[1,0\ldots,0]^T,\qquad {\rm{if}}\;{i_{1,1}d+i} = 0},\\
&{[0,0,\ldots,1]^T,\;\; \quad  {\rm{if}}\;{i_{1,1}d+i} = \frac{{P_{{\text{CSI-RS}}}} }{2}- 1,
}\\
&{[0,\ldots,1,\ldots,0]^T,\; {\rm{otherwise,}}}
\end{aligned}} \right.
\end{equation}
where ${\bf{v}}_{i_{1,1}d+i}$ is a ${P_{{\text{CSI-RS}}}}/2$-element vector containing a value of $1$ in element $m+1$ and zeros elsewhere.
The supported number of CSI-RS ports is given by ${P_{{\text{CSI-RS}}}} \in \left\{ {4,8,12,16,24,32} \right\}$ as configured by higher layer parameter \emph{nrofPorts}. Although parameter $L$ refers to $L$ ports, it is still configured with the higher layer parameter \emph{numberOfBeams}. The value of $d$  is configured with the higher layer parameter \emph{portSelectionSamplingSize}, where $d \in \left\{ {1,2,3,4} \right\}$ and 
$d \le \min \left( {{P_{{\text{CSI-RS}}}}/2,L} \right)$. The $L$ consecutive antenna ports per polarization are selected by 
\begin{equation}
{i_{1,1}} \in \left\{ {0,1, \ldots ,\left\lceil {\frac{{{P_{{\text{CSI-RS}}}}}}{{2d}}} \right\rceil  - 1} \right\},
\end{equation}
where $\left\lceil \cdot \right\rceil$ is the ceiling operator.
The protocol supports a feedback RI value \(v \in \{1,2\}\), where each PMI value corresponding to $v$ is given by
\[\begin{aligned}
&{i_1} = \left\{ {\begin{aligned}
&{\left[ 
{{i_{1,1}}}\;\;{{i_{1,3,1}}}\;\;{{i_{1,4,1}}}
 \right]},\quad {v  = 1},\\
&{\left[ 
{{i_{1,1}}}\;\;{{i_{1,3,1}}}\;\;{{i_{1,4,1}}}\;\;{{i_{1,3,2}}}\;\;{{i_{1,4,2}}}
\right]},\quad {v  = 2},
\end{aligned}} \right.\\
&{i_2} = \left\{ {\begin{aligned}
&{\left[ {{i_{2,1,1}}} \right]},\quad {{\rm{SA}} = {\rm{false}},\;v  = 1},\\
&{\left[ 
{{i_{2,1,1}}}\;\;{{i_{2,1,2}}}
 \right]},\quad {{\rm{SA}} = {\rm{false}},\;v  = 2},\\
&{\left[ 
{{i_{2,1,1}}}\;\;{{i_{2,2,1}}}
\right]},\quad {{\rm{SA}} = {\rm{true}},\;v  = 1},\\
&{\left[ 
{{i_{2,1,1}}}\;\;{{i_{2,2,1}}}\;\;{{i_{2,1,2}}}\;\;{{i_{2,2,2}}}
 \right]},\quad  {{\rm{SA}} = {\rm{true}},\;v  = 2},
\end{aligned}} \right.
\end{aligned}\]
where SA represents the higher layer parameter \emph{subbandAmplitude}, which can be set to true or false. The remaining parameters follow the same definitions as those in the Type II Regular Codebook.

\begin{table}[t]   
\begin{center} 
\centering  
\captionsetup{
  labelsep=newline, % 让标题另起一行
}
\renewcommand{\arraystretch}{1.5}  %每行的高度为1.5倍
\caption{New Parameters in Type II Port-Selection Codebook.}  
\label{paratypeiips} 
\begin{tabular}{|c|m{6cm}|}   
\hline   \textbf{Parameters} &\textbf{Interpretation} \\  
\hline   $i_{1,1}$ & The initial index of the uniformly sampled ports. \\
\hline   $d$ & The sampling interval of uniformly sampled ports.\\
\hline   ${\bf{v}}_{m}$ & The beam basis generated by the PEB on the $m$-th CSI\textendash RS port. \\
\hline   $\overline \beta  _l$ &  The power scaling coefficient on layer \(l\).  \\
\hline   $P_{\rm{CSI-RS}}$ & The number of CSI-RS ports, which is configured by gNB and is independent of the logical antenna array structure, i.e., $N_1$ and $N_2$. \\ 
\hline   
\end{tabular}   
\end{center}   
\vspace{-12pt}
\end{table}

\subsection{R16 Enhanced Type II Port-Selection Codebook} \label{r16ps}
To comprehend the design logic of the Enhanced Type II Port-Selection Codebook, one must first be acquainted with the Enhanced Type II Regular Codebook detailed in Sec. \ref{enhanceregularII} above. The Port-Selection Codebook is tailored for BSs using a fully-connected PEB. This PEB leverages the inherent properties of the channel to convert the channel in the space domain into one in the angular domain, which is then measured as a sparse channel by the CSI-RS ports.

\begin{table*}[t]  
\centering   
\captionsetup{
  labelsep=newline, % 让标题另起一行
}
\renewcommand{\arraystretch}{2}  
\caption{R16 Enhanced Type II Port-Selection Codebook (TS 38.214 Table 5.2.2.2.6-2)}  \label{tabesps}
\begin{tabular}{|Sc|Sc|}   
\hline  
\rowcolor{gray!20}
\textbf{Layers}  &  \\
\hline   
$v=1$ & ${\bf{w}}^{(1)}_{i_{1,1},n_{3,1},p^{(1)}_1,p^{(2)}_1,i_{2,5,1},t}
={\bf{w}}^1_{i_{1,1},n_{3,1},p^{(1)}_1,p^{(2)}_1,i_{2,5,1},t}$  \\
\hline
% $v=2,3,4$& ... \\
$v=2$ & $ {\bf{W}}^{(2)}_{i_{1,1},n_{3,1},p^{(1)}_1,p^{(2)}_1,i_{2,5,1},n_{3,2},p^{(1)}_2,p^{(2)}_2,i_{2,5,2},t}
= \frac{1}{\sqrt{2}}
\begin{bmatrix} {\bf{w}}^1_{i_{1,1},n_{3,1},p^{(1)}_1,p^{(2)}_1,i_{2,5,1},t} \ \ {\bf{w}}^2_{i_{1,1},n_{3,2},p^{(1)}_2,p^{(2)}_2,i_{2,5,2},t}
\end{bmatrix}$ \\
\hline
$v=3$ & {\linespread{2}\selectfont\makecell{$ {\bf{W}}^{(3)}_{i_{1,1},n_{3,1},p^{(1)}_1,p^{(2)}_1,i_{2,5,1},n_{3,2},p^{(1)}_2,p^{(2)}_2,i_{2,5,2},n_{3,3},p^{(1)}_3,p^{(2)}_3,i_{2,5,3},t}$ \qquad \qquad \qquad \qquad \qquad \qquad\qquad \qquad \qquad \qquad \qquad \qquad \qquad \qquad \\
$\qquad \qquad \qquad \qquad= \frac{1}{\sqrt{3}}
\begin{bmatrix} {\bf{w}}^1_{i_{1,1},n_{3,1},p^{(1)}_1,p^{(2)}_1,i_{2,5,1},t}  \ \ {\bf{w}}^2_{i_{1,1},n_{3,2},p^{(1)}_2,p^{(2)}_2,i_{2,5,2},t} \ \
{\bf{w}}^3_{i_{1,1},n_{3,3},p^{(1)}_3,p^{(2)}_3,i_{2,5,3},t}
\end{bmatrix}$} }\\
\hline
$v=4$ & {\linespread{2}\selectfont\makecell{$ {\bf{W}}^{(4)}_{i_{1,1},n_{3,1},p^{(1)}_1,p^{(2)}_1,i_{2,5,1},n_{3,2},p^{(1)}_2,p^{(2)}_2,i_{2,5,2},n_{3,3},p^{(1)}_3,p^{(2)}_3,i_{2,5,3},n_{3,4},p^{(1)}_4,p^{(2)}_4,i_{2,5,4},t}$ \qquad \qquad \qquad \qquad \qquad \qquad \qquad \qquad \qquad \qquad\\
$= \frac{1}{2}
\begin{bmatrix} {\bf{w}}^1_{i_{1,1},n_{3,1},p^{(1)}_1,p^{(2)}_1,i_{2,5,1},t}  \ \ {\bf{w}}^2_{i_{1,1},n_{3,2},p^{(1)}_2,p^{(2)}_2,i_{2,5,2},t}  \ \cdots \
{\bf{w}}^4_{i_{1,1},n_{3,4},p^{(1)}_4,p^{(2)}_4,i_{2,5,4},t}
\end{bmatrix}$} }\\
\hline
\multicolumn{2}{|Sc|}{\linespread{2}\selectfont  
\makecell{  Precoding vector: ${\bf{w}}^l_{i_{1,1},n_{3,l},p^{(1)}_l,p^{(2)}_l,i_{2,5,l},t}
=\frac{1}{\sqrt{\gamma_{t,l}}}
\begin{bmatrix}\sum^{L-1}_{i=0}{\bf{v}}_{i_{1,1}d+i}p^{(1)}_{l,0}\sum^{M_v-1}_{f=0}y^{(f)}_{t,l}p^{(2)}_{l,i,f}\varphi_{l,i,f}\\
\sum^{L-1}_{i=0}{\bf{v}}_{i_{1,1}d+i}p^{(1)}_{l,1}\sum^{M_v-1}_{f=0}y^{(f)}_{t,l}p^{(2)}_{l,i+L,f}\varphi_{l,i+L,f}
\end{bmatrix},\;\; l=1,2,3,4,$ \\
$\gamma_{t,l}=\sum^{2L-1}_{i=0}(p_{l,\left\lfloor \frac{i}{L}\right\rfloor}^{(1)})^2\begin{vmatrix}\sum^{M_v-1}_{f=0}y^{(f)}_{t,l}p^{(2)}_{l,i,f}\varphi_{l,i,f}\end{vmatrix}^2$.\\
  }
}\\ 
\hline
\end{tabular}  
\vspace{-8pt}
\end{table*} 

The protocol details the Enhanced Type II Port-Selection Codebook in Table \ref{tabesps}. The Regular Codebook of Type II chooses $L$ beams per layer to construct a precoding vector. In contrast, the Port-Selection Codebook of Type II opts for $L$ CSI-RS ports per layer for the same purpose. Given the PEB, selecting $L$ CSI-RS ports effectively equates to choosing $L$ external beams. Consequently, when combined with the PEB, the Port-Selection Codebook's precoding vector is functionally identical to that of the Regular Codebook. 

The compression methods for the Enhanced Type II Port-Selection Codebook in the spatial-frequency domains are identical to those of the Enhanced Type II Regular Codebook. As a result, the mechanisms for indicating beams, taps, bimaps, and so on, are also exactly the same. By comparing Table \ref{tabesps} to Table \ref{tabesII}, it can be observed that the PMI information omits $q_1$, $q_2$, $n_1$, and $n_2$, but includes an explicit $i_{1,1}$.\footnote{In Table \ref{tabesII}, the PMI information $i_{1,1}$ is implicit as $i_{1,1} = [q_1,q_2]$ .} Thus, the PMI value can be expressed by 
\begin{align}\label{a86}
&{i_1} = \Big[ {{i_{1,1}}\;\;{i_{1,5}}\;\;\{ {i_{1,6,l}}\} _{l = 1}^v\;\;\{ {i_{1,7,l}}\} _{l = 1}^v\;\;\{ {i_{1,8,l}}\} _{l = 1}^v} \Big],\notag\\
&{i_2} = \Big[ \{{{i_{2,3,l}}\} _{l = 1}^v\;\;\{ {i_{2,4,l}}\} _{l = 1}^v\;\;\{ {i_{2,5,l}}\} _{l = 1}^v} \Big].
\end{align}
Upon reviewing the protocol description, there are two primary differences between the Enhanced Type II Port-Selection Codebook and the Enhanced Type II Regular Codebook:

\begin{itemize}
    \item The spatial domain basis has shifted from the two-dimensional DFT vector ${\bf{v}}_{m_1^{(i)}m_2^{(i)}}$ to the standard basis vector ${\bf{v}}_{i_{1,1}d+i}$ that has been defined in (\ref{vd}).

    \item  The supported values for $L$, $\beta$, and $p_\upsilon$ are listed in a new table, see TS 38.214 Table 5.2.2.2.6-1, which differs from the parameters provided in Table \ref{tabp}.    
\end{itemize}
Moreover, the UE configured with a higher layer parameter \emph{codebookType} will be set to \emph{typeII-PortSelection-r16}. The supported number of CSI-RS ports is given by ${P_{{\text{CSI-RS}}}} \in \left\{ {4,8,12,16,24,32} \right\}$ as configured by higher layer parameter \emph{nrofPorts}. Although parameter $L$ refers to $L$ ports, it is still configured with the higher layer parameter \emph{numberOfBeams}. The value of $d$  is configured with the higher layer parameter \emph{portSelectionSamplingSize}, where $d \in \left\{ {1,2,3,4} \right\}$ and 
$d \le \min \left( {{P_{{\text{CSI-RS}}}}/2,L} \right)$. The $L$ consecutive antenna ports per polarization are selected by 
\begin{equation}
{i_{1,1}} \in \left\{ {0,1, \ldots ,\left\lceil {\frac{{{P_{{\text{CSI-RS}}}}}}{{2d}}} \right\rceil  - 1} \right\}.
\end{equation}

\end{appendices}

\section*{Acknowledgement}
\small Boyu Ning is an independent researcher, Chengdu 611731, China (e-mails: boydning@outlook.com).

Haifan Yin and Sixu Liu are with Huazhong University of Science and Technology, 430074 Wuhan, China (e-mail: yin@hust.edu.cn; sxliu@hust.edu.cn).

Hao Deng, Songjie Yang, and Weidong Mei are with National Key Laboratory of Wireless Communications, University of Electronic Science and Technology of China, Chengdu 611731, China (e-mails: haod@std.uestc.edu.cn; yangsongjie@std.uestc.edu.cn; wmei@uestc.edu.cn).

Yuchen Zhang is with the Electrical and Computer Engineering Program, Computer, Electrical and Mathematical Sciences and Engineering (CEMSE), King Abdullah University of Science and Technology (KAUST), Thuwal 23955-6900, Kingdom of Saudi Arabia (e-mail: yuchen.zhang@kaust.edu.sa).

David Gesbert is with the EURECOM, 06410 Sophia Antipolis, France (e-mail: david.gesbert@eurecom.fr).

Jaebum Park and Robert W. Heath Jr. are with the Department of Electrical and Computer Engineering, University of California at San Diego, San Diego, CA 92093 USA (e-mail: jp261@ucsd.edu; rwheathjr@ucsd.edu).

Emil Bj{\"o}rnson is with Department of Computer Science, KTH Royal Institute of Technology, Stockholm, Sweden (email: emilbjo@kth.se). He was supported by the Grant 2022-04222 from the Swedish Research Council.

\bibliography{main}
\bibliographystyle{IEEEtran}

\begin{IEEEbiography}[{\includegraphics[width=1in,clip]{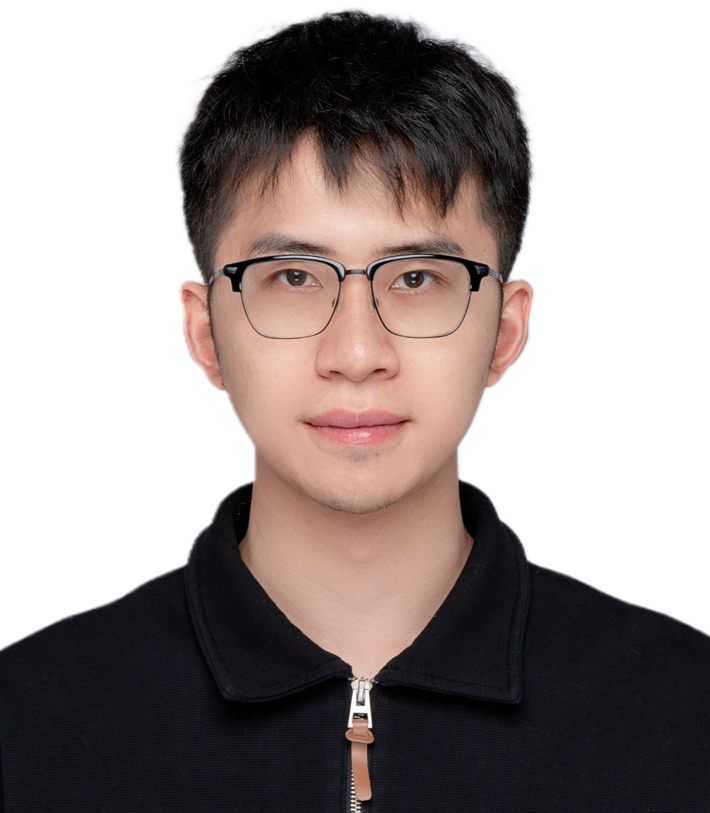}}]{Boyu Ning} (Member, IEEE) received the B.Eng. degree in communication engineering and the Ph.D. degree from the National Key Laboratory of Science and Technology on Communications, University of Electronic Science and Technology of China (UESTC), in 2018 and 2023, respectively. His research interests include 5G/6G networks, all MIMO-related technologies (such as beamforming and codebook design), deep learning, terahertz communications, movable antennas, intelligent reflecting surfaces, Bayesian estimation, and convex optimization. He has published over 100 journal and conference papers and secured over 30 national patents. He has experience in the communication industry, including the evolution of 5G base station architectures, the baseband algorithms design for 5G NR, and the standardization for 3GPP Release 19/20. He was a recipient of National Scholarship from China in 2021. He was a recipient of the Top 10 Outstanding Students from UESTC in 2022, and the Outstanding Graduate Award from the Sichuan Provincial Department of Education in 2023. From 2022 to 2023, she was a Visiting Research Scholar with National University of Singapore, Singapore. He was honored as the Member in Huawei Top Minds Program in 2023. He was a recipient of the Outstanding Ph.D Thesis Award from the China Education Society of Electronics in 2024. He has been listed in World's Top 2\% Scientists by Stanford University since 2025. He has served on multiple occasions as a General/TPC Co-Chair and Moderator for workshops and industry panels at leading IEEE conferences, including Globecom, ICC, VTC, and ICCC.
\end{IEEEbiography}

\begin{IEEEbiography}[{\includegraphics[width=1in,clip]{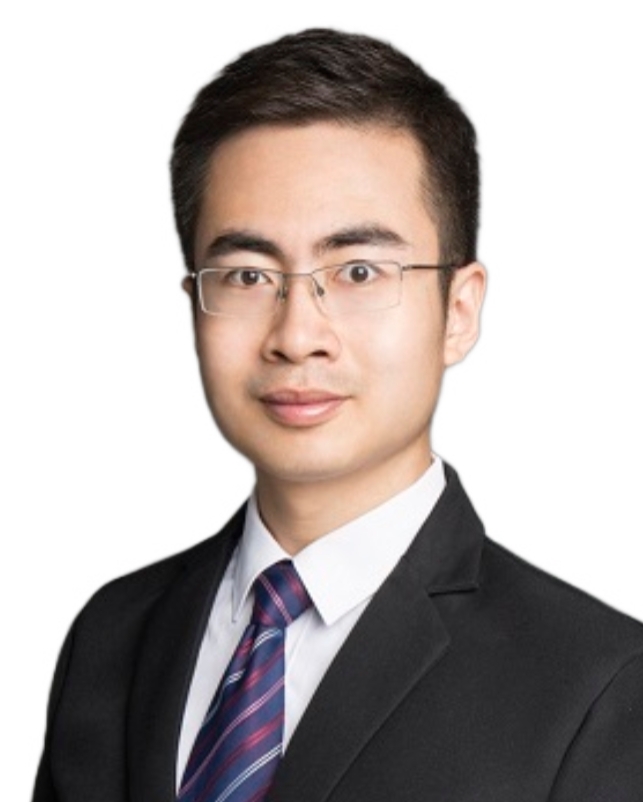}}]{Haifan Yin} (Senior Member, IEEE) received the B.Sc. degree in electrical and electronic engineering and the M.Sc. degree in electronics and information engineering from the Huazhong University of Science and Technology, Wuhan, China, in 2009 and 2012, respectively, and the Ph.D. degree from Télécom ParisTech in 2015. From 2009 to 2011, he was a Research and Development Engineer with the Wuhan National Laboratory for Optoelectronics, Wuhan, working on the implementation of TD-LTE systems. From 2016 to 2017, he was a DSP Engineer at Sequans Communications (IoT chipmaker), Paris, France. From 2017 to 2019, he was a Senior Research Engineer working on 5G standardization at Shanghai Huawei Technologies Company Ltd., where he has made substantial contributions to 5G standards, particularly the 5G codebooks. Since May 2019, he has been a Full Professor with the School of Electronic Information and Communications, Huazhong University of Science and Technology. His current research interests include 5G and 6G networks, signal processing, machine learning, and massive MIMO systems. He was the National Champion of 2021 High Potential Innovation Prize awarded by the Chinese Academy of Engineering, a recipient of the China Youth May Fourth Medal (the top honor for young Chinese), and a recipient of the 2024 Stephen O. Rice Prize.
\end{IEEEbiography}

\begin{IEEEbiography}[{\includegraphics[width=1in,clip]{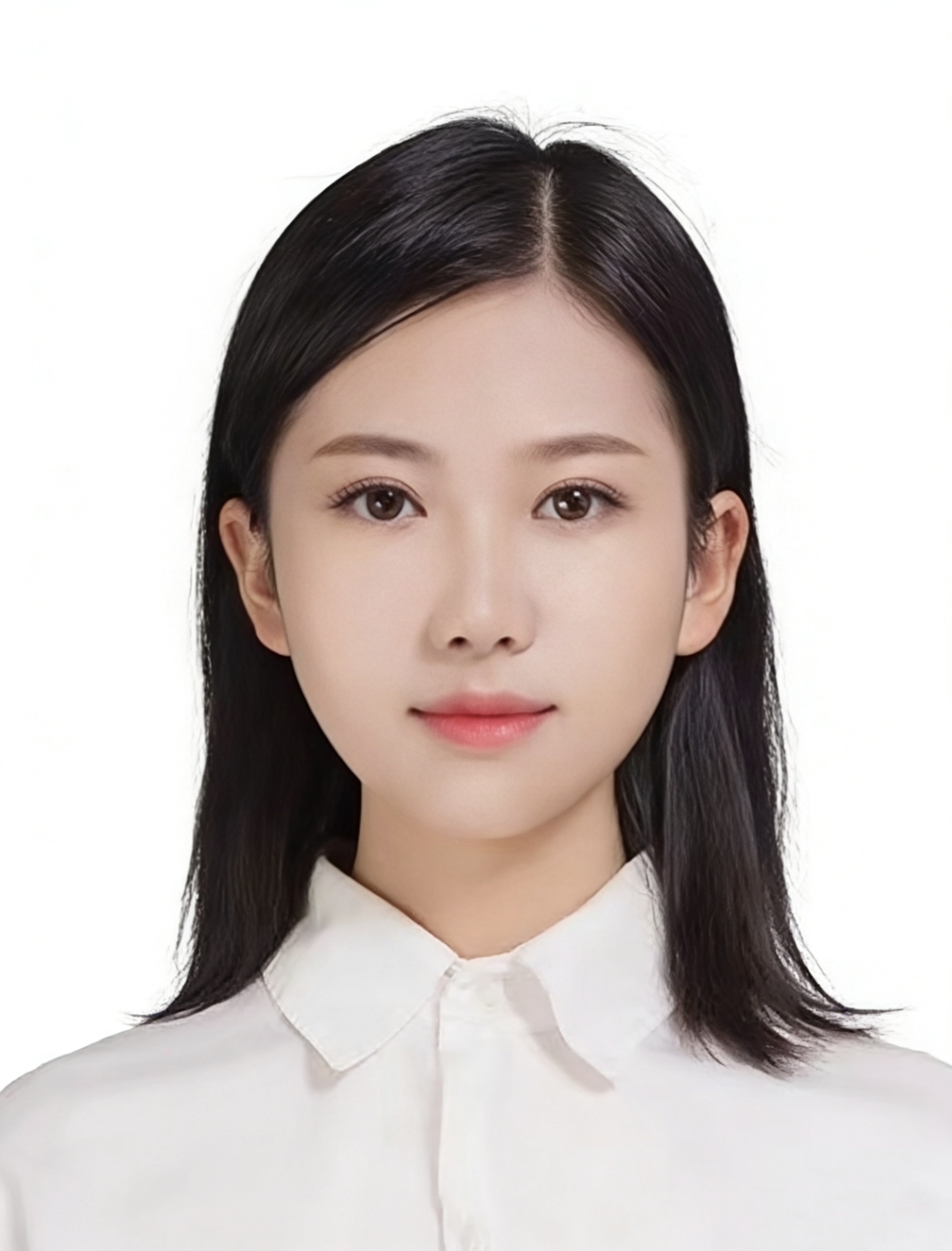}}]{Sixu Liu} received the B.Sc. degree in electronic and information engineering from the Huazhong University of Science and Technology, Wuhan, China, in 2025, where she is currently pursuing the M.E. degree in information and communications engineering. Her current research interests include channel estimation, signal processing, reconfigurable intelligent surfaces, codebook design.
\end{IEEEbiography}

\begin{IEEEbiography}[{\includegraphics[width=1in,clip]{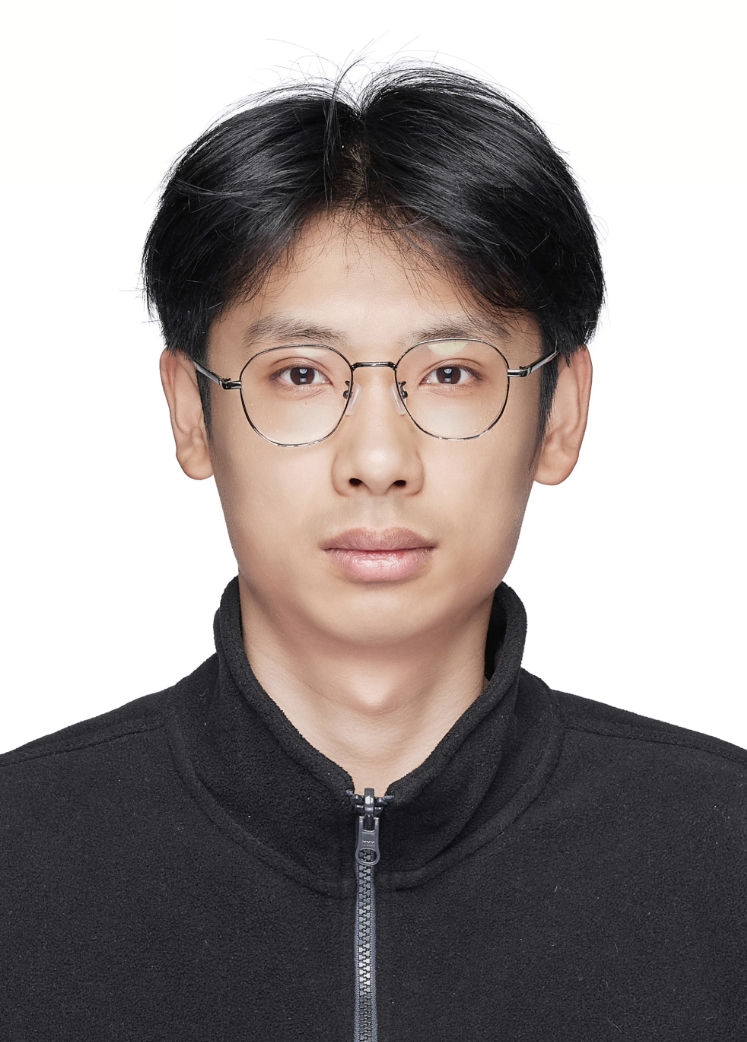}}]{Hao Deng} received the B.S. degree in communication engineering from the University of Electronic Science and Technology of China in 2024, where he is currently working toward the M.S. degree in electronic information. His research interests include 5G NR, beamforming, codebook design and terahertz spectrum situational awareness.
\end{IEEEbiography}

\begin{IEEEbiography}[{\includegraphics[width=1in,clip]{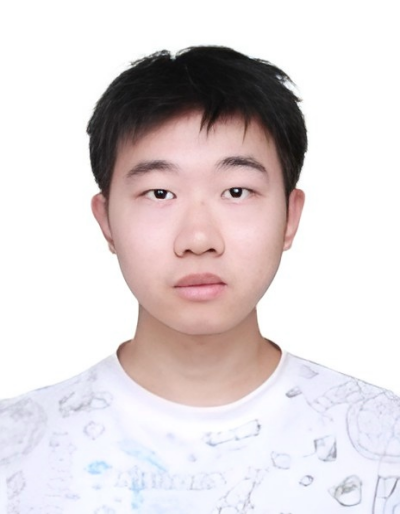}}]{Songjie Yang} received the B.E. degree in Communication Engineering from Shanghai University in 2020. He is currently pursuing his Ph.D. degree with the National Key Laboratory of Wireless Communications, University of Electronic Science and Technology of China, and he was also a visiting Ph.D. student under a joint training program at Singapore University of Technology and Design and Nanyang Technological University, Singapore. He served as a TPC Member for the IEEE International Conference on Communications, IEEE Wireless Communications and Networks, and IEEE Global Communications Conference. His research interests include reconfigurable MIMO, near-field extremely large-scale antenna array communications, integrated sensing and communications, and sparse signal processing.
\end{IEEEbiography}

\begin{IEEEbiography}[{\includegraphics[width=1in,clip]{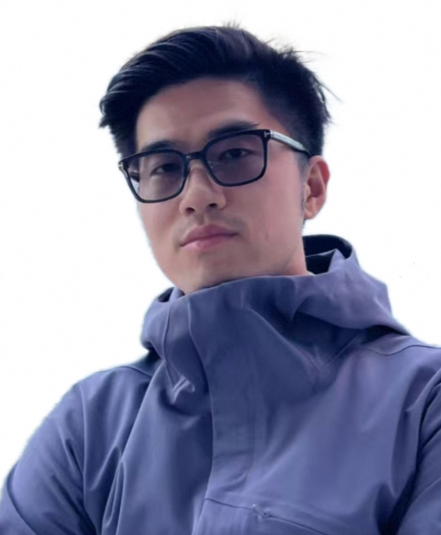}}]{Yuchen Zhang} (Member, IEEE) received the B.E. and Ph.D. degrees in communication engineering from the University of Electronic Science and Technology of China in 2018 and 2024, respectively. His Ph.D. research was supervised by Prof. Wanbin Tang, Head of the National Key Laboratory of Wireless Communications. From 2022 to 2023, he was a visiting Ph.D. student at the Weizmann Institute of Science, Israel, under the supervision of Prof. Yonina C. Eldar. He joined King Abdullah University of Science and Technology (KAUST), Saudi Arabia, in 2024 as a Postdoctoral Researcher with Prof. Tareq Y. Al-Naffouri, and was elevated to Postdoctoral Global Fellow in 2025 under the KAUST Global Fellowship program. His current research interests focus on non-terrestrial networks, particularly low-Earth-orbit satellite systems, and reconfigurable antennas for 6G-and-beyond communications, positioning, and sensing. 
\end{IEEEbiography}

\begin{IEEEbiography}[{\includegraphics[width=1in,clip]{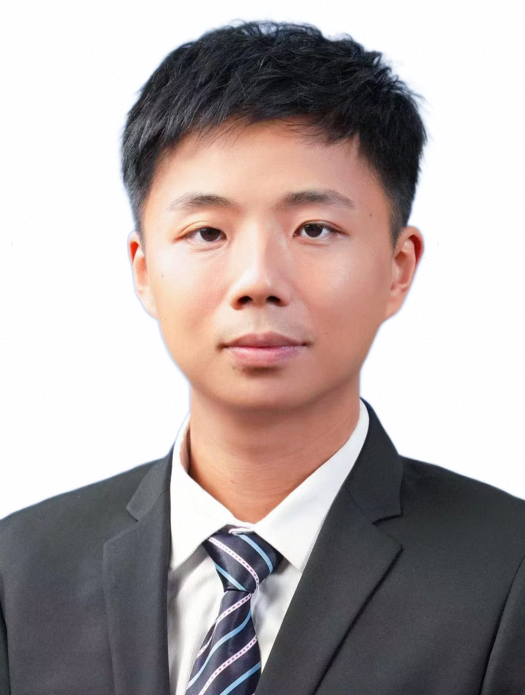}}]{Weidong Mei} (Member, IEEE) received the B.Eng. degree in communication engineering and the M.Eng. degree in communication and information systems from the University of Electronic Science and Technology of China, Chengdu, China, in 2014 and 2017, respectively, and the Ph.D. degree from the NUS Graduate School, National University of Singapore, in 2021 under the Integrative Sciences and Engineering Programme (ISEP) Scholarship. 

He was a Research Fellow with the Department of Electrical and Computer Engineering, National University of Singapore, from July 2021 to January 2023. He is currently a Professor with the University of Electronic Science and Technology of China. His research interests include reconfigurable MIMO, intelligent reflecting surface, wireless drone communications, and convex optimization techniques. 

Dr. Mei has been listed in World's Top 2\% Scientists by Stanford University since 2021. He was the recipients of the Best Paper Award from the IEEE International Conference on Communications in 2021, the Best Student Paper Award from the International Conference on Future Communications and Networks in 2025, and the Outstanding Master's Thesis Award from the Chinese Institute of Electronics in 2017. He mentored his students to win the IEEE ComSoc SPCC Technical Committee Student Challenge and Video Contest Award in 2024. He was honored as the Exemplary Editor of the \textsc{IEEE Open Journal of the Communications Society} in 2024, the Exemplary Reviewer of the \textsc{IEEE Transactions on Communications} in 2019 and 2020, the \textsc{IEEE Open Journal of the Communications Society} in 2021 and 2024, the \textsc{IEEE Wireless Communications Letters} in 2019, 2021, 2022, and 2023, the \textsc{IEEE Communications Letters} in 2021 and 2022, and the {\it SCIENCE CHINA Information Sciences} in 2024. He serves as an Associate Editor for the \textsc{IEEE Transactions on Communications}, \textsc{IEEE Wireless Communications Letters}, and \textsc{IEEE Open Journal of the Communications Society}, and as a Co-Chair of the Workshop on {\it Intelligent Movable and Reconfigurable Antennas for Future Wireless Communication and Sensing} in IEEE Globecom from 2024 to 2025 and IEEE ICC from 2025 to 2026.
\end{IEEEbiography}

\begin{IEEEbiography}[{\includegraphics[width=1in,clip]{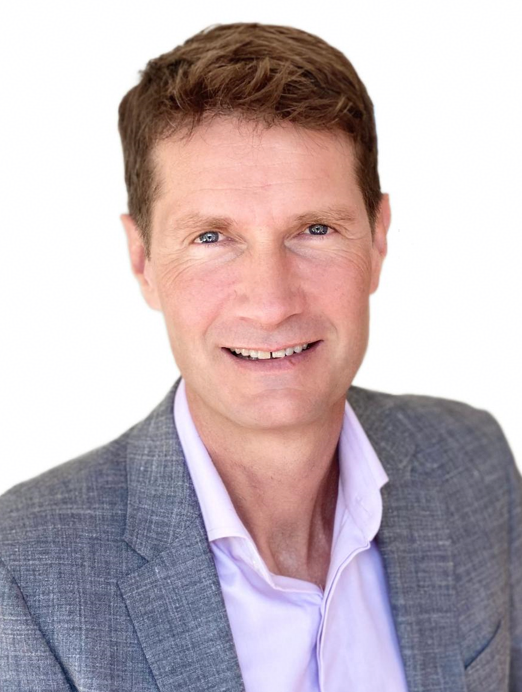}}] {David Gesbert} (Fellow, IEEE)  is serving as Director of EURECOM, Sophia Antipolis, France. He received the Ph.D. degree from TelecomParis, France, in 1997. From 1997 to 1999, he was with the Information Systems Laboratory, Stanford University. He was a founding engineer of Iospan Wireless Inc., a Stanford spin off pioneering MIMO-OFDM (currently Intel). Before joining EURECOM in 2004, he was with the Department of Informatics, University of Oslo. He has published about 350 articles and 25 patents, 7 of them winning  IEEE Best paper awards. He has been the Technical Program Co-Chair for ICC 2017 and has been named a Thomson-Reuters Highly Cited Researchers in computer science.  He is a Board Member for the OpenAirInterface (OAI) Software Alliance. He was a previous awardee of an ERC Advanced Grant in the area of future networks. In 2020, he was also awarded funding by the French Interdisciplinary Institute on Artificial Intelligence for a Chair in the area of AI for the future IoT. In 2021, he received the Grand Prix in Research jointly from IMT and the French Academy of Sciences.
\end{IEEEbiography}

\begin{IEEEbiography}[{\includegraphics[width=1in,clip]{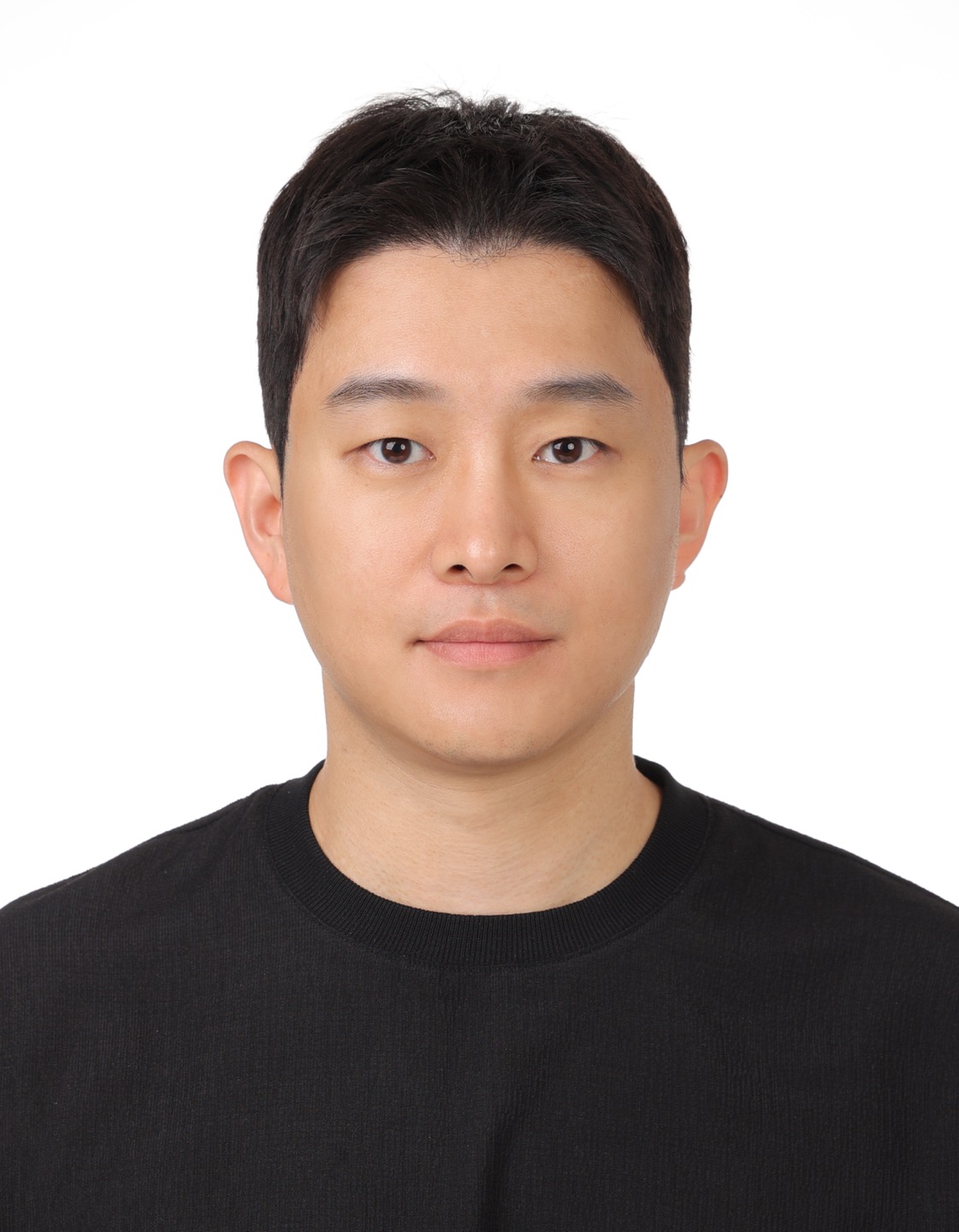}}] {Jaebum Park} (Student Member, IEEE) received the B.S. degree in Electrical and Electronic Engineering from Yonsei University, Seoul, South Korea in 2015 and the M.S. degree in Electrical and Electronic Engineering from the Korea Advanced Institute of Science and Technology (KAIST), Daejeon, South Korea in 2017. From 2017 to 2024, he worked in the 6G Research Team at Samsung Research, Seoul, South Korea. He is currently pursuing the Ph.D. degree with the Department of Electrical and Computer Engineering, University of California at San Diego, CA, USA. His research interests include advanced MIMO transceiver algorithms, energy-efficient MIMO techniques for next-generation RAN architectures, and wireless communications in upper-midband frequency bands.  
\end{IEEEbiography}

\begin{IEEEbiography}[{\includegraphics[width=1in,clip]{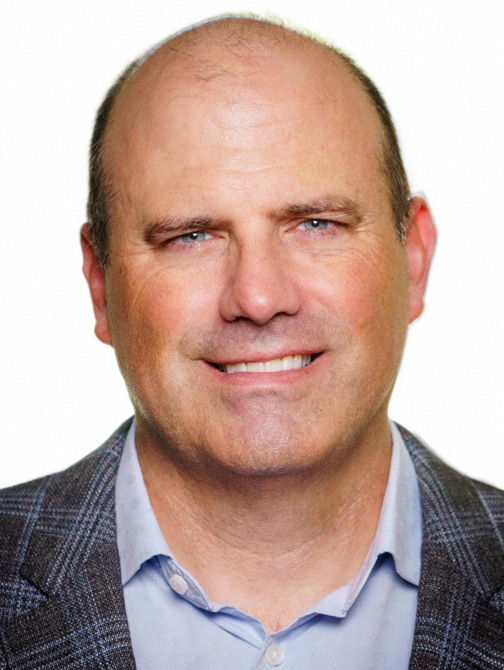}}] {Robert W. Heath Jr.} (S'96 - M'01 - SM'06 - F’11) is the Charles Lee Powell Chair in Wireless Communications in the Department of Electrical and Computer Engineering at the University of California, San Diego. He is also President and CEO of MIMO Wireless Inc. From 2020-2023 he was the Lampe Distinguished Professor at North Carolina State University and co-founder of 6GNC. From 2002-2020 he was with The University of Texas at Austin, most recently as Cockrell Family Regents Chair in Engineering and Director of UT SAVES.  He authored ``Introduction to Wireless Digital Communication'' (Prentice Hall, 2017) and ``Digital Wireless Communication: Physical Layer Exploration Lab Using the NI USRP'' (National Technology and Science Press, 2012), and co-authored ``Millimeter Wave Wireless Communications'' (Prentice Hall, 2014) and ``Foundations of MIMO Communication'' (Cambridge University Press, 2018).
Dr. Heath has been a co-author of a number award winning conference and journal papers including recently the 2017 Marconi Prize Paper Award,  the 2019 IEEE Communications Society Stephen O. Rice Prize, the 2020 IEEE Signal Processing Society Donald G. Fink Overview Paper Award, the 2021 IEEE Vehicular Technology Society Neal Shepherd Memorial Best Propagation Paper Award, and the 2022 IEEE Vehicular Technology Society Best Vehicular Electronics Paper Award. Other notable awards include the 2017 EURASIP Technical Achievement award, the 2019 IEEE Kiyo Tomiyasu Award, the 2021 IEEE Vehicular Technology Society James Evans Avant Garde Award, and the 2025 IEEE/RSE James Clerk Maxwell Medal.  In 2017, he was selected as a Fellow of the National Academy of Inventors. In 2024, he was selected as a Fellow of the American Association for the Advancement of Science. He was a member-at-large on the IEEE Communications Society Board-of-Governors (2020-2022) and the IEEE Signal Processing Society Board-of-Governors (2016-2018). He was Editor-in-Chief of IEEE Signal Processing Magazine from 2018-2020. He is also a licensed Amateur Radio Operator, a Private Pilot, a registered Professional Engineer in Texas. He is an elected member of the National Academy of Engineers.
\end{IEEEbiography}

\begin{IEEEbiography}[{\includegraphics[width=1in,clip]{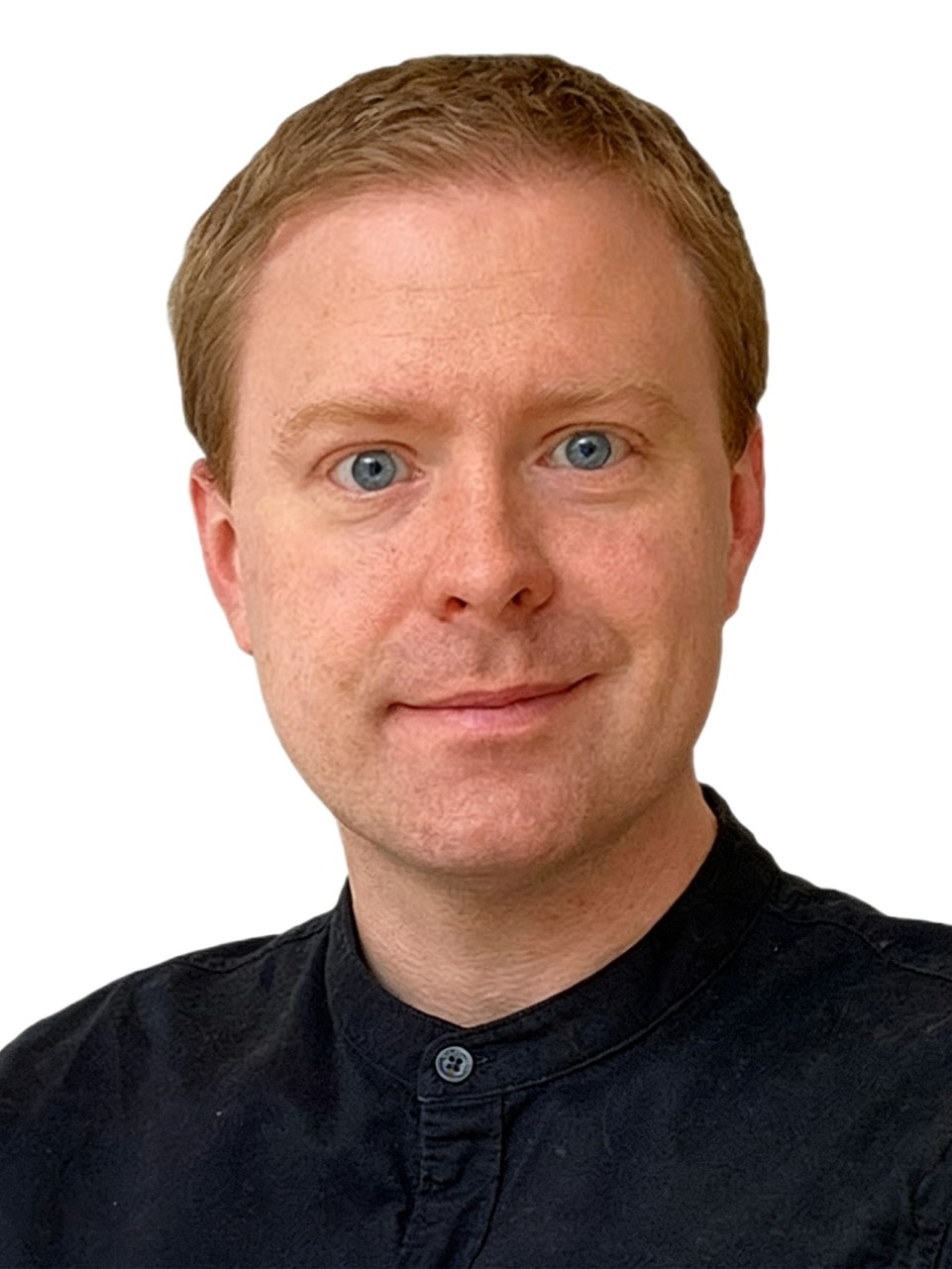}}] {Emil Bj\"ornson} (Fellow, IEEE) received the M.S. degree in engineering mathematics from Lund University, Sweden, in 2007, and the Ph.D. degree in telecommunications from the KTH Royal Institute of Technology, Sweden, in 2011.

From 2012 to 2014, he was a Post-Doctoral Researcher with the Alcatel-Lucent Chair on Flexible Radio, SUPELEC, France. From 2014 to 2021, he held different professor positions at Link\"oping University, Sweden. He has been a Full Professor of Wireless Communication at KTH since 2020 and the Head of the Communication Systems division since 2024. He has authored the textbooks \emph{Optimal Resource Allocation in Coordinated Multi-Cell Systems} (2013), \emph{Massive MIMO Networks: Spectral, Energy, and Hardware Efficiency} (2017), \emph{Foundations of User-Centric Cell-Free Massive MIMO} (2021), and \emph{Introduction to Multiple Antenna Communications and Reconfigurable Surfaces} (2024). He is dedicated to reproducible research and has published much simulation code. He researches multi-antenna communications, reconfigurable intelligent surfaces, radio resource allocation, machine learning for communications, and energy efficiency.

Dr. Bj\"ornson has performed MIMO research since 2006. His papers have received more than 40000 citations, he has filed more than 30 patent applications, and he is recognized as a Clarivate Highly Cited Researcher. He co-hosts the podcast Wireless Future and has a popular YouTube channel with the same name. He is a Wallenberg Academy Fellow, a Digital Futures Fellow, and an SSF Future Research Leader. He has received the 2014 Outstanding Young Researcher Award from IEEE ComSoc EMEA, the 2015 Ingvar Carlsson Award, the 2016 Best Ph.D. Award from EURASIP, the 2018 and 2022 IEEE Marconi Prize Paper Awards in Wireless Communications, the 2019 EURASIP Early Career Award, the 2019 IEEE ComSoc Fred W. Ellersick Prize, the 2019 IEEE Signal Processing Magazine Best Column Award, the 2020 Pierre-Simon Laplace Early Career Technical Achievement Award, the 2020 CTTC Early Achievement Award, the 2021 IEEE ComSoc RCC Early Achievement Award, the 2023 IEEE ComSoc Outstanding Paper Award, and the 2024 IEEE ComSoc Stephen O. Rice Prize. He also coauthored papers that received best paper awards at five conferences.
\end{IEEEbiography}

%\vfill
\end{document}